\documentclass{aa}

\usepackage{graphicx,subfigure}
\usepackage{pstricks}
\usepackage{txfonts}

  \usepackage{natbib}
  \bibpunct{[}{]}{;}{a}{}{,}

\graphicspath{{figeps/},{figeps/shapes/}}

\newcommand{\tbd}[1]{#1}

\usepackage{hyperref}
\hypersetup{
    unicode=false,          
    pdftoolbar=true,        
    pdfmenubar=true,        
    pdffitwindow=false,     
    pdftitle={Physical properties of (21) Lutetia},
    pdfauthor={Benoit Carry},
    pdfsubject={Planetary Science},
    pdfkeywords={},         
    pdfnewwindow=true,      
    colorlinks=true,        
    linkcolor=gray,         
    citecolor=blue,         
    filecolor=gray,         
    urlcolor=gray           
}

\begin{document}
  \title{Physical properties of ESA Rosetta target asteroid (21)~Lutetia:
    Shape and flyby geometry\thanks{Based on observations
      collected at
      the W. M. Keck Observatory and at
      European Southern Observatory
      Very Large Telescope
      (program ID:
      \href{http://archive.eso.org/wdb/wdb/eso/eso_archive_main/query?prog_id=079.C-0493\%28A\%29\&max_rows_returned=1000}{079.C-0493},
      PI: E.~Dotto).
      The W. M. Keck Observatory is operated as a scientific
      partnership among the California Institute of Technology, the
      University of California, and the National Aeronautics and Space
      Administration. The Observatory
      was made possible by the generous financial support of the W. M. Keck
      Foundation.}
  }

  \author{%
    Beno\^{i}t Carry\inst{1,2}
    \and
    Mikko Kaasalainen\inst{3}
    \and
    C\'edric Leyrat\inst{1}
    \and
    Willam J. Merline\inst{4}
    \and
    Jack D. Drummond\inst{5}
    \and
    Al Conrad\inst{6}
    \and
    Harold A. Weaver\inst{7}
    \and
    Peter M. Tamblyn\inst{4}
    \and
    Clark R. Chapman\inst{4}
    \and
    Christophe Dumas\inst{8}
    \and
    Fran\c{c}ois Colas\inst{9}
    \and
    Julian C. Christou\inst{10}
    \and
    Elisabetta Dotto\inst{11}
    \and
    Davide Perna\inst{1,11,12}
    \and
    Sonia Fornasier\inst{1,2}
    \and
    Laurent Bernasconi\inst{13}
    \and
    Raoul Behrend\inst{14}
    \and
    Fr\'ed\'eric Vachier\inst{9}
    \and
    Agnieszka Kryszczynska\inst{15}
    \and
    Magdalena Polinska\inst{15}
    \and
    Marcello Fulchignoni\inst{1,2}
    \and
    Ren\'e Roy\inst{16}
    \and
    Ramon Naves\inst{17}
    \and
    Raymond Poncy\inst{18}
    \and
    Patrick Wiggins\inst{19}
  }

  \offprints{B. Carry: benoit.carry@obspm.fr}

  \institute{%
    LESIA, Observatoire de Paris,
    5 place Jules Janssen,
    92190 Meudon, France.
    \email{benoit.carry@obspm.fr}
    \and
    Universit\'e Paris 7 Denis-Diderot,
    5 rue Thomas Mann,
    75205 Paris CEDEX, France.
    \and
    Tampere University of Technology, P.O. Box 553, 33101 Tampere, Finland.
    \and
    Southwest Research Institute, 1050 Walnut St. \#300, Boulder, CO
    80302, USA.
    \and
    Starfire Optical Range, Directed Energy Directorate, Air Force
    Research Laboratory, Kirtland AFB, NM 87117-577, USA.
    \and
    W.M. Keck Observatory, 65-1120 Mamalahoa Highway, Kamuela, HI
    96743, USA.
    \and
    Johns Hopkins University Applied Physics Laboratory, Laurel, MD 20723-6099, USA.
    \and
    European Southern Observatory, Alonso de C\'ordova 3107, Vitacura,
    Casilla 19001, Santiago de Chile, Chile.
    \and
    IMCCE, Observatoire de Paris, 14 bvd de l'Observatoire, 75014 Paris, France.
    \and
    Gemini Observatory, Northern Operations Center, 670 N. A'ohoku
    Place, Hilo, HI, 96720, USA.
    \and
    INAF, Osservatorio Astronomico di Roma, via Frascati 33, 00040
    Monteporzio Catone (Roma), Italy.
    \and
    Dipartimento di Fisica, Universit\`a di Roma Tor Vergata, Via della
    Ricerca Scientifica 1, 00133 Roma, Italy.
    \and
    Les Engarouines Observatory, 84570 Mallemort-du-Comtat, France.
    \and
    Geneva Observatory, 1290 Sauverny, Switzerland.
    \and
    Astronomical Observatory, Adam Mickiewicz University, Sloneczna 36, 60-286 Poznan, Poland.
    \and
    Blauvac Observatory, 84570 St-Est\`{e}ve, France.
    \and
    Observatorio Montcabre, C/Jaume Balmes 24, 08348 Cabrils,
    Barcelona, Spain.
    \and
    Le Cres Observatory, 2 Rue des \'Ecoles, 34920 le Cres, France.
    \and
    Wiggins Observatory, 472 Country Club, Tooele Utah 84074, USA.
  }

  \date{Received \tbd{Month dd, yyyy}; accepted \tbd{Month dd, yyyy}}


  \abstract
      {}
      {We determine the physical properties (spin state and 
        shape) of asteroid (21) Lutetia,
        target of the ESA Rosetta mission, to help in preparing for
        observations during the flyby on 2010 July 10
        by predicting the orientation of Lutetia as seen from
        Rosetta.} 
      {We use 
        our novel KOALA inversion algorithm to 
        determine the physical properties
        of asteroids from a combination of optical lightcurves,
        disk-resolved images, and stellar occultations, although the
        latter are not available for (21) Lutetia.}
      {We find the spin axis of (21) Lutetia to lie within 5\degr~of
        ($\lambda = 52$\degr,
        $\beta = -6$\degr) in
        Ecliptic J2000 reference frame
        (equatorial
        $\alpha = 52$\degr,
        $\delta = +12$\degr),
        and determine an improved 
        sidereal period of 8.168\,270 $\pm$ 0.000\,001 h.
        This pole solution implies the southern hemisphere of 
        Lutetia will be in
        ``\textsl{seasonal}''
        shadow at the time of the flyby.
        The apparent cross-section
        of Lutetia is triangular 
        as seen ``\textsl{pole-on}'' and
        more rectangular as seen ``\textsl{equator-on}''.
        The best-fit model suggests the presence of 
        several concavities. The largest of these
        is close to the north pole and may be
        associated with large impacts.
      }
      {}


   \maketitle
%

\section{Introduction}
  \indent The origin and evolution of the Solar
  System and its implications for early planetesimal formation
  are key questions in planetary science.
  Unlike terrestrial planets, which have experienced significant
  mineralogical evolution, through endogenic activity, since their accretion,
  small Solar System bodies have remained essentially unaltered.
  Thus, a considerable amount of
  information regarding the primordial planetary
  processes that occurred during and immediately after the accretion of
  the early planetesimals is still present
  among this population.
  Consequently, studying asteroids is of prime
  importance in understanding the planetary formation processes
  \citep{2002-AsteroidsIII-1-Bottke} and, first and
  foremost, requires a reliable knowledge of their physical properties
  (size, shape, spin, mass, density, internal
  structure, etc.) in addition to their compositions and dynamics.
  Statistical analyses of these parameters for a wide range of
  asteroids can provide relevant
  information about inter-relationships and formation scenarios. \\
  \indent In this respect, our observing program with adaptive
  optics, allowing
  diffraction-limited observations from the ground with 10 m-class
  telescopes, has now broken the barrier
  which separated asteroids from real planetary worlds
  \citep[\textsl{e.g.,}][]{2007-Icarus-191-Conrad, 2008-AA-478-Carry,
      2009-Icarus-202-Drummond, 2010-Icarus-205-Carry-a, 2010-AA--Drummond}.
  Their shapes,
  topography, sizes, spins, surface
  features, albedos, and color variations can now be directly observed
  from the ground.
  This opens these objects to
  geological, rather than astronomical-only,
  study. While such surface detail is only possible for the largest
  asteroids, our main focus is on
  determining accurately the size, shape, and pole.
  Among them, we have observed (21) Lutetia, an asteroid that will
  be observed in-situ by the ESA Rosetta mission.\\
  \indent The Rosetta Mission will encounter its principal target, the
  comet 67P/Churyumov-Gerasimenko, in 2014.
  However, its interplanetary journey was designed to allow close
  encounters with two main-belt asteroids: (2867) \v{S}teins and (21)
  Lutetia.
  The small asteroid (2867) \v{S}teins was
  visited on 2008 September 5
  at a minimum distance of about 800 km
  \citep{2009-DPS-40-Schulz} and
  (21) Lutetia will be encountered on 2010 July 10.
  Knowing the geometry of the flyby
  (\textsl{e.g.}, visible hemisphere, sub-spacecraft
  coordinates as function of time, and distance)
  before the encounter is
  crucial to
  optimize the observation sequence
  and schedule the on-board operations.
  The size of Lutetia
  \citep[estimated at $\sim$100 km, see][]{
    2002-AJ-123-Tedesco,
    PDSSBN-Iras,
    2006-AA-447-Mueller}
  allows its apparent disk to be spatially resolved from Earth.
  Our goal is therefore to
  improve knowledge of its physical properties to
  prepare for the spacecraft flyby. \\
  \indent Lutetia, the Latin name for the city of Paris, is a
  main-belt asteroid (semi-major axis 2.44 AU) that has been
  studied extensively from the ground
  \citep[see][for a review, primarily of recent observations]{2007-SSRv-128-Barucci}.
  Numerous studies have estimated indirectly
  its spin
  \citep[by lightcurve, \textsl{e.g.,}][]{
    1987-KFNT-3-Lupishko-a,
    1992-AA-95-Dotto,
    2003-Icarus-164-Torppa}.
  Size and albedo
  were reasonably well determined in the 1970s by
  \citet{1977-Icarus-31-Morrison} using
  thermal radiometry (108--109 km), and by
  \citet{1976-AJ-81-Zellner}
  using polarimetry (110 km).
  Five somewhat scattered IRAS scans
  \citep[\textsl{e.g.,}][]{2002-AJ-123-Tedesco,
    PDSSBN-Iras}
  yielded a higher albedo and
  smaller size than the dedicated observations in the 1970s.
  \citet{2006-AA-447-Mueller}
  derived results from new radiometry that are roughly compatible with
  the earlier results or with the IRAS results, depending on which
  thermal model is used.
  \citet{2008-AA-479-Carvano} later derived a lower albedo from 
  ground-based observations, seemingly incompatible with previous works.
  Radar data analyzed by
  \citet{1999-Icarus-140-Magri, 2007-Icarus-186-Magri}
  yielded an effective diameter for Lutetia of 116 km;
  reinterpretation of those data and new radar observations
  \citep{2008-Icarus-195-Shepard}
  suggest an effective diameter of 100 $\pm$ 11 km and an
  associated visual albedo of 0.20.  
  Recent HST observations of Lutetia \citep{2009-arXiv-Weaver}
  indicate a visual albedo of about 16\%, a result based
  partly on the size/shape/pole determinations from our work in
  the present paper and \citet{2010-AA--Drummond}.\\
  \indent Lutetia has been extensively studied using spectroscopy in the
  visible, near- and mid-infrared and its albedo measured by polarimetry
  and thermal radiometry
  \citep{1975-ApJ-195-McCord,
    1975-Icarus-25-Chapman, 
    1976-AJ-81-Zellner,
    1978-Icarus-35-Bowell,
    1995-Icarus-117-Rivkin, 
    1999-Icarus-140-Magri,
    2000-Icarus-145-Rivkin,
    2004-AA-425-Lazzarin,
    2005-AA-430-Barucci,
    2006-AA-454-Birlan,
    2007-AA-470-Nedelcu,
    2008-AA-477-Barucci,
    2008-Icarus-195-Shepard, 
    2009-Icarus-202-DeMeo,
    2009-Icarus-202-Vernazza,
    2009-AA-498-Lazzarin,
    lazzarin-2010,
    2010-AA-513-Perna, 
    2010-AA-Belskaya}.
  We present a discussion on Lutetia's taxonomy and composition in a
  companion paper \citep{2010-AA--Drummond}.\\
  \indent Thermal infrared observations used to determine the
  size and albedo of  Lutetia were initially inconsistent, with
  discrepancies in diameters and visible albedos reported
  \citep[\textsl{e.g.,}][]{%
    1976-AJ-81-Zellner,
    1989-Icarus-78-Lupishko,
    1996-PSS-44-Belskaya,
    2002-AJ-123-Tedesco,
    2006-AA-447-Mueller,
    2008-AA-479-Carvano}.
  \citet{2006-AA-447-Mueller} and
  \citet{2008-AA-479-Carvano}, however,
  interpreted these variations as an indication of surface
  heterogeneity, inferring that the terrain
  roughness of Lutetia increased toward northern
  latitudes\footnote{our use of ``northern hemisphere'' refers to 
    the hemisphere in the direction of the positive pole as defined by
    the right-hand rule 
    from IAU recommendations \citep{2007-CeMDA-98-Seidelmann}}, that
  the crater distribution is different over the northern/southern 
  hemispheres, and includes a possibility of one or several large
  craters in Lutetia's northern hemisphere.
  Indeed, the \textsl{convex} shape model derived from the inversion
  of 32 optical lightcurves 
  \citep{2003-Icarus-164-Torppa} displays
  a flat top near the north pole of Lutetia.
  \citet{2002-AA-383-Kaasalainen} have shown that large flat
  regions in these convex models 
  could be a site of concavities.
  The southern hemisphere is not expected to be free from craters however,
  as \citet{2010-AA-513-Perna}
  detected a slight variation of the visible spectral 
  slope, possibly due to the presence of large 
  craters or albedo spots in the 
  southern hemisphere.\\
  \indent In this paper, we present simultaneous analysis of
  adaptive-optics images obtained at the W. M. Keck and
  the European Southern Observatory (ESO)
  Very Large Telescope (VLT) observatories,
  together with lightcurves, and
  we determine the shape and spin state of Lutetia.
  In section~\ref{sec: obs}, we present the observations,
  in section~\ref{sec: shape} the shape of
  Lutetia, and finally, we describe the geometry of the upcoming
  Rosetta flyby in section~\ref{sec: flyby}.

%
%
%
\section{Observations and data processing\label{sec: obs}}
  \subsection{Disk-resolved imaging observations}
    \indent We have obtained high angular-resolution
    images of the apparent disk of (21) Lutetia
    over six nights during its last opposition (late 2008 - early
    2009)
    at the W. M. Keck Observatory
    with the
    adaptive-optics-fed
    NIRC2 camera
    \citep[$9.942 \pm 0.050$ milli-arcsecond per pixel,][]{2004-AppOpt-43-vanDam}.
    We also obtained
    data in 
    2007\footnote{program ID:
      \href{http://archive.eso.org/wdb/wdb/eso/eso_archive_main/query?prog_id=079.C-0493\%28A\%29\&max_rows_returned=1000}{079.C-0493}}
    \citep{2007-DPS-39-Perna}
    at the ESO VLT
    with the adaptive-optics-assisted
    NACO camera
    \citep[$13.27 \pm 0.050$ milli-arcsecond per pixel,
    ][]{2003-SPIE-4839-Rousset,2003-SPIE-4841-Lenzen}.
    We list observational circumstances for all epochs in
    Table~\ref{tab: obs}.
    Although the AO data used here are the same as in 
    \citet{2010-AA--Drummond},
    we analyze them with an independent approach.
    We do not use our 2000 epoch, however, from Keck
    (NIRSPEC instrument) because 
    those data were taken for the purpose of a search for satellites
    and therefore the
    Point-Spread Function (PSF) calibrations were not adequate for shape recovery.\\

\begin{table*}
\caption[Observing log]{%
  Heliocentric distance ($r$),
  range to observer ($\Delta$),
  solar phase angle ($\alpha$),
  apparent visual magnitude (m$_V$),
  angular diameter ($\phi$),
  coordinates (longitude $\lambda$ and latitude $\beta$)
  of the
  Sub-Earth Point (SEP) and
  Sub-Solar Point (SSP), for each epoch 
  (mean time listed in UT, without light-time correction).
  All the data were obtained at W. M. Keck observatory, except the 2007
  epochs, which were obtained at ESO Very Large Telescope.
}
\label{tab: obs}
\centering
\begin{tabular}{cccccccccc}
\hline\hline
  Date  & $r$ & $\Delta$ & $\alpha$ & m$_V$ & $\phi$  & SEP$_\lambda$
  & SEP$_\varphi$ & SSP$_\lambda$ & SSP$_\varphi$ \\
  (UT) & (AU) & (AU) & (\degr) & (mag) & (\arcsec) & (\degr) & (\degr)  & (\degr) & (\degr) \\
\hline
2007-06-06T00:19 & 2.30 & 1.30 &  3.2 & 10.1 & 0.14 & 339 &  73 & 337 &  70 \\%
2007-06-06T02:56 & 2.30 & 1.30 &  3.2 & 10.1 & 0.14 & 223 &  73 & 221 &  70 \\%
2007-06-06T06:45 & 2.30 & 1.30 &  3.3 & 10.1 & 0.14 &  55 &  73 &  53 &  70 \\%
2007-06-06T08:08 & 2.30 & 1.30 &  3.3 & 10.1 & 0.14 & 354 &  73 & 352 &  70 \\%
2007-06-06T08:16 & 2.30 & 1.30 &  3.3 & 10.1 & 0.14 & 348 &  73 & 346 &  70 \\%
2007-06-06T08:22 & 2.30 & 1.30 &  3.3 & 10.1 & 0.14 & 344 &  73 & 342 &  70 \\%
2007-06-06T08:27 & 2.30 & 1.30 &  3.3 & 10.1 & 0.14 & 340 &  73 & 338 &  70 \\%
2008-10-22T15:14 & 2.36 & 1.55 & 17.9 & 11.1 & 0.12 & 267 & -65 & 298 & -82 \\%
2008-10-22T15:20 & 2.36 & 1.55 & 17.9 & 11.1 & 0.12 & 263 & -65 & 294 & -82 \\%
2008-10-22T15:25 & 2.36 & 1.55 & 17.9 & 11.1 & 0.12 & 259 & -65 & 290 & -82 \\%
2008-10-22T15:33 & 2.36 & 1.55 & 17.9 & 11.1 & 0.12 & 253 & -65 & 284 & -82 \\%
2008-11-21T10:39 & 2.41 & 1.43 &  4.7 & 10.5 & 0.13 &  61 & -70 &  68 & -75 \\%
2008-12-02T07:05 & 2.43 & 1.44 &  1.1 & 10.2 & 0.13 & 106 & -73 & 106 & -72 \\%
2008-12-02T07:12 & 2.43 & 1.44 &  1.1 & 10.2 & 0.13 & 100 & -73 & 101 & -72 \\%
2008-12-02T07:29 & 2.43 & 1.44 &  1.1 & 10.2 & 0.13 &  89 & -73 &  89 & -72 \\%
2008-12-02T07:35 & 2.43 & 1.44 &  1.1 & 10.2 & 0.13 &  84 & -73 &  84 & -72 \\%
2008-12-02T07:49 & 2.43 & 1.44 &  1.1 & 10.2 & 0.13 &  74 & -73 &  74 & -72 \\%
2008-12-02T07:54 & 2.43 & 1.44 &  1.1 & 10.2 & 0.13 &  70 & -73 &  70 & -72 \\%
2008-12-02T08:07 & 2.43 & 1.44 &  1.1 & 10.2 & 0.13 &  61 & -73 &  61 & -72 \\%
2008-12-02T08:12 & 2.43 & 1.44 &  1.1 & 10.2 & 0.13 &  57 & -73 &  57 & -72 \\%
2008-12-02T08:18 & 2.43 & 1.44 &  1.1 & 10.2 & 0.13 &  53 & -73 &  53 & -72 \\%
2008-12-02T08:23 & 2.43 & 1.44 &  1.1 & 10.2 & 0.13 &  49 & -73 &  49 & -72 \\%
2008-12-02T08:28 & 2.43 & 1.44 &  1.1 & 10.2 & 0.13 &  45 & -73 &  46 & -72 \\%
2008-12-02T08:34 & 2.43 & 1.44 &  1.1 & 10.2 & 0.13 &  41 & -73 &  41 & -72 \\%
2008-12-02T08:39 & 2.43 & 1.44 &  1.1 & 10.2 & 0.13 &  37 & -73 &  37 & -72 \\%
2008-12-02T08:45 & 2.43 & 1.44 &  1.1 & 10.2 & 0.13 &  33 & -73 &  33 & -72 \\%
2008-12-02T08:50 & 2.43 & 1.44 &  1.1 & 10.2 & 0.13 &  29 & -73 &  29 & -72 \\%
2008-12-02T08:56 & 2.43 & 1.44 &  1.1 & 10.2 & 0.13 &  25 & -73 &  25 & -72 \\%
2008-12-02T09:01 & 2.43 & 1.44 &  1.1 & 10.2 & 0.13 &  21 & -73 &  21 & -72 \\%
2008-12-02T09:07 & 2.43 & 1.44 &  1.1 & 10.2 & 0.13 &  17 & -73 &  17 & -72 \\%
2008-12-02T09:12 & 2.43 & 1.44 &  1.1 & 10.2 & 0.13 &  13 & -73 &  13 & -72 \\%
2009-01-23T06:24 & 2.52 & 1.89 & 20.1 & 11.8 & 0.10 & 232 & -78 & 209 & -59 \\%
2009-01-23T09:17 & 2.52 & 1.90 & 20.1 & 11.8 & 0.10 & 105 & -78 &  82 & -59 \\%
2009-02-02T08:35 & 2.54 & 2.03 & 21.6 & 12.0 & 0.09 & 357 & -77 & 336 & -57 \\%
2009-02-02T08:41 & 2.54 & 2.03 & 21.6 & 12.0 & 0.09 & 352 & -77 & 331 & -57 \\%
2009-02-02T08:45 & 2.54 & 2.03 & 21.6 & 12.0 & 0.09 & 350 & -77 & 328 & -57 \\%
\hline
\end{tabular}
\end{table*}

\begin{figure}
  \centering
  \includegraphics[width=.5\textwidth]{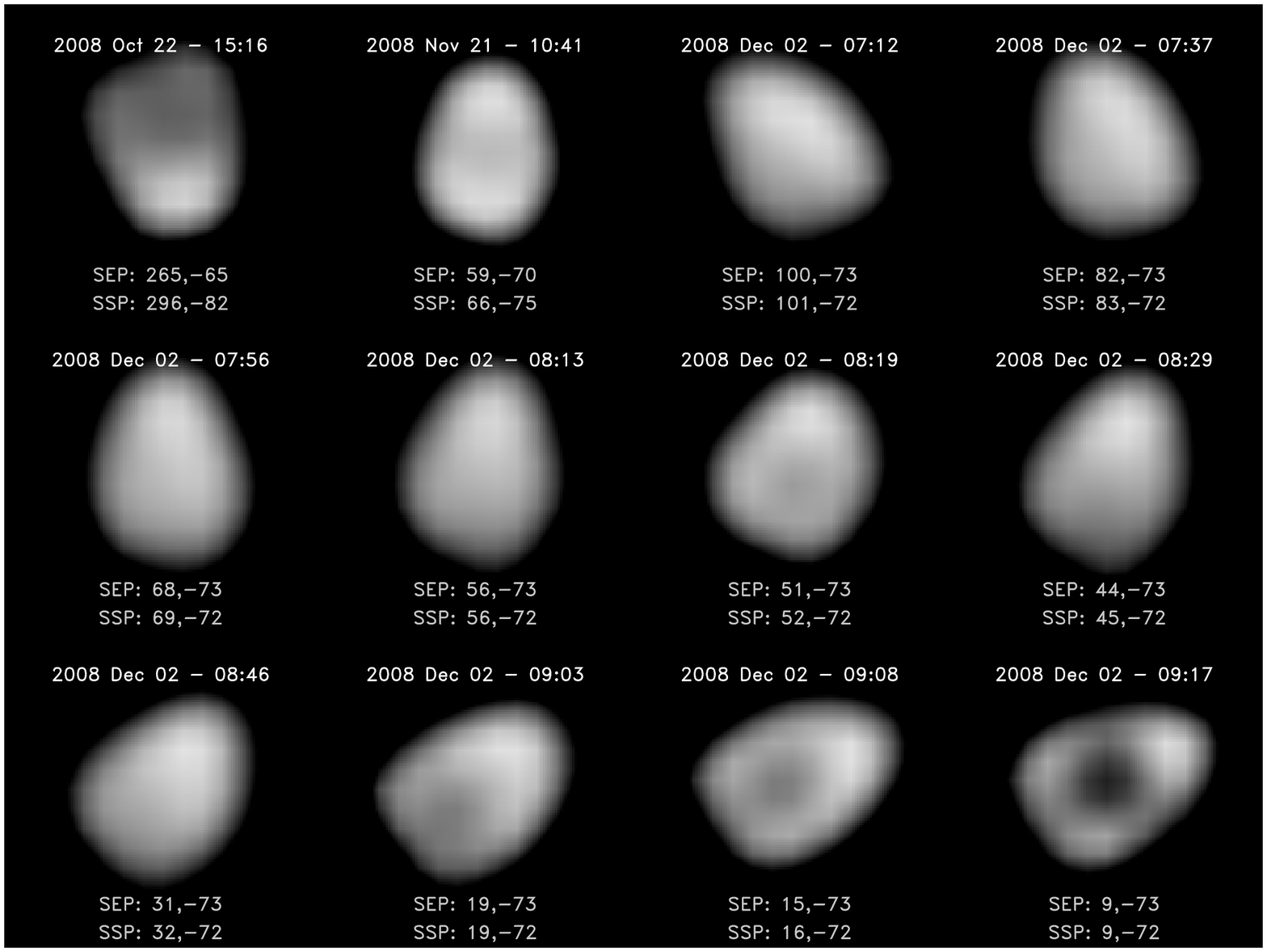}
  \caption[Selected views of (21) Lutetia]{%
    Selected views of (21) Lutetia in the near-infrared.
    All the images have been deconvolved with \textsc{Mistral} 
    to enhance the definition of the edges.
    No effort has been made to restore photometric accuracy over the apparent
    disk and some ringing effects can be seen in the frames. However,
    this does not influence the shape of the derived contours. 
    All images have been scaled to display the
    same apparent size and rotated so that
    the projection of the rotation axis is
    directed toward the top of the page (the
    observer is still viewing from above high
    southern latitudes, \textsl{i.e.}, largely pole-on,
    and so, in 3D, the spin axis is
    mostly directed out of the page).
    Observing time (UT), Sub-Earth Point (SEP), and Sub-Solar Point
    (SSP) coordinates are
    listed on each frame.}
  \label{fig: obs}
\end{figure}

    \indent We reduced the data using usual procedures for
    near-infrared images, including bad pixel removal,
    sky subtraction, and flat-fielding \citep[see][for a more detailed
      description]{2008-AA-478-Carry}.
    We then restored the images to optimal angular-resolution using 
    the \textsc{Mistral} deconvolution algorithm
    \citep{2000-Msn-99-Conan,
      2004-JOSAA-21-Mugnier}.
    The validity of this approach (real-time Adaptive-Optics correction
    followed by \textsl{a posteriori} deconvolution) has already been
    demonstrated elsewhere
    \citep{2002-Icarus-160-Marchis,2006-JGR-111-Witasse}.
    Although PSF observations were not available close in time
    to each Lutetia observation and could lead to a possible bias on
    the apparent size of Lutetia, two lines of evidence provide
    confidence in our results.
    First, we note that the Next-Generation Wave-Front Controller
    \citep[NGWFC,][]{2007-KAON-489-VanDam}
    of NIRC2 provides stable correction and 
    therefore limits such biases.
    Second, the image analysis presented in \citet{2010-AA--Drummond},
    which does not rely on separately measured PSF profiles
    \citep[Parametric Blind Deconvolution, see][]{2000-LGSAO-Drummond},
    confirms our overall size and orientation of Lutetia on the plane
    of the sky at each epoch.
    We are thus confident in the
    large scale features presented by the shape model derived
    below.\\
    \indent In total, we obtained 324 images of 
    (21) Lutetia on 7 nights
    over 2007-2009
    (Table~\ref{tab: obs}).
    A subset of the restored images is presented
    in Fig.~\ref{fig: obs}.

  \subsection{Optical lightcurve observations}
    \indent We utilized all 32 optical lightcurves from
    \citet{2003-Icarus-164-Torppa} to derive the \textsl{convex} shape
    of (21) Lutetia from lightcurve inversion
    \citep{2001-Icarus-153-Kaasalainen-a,2001-Icarus-153-Kaasalainen-b}.
    We present these lightcurves in Fig.~\ref{fig: comp: lc},
    together with
    18
    additional lightcurves
    acquired subsequent to ESA's
    decision to target Lutetia. 
    Some of the new data were taken
    in 2007 January by the 
    OSIRIS camera on-board Rosetta during its interplanetary journey 
    \citep{2009-DPS-40-Faury}.
    Eight lightcurves come from the CDR-CDL group led by 
    Raoul Behrend at the Geneva
    observatory\footnote{\href{http://obswww.unige.ch/~behrend/page_cou.html}
      {http://obswww.unige.ch/~behrend/page\_cou.html}}. 
    The aim of this group is to organize photometric observations
    (including those from many amateurs) for
    selected asteroids and to search for binary objects
    \citep{2006-AA-446-Behrend}.
    The result
    is two full composite lightcurves in 2003 and 2010 covering Lutetia's
    period.
    Six other lightcurves come from the Pic du Midi 1m telescope
    in 2006 \citep{2007-AA-470-Nedelcu}
    and 2009 (new data presented here).
    See Fig.~\ref{fig: comp: lc} for a detailed listing of the observations.
    In total we used
    50
    lightcurves spread over
    years 1962-2010.

  \subsection{The KOALA method}
    \indent We use
    a novel method to derive physical properties of
    asteroids from a combination of disk-resolved images,
    stellar occultation chords and optical
    lightcurves, called KOALA (for Knitted Occultation,
    Adaptive-optics, and Lightcurve Analysis).
    A complete description of the method can be found elsewhere
    \citep{2009-IPI-Kaasalainen}, as well as an
    example
    of its
    application on (2) Pallas
    \citep{2010-Icarus-205-Carry-a}.\\
    \indent We first extracted the contour of (21) Lutetia on each
    image using a Laplacian of Gaussian wavelet
    \citep[see][for more detail about this method]{2008-AA-478-Carry}.
    These contours provide a direct measurement of
    Lutetia's size and
    shape at each epoch (Table~\ref{tab: obs}).
    Stellar occultations also provide similar constraints
    if several chords per event are observed.
    Unfortunately, there are only two archived stellar occultations
    by Lutetia
    \citep{2009-Occultations}
    with only one chord each that do not provide useful constraints.
    The optical lightcurves 
    bring indirect constraints on the shape of Lutetia,
    provided the albedo
    is homogeneous over its surface.
    Indeed, lightcurves are influenced by a
    combination of the asteroid
    shape\footnote{through a surface reflectance law, taken
      here as a combination of the Lommel-Seelinger ($\mathcal{LS}$) and
      Lambert ($\mathcal{L}$) diffusion laws: $0.9 \times \mathcal{LS}
      + 0.1 \times \mathcal{L}$, following
    \citet{2001-Icarus-153-Kaasalainen-a}}
    and albedo variation
    \citep[see the discussion by][regarding the effect of albedo features
      on the shape reconstruction]{2010-Icarus-205-Carry-a}.\\
    \indent Slight spectral heterogeneity has been reported from visible and
    near-infrared spectroscopy
    \citep{2007-AA-470-Nedelcu, 
      2010-AA-513-Perna,
      lazzarin-2010},
    spanning several oppositions and hence Sub-Earth
    Point (SEP) latitudes and longitudes.
    Although \citet{2010-AA-Belskaya} claim that
    Lutetia's surface is highly heterogeneous,
    they indicate that there is no strong evidence
    for large variations in albedo over the surface.
    They argue that the observed level of albedo
    variation is consistent with variations in
    regolith texture or minerology. 
    We therefore assume homogeneously distributed albedo features
    on the surface (valid for variations of small amplitude)
    and that lightcurves are influenced by the shape of
    Lutetia only.\\

\section{Shape and spin of (21) Lutetia\label{sec: shape}}
\begin{figure}
  \centering
  \includegraphics[width=.24\textwidth]{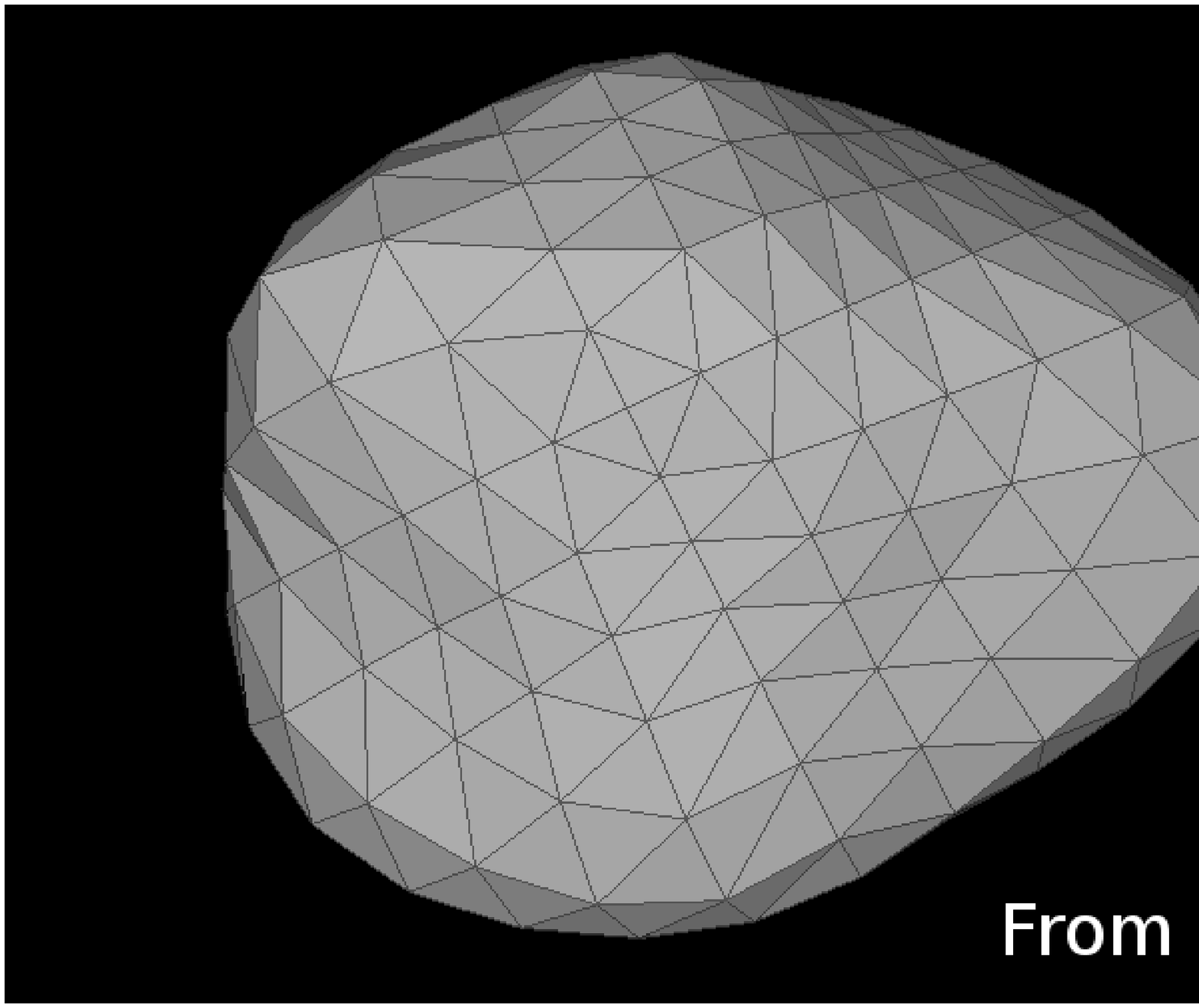}\hspace*{.02\textwidth}%
  \includegraphics[width=.24\textwidth]{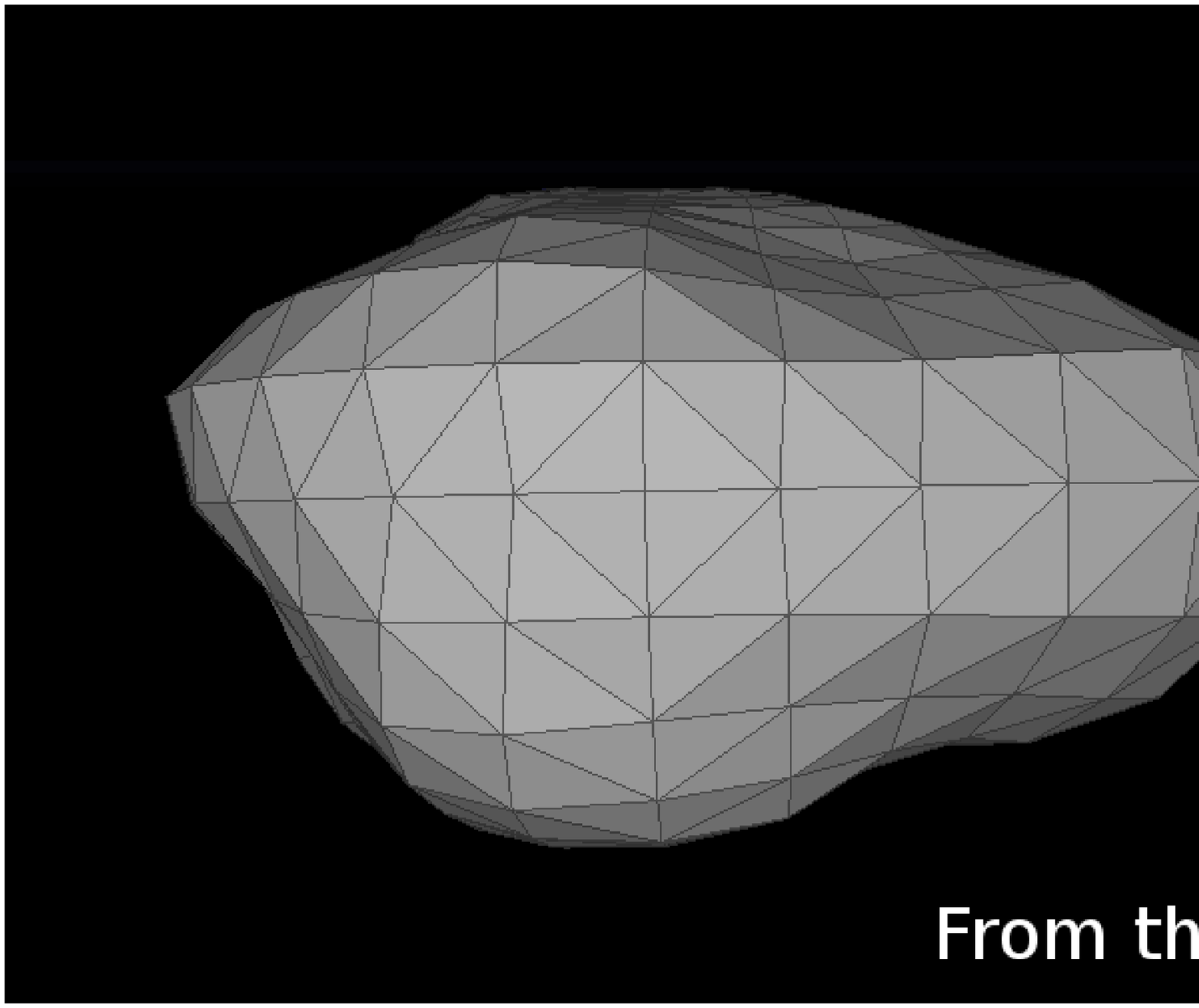}\\
  \includegraphics[width=.5\textwidth]{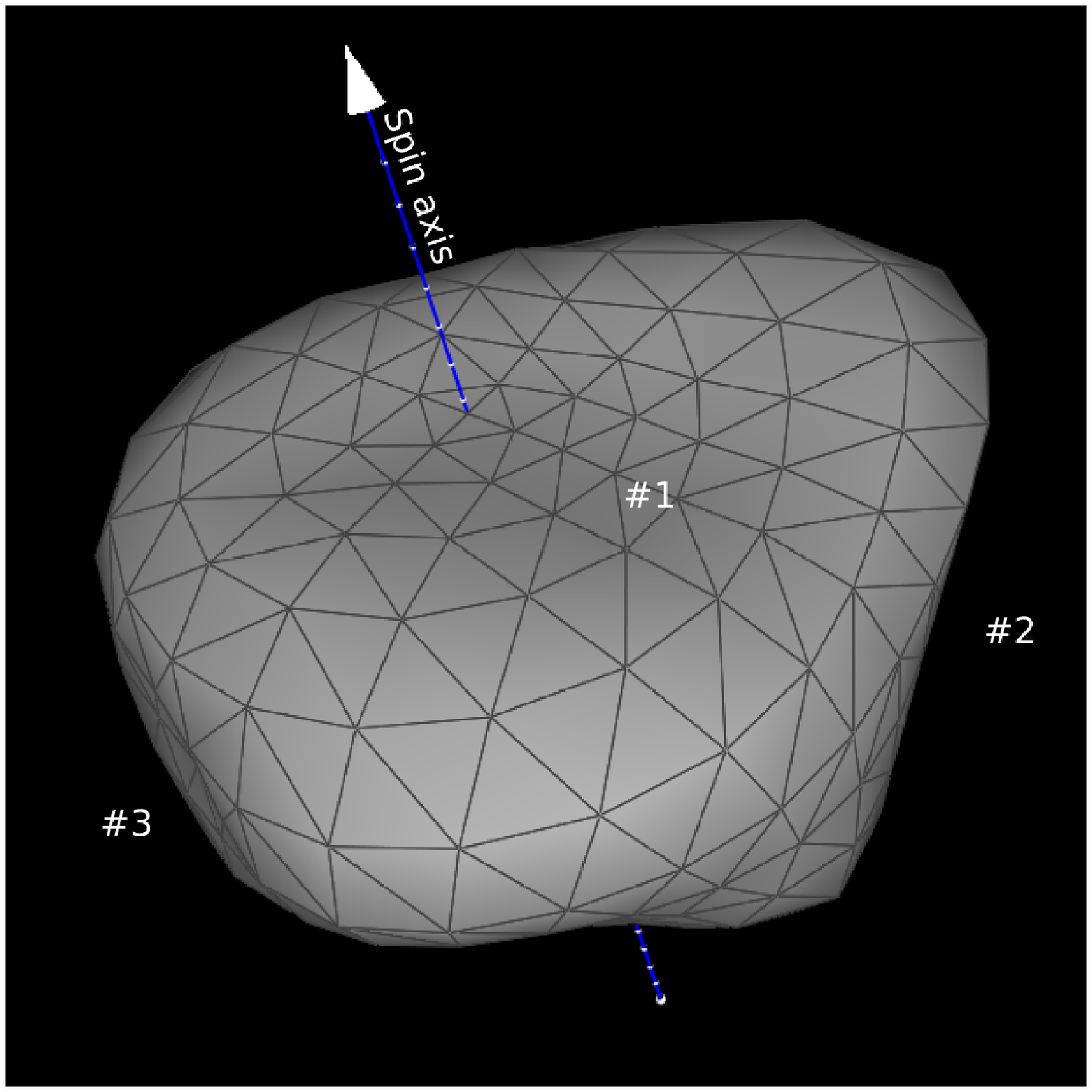}
  \caption{%
    A view of Lutetia's shape model obtained with the KOALA algorithm 
    The three possible concavities listed in the text are labeled
    here.
    The view from the "top" means a view from above the north pole.
    }
  \label{fig: model}
\end{figure}

\begin{figure*}[!t]
\begin{center}
  \subfigure[First set of Lutetia contours]{%
    \begin{tabular}{ccc}
    \includegraphics[width=.3\textwidth]{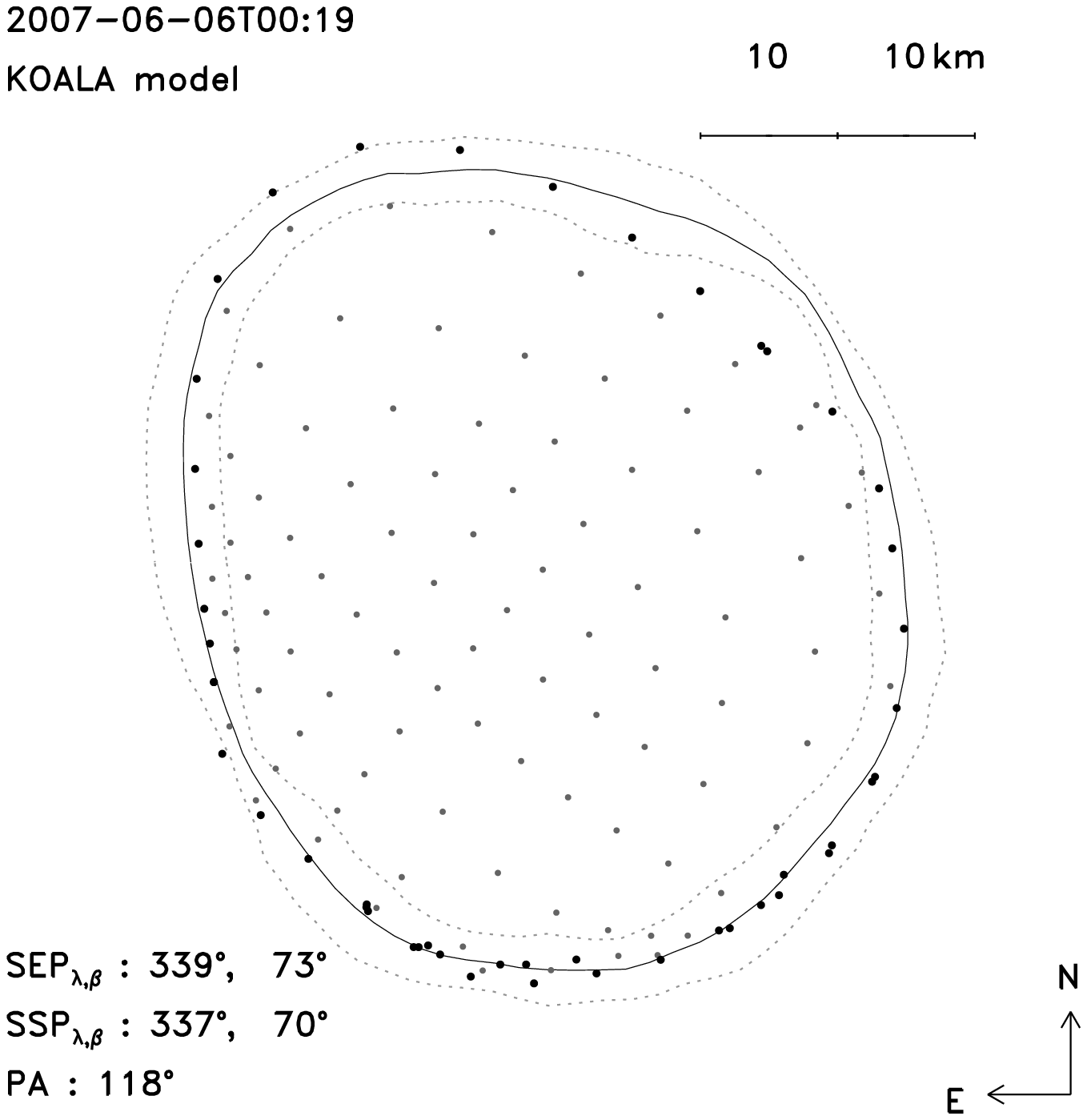}&
    \includegraphics[width=.3\textwidth]{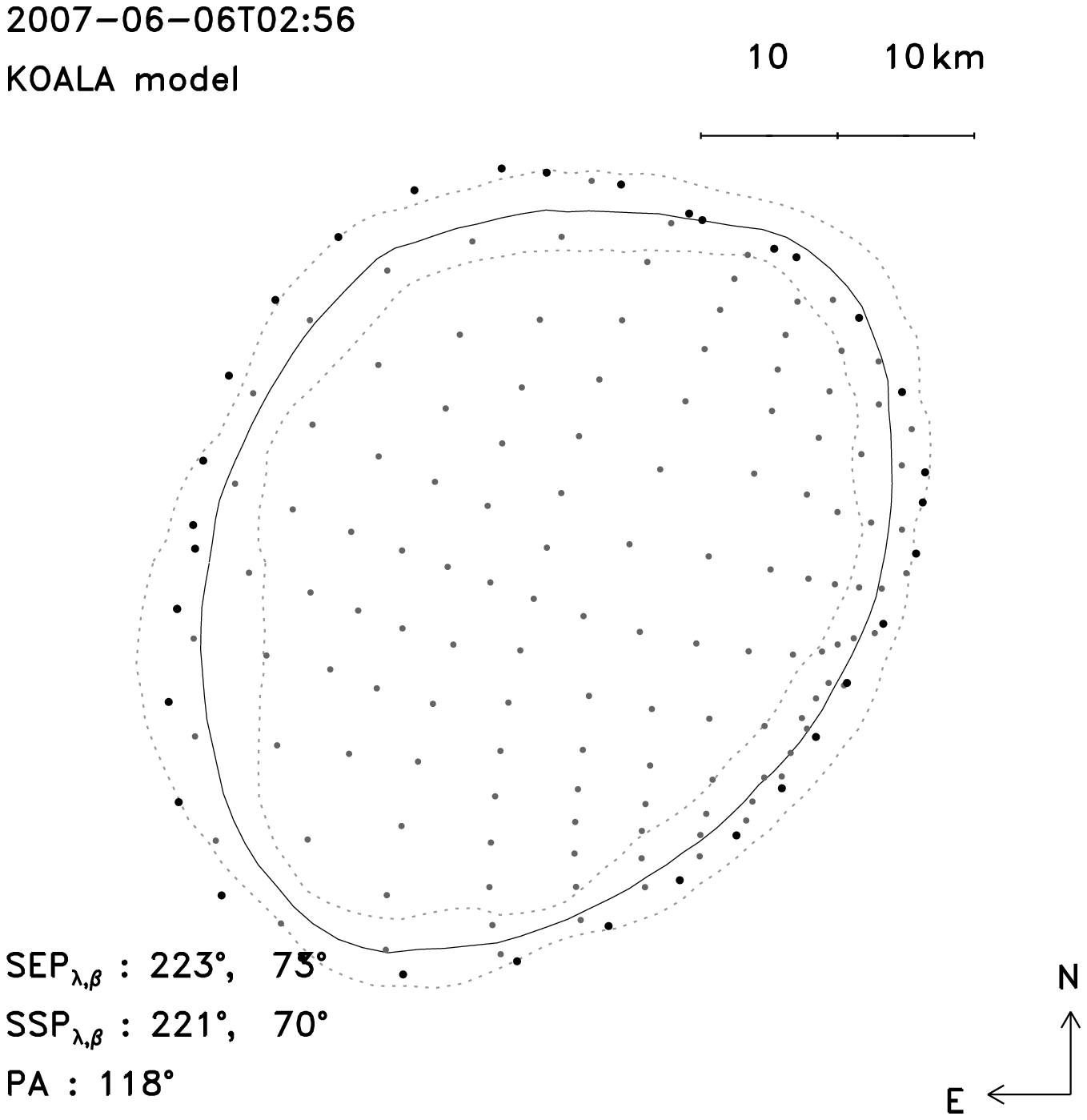}&
    \includegraphics[width=.3\textwidth]{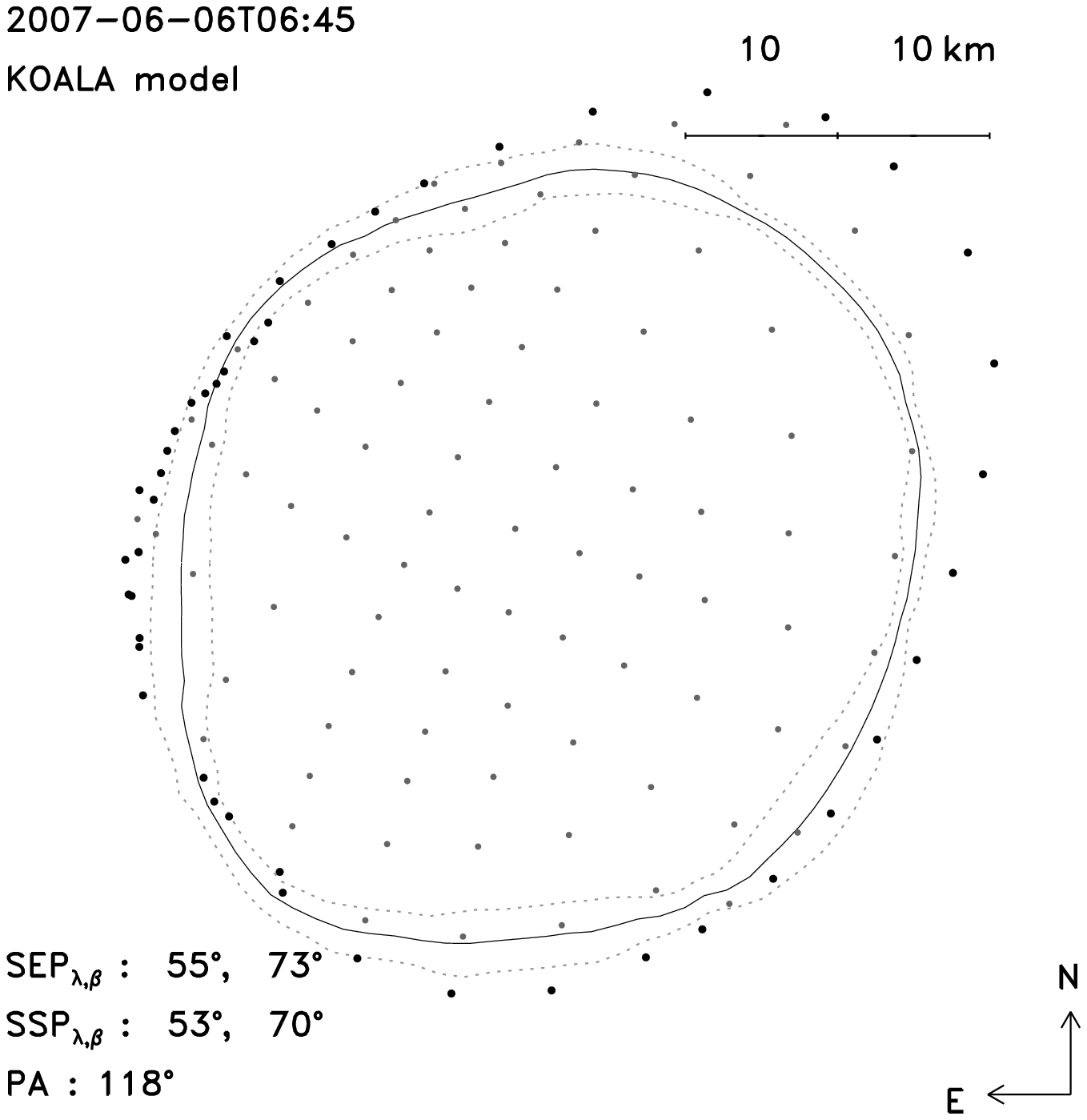}\\
    \includegraphics[width=.3\textwidth]{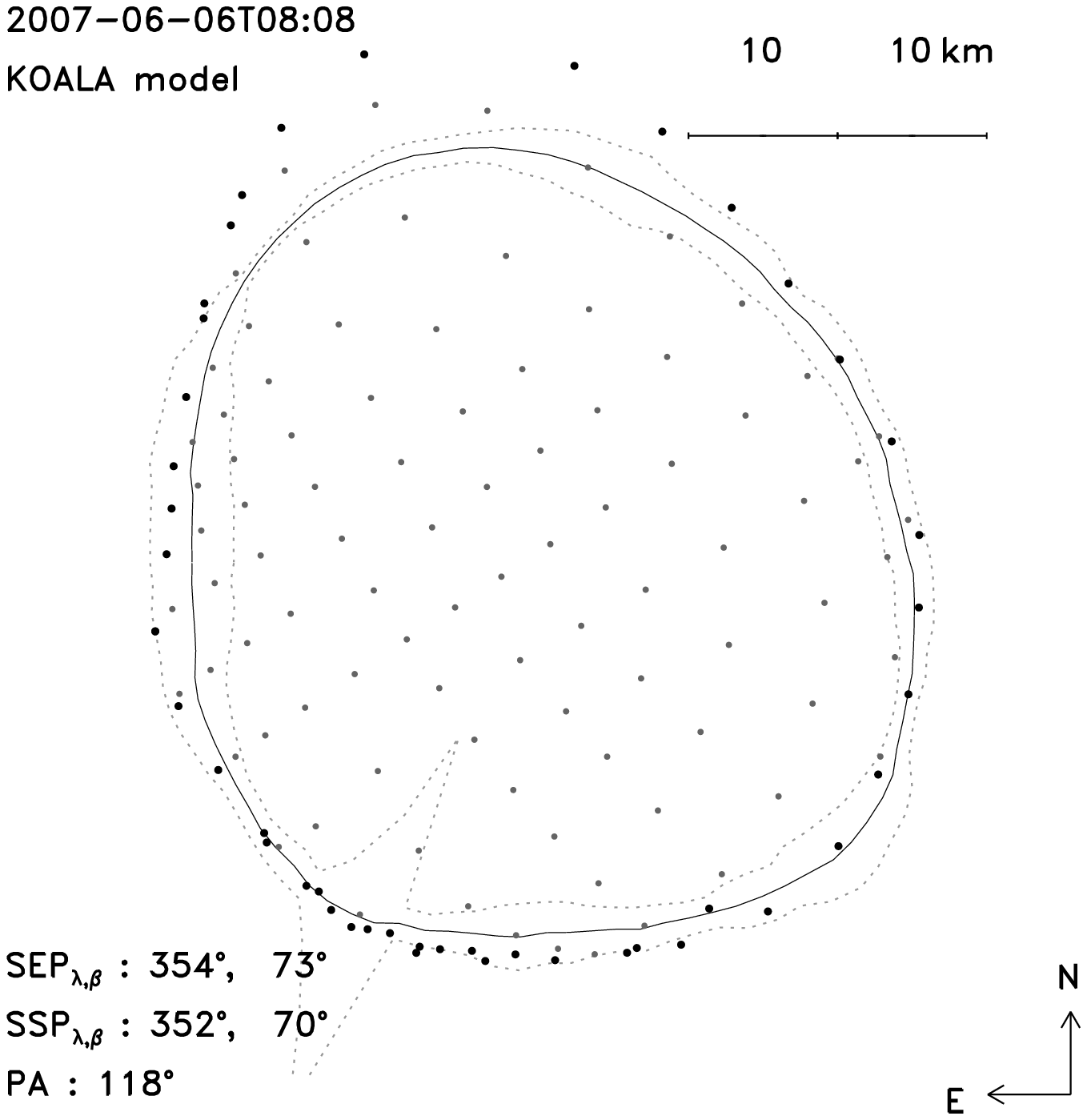}&
    \includegraphics[width=.3\textwidth]{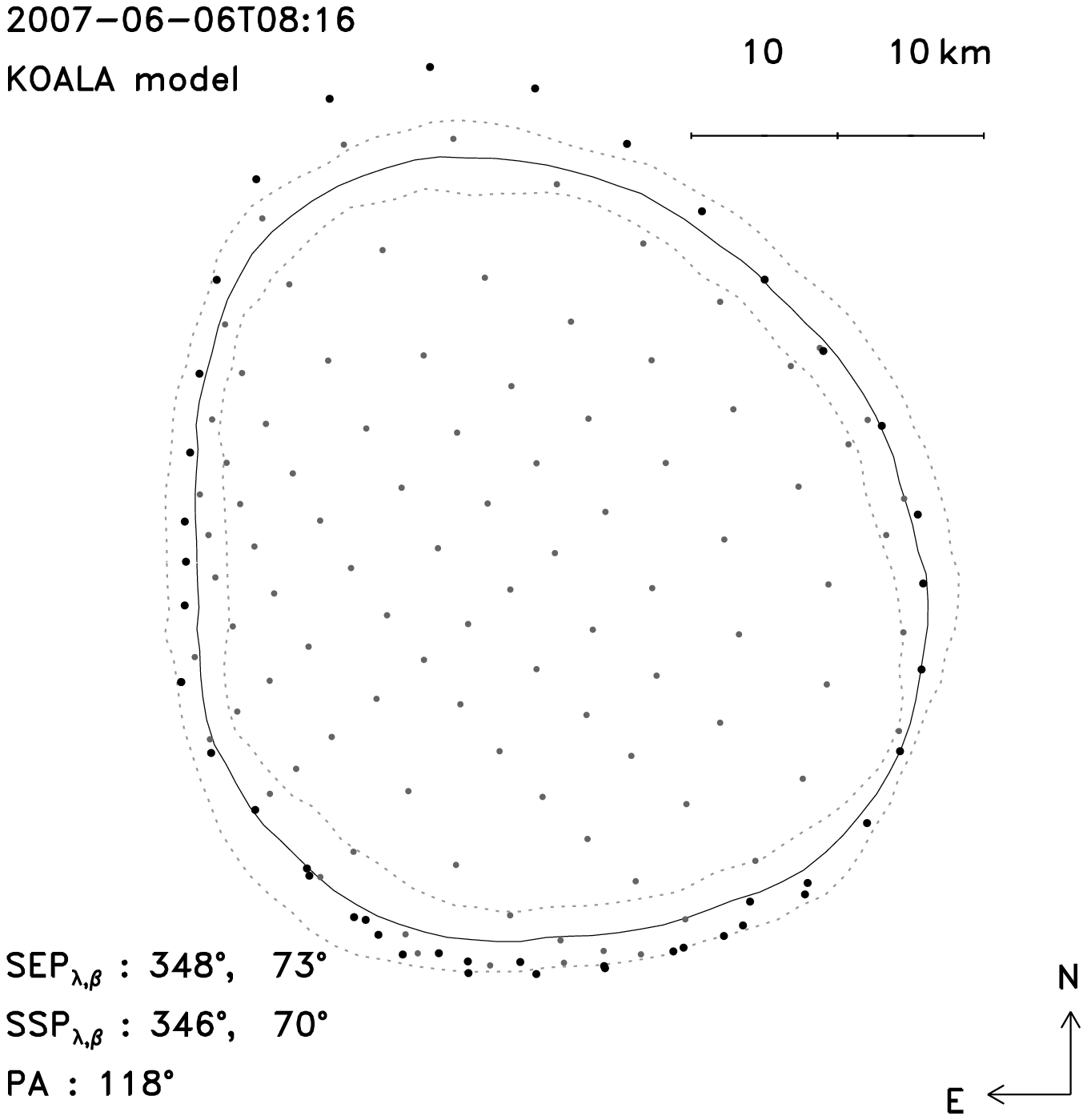}&
    \includegraphics[width=.3\textwidth]{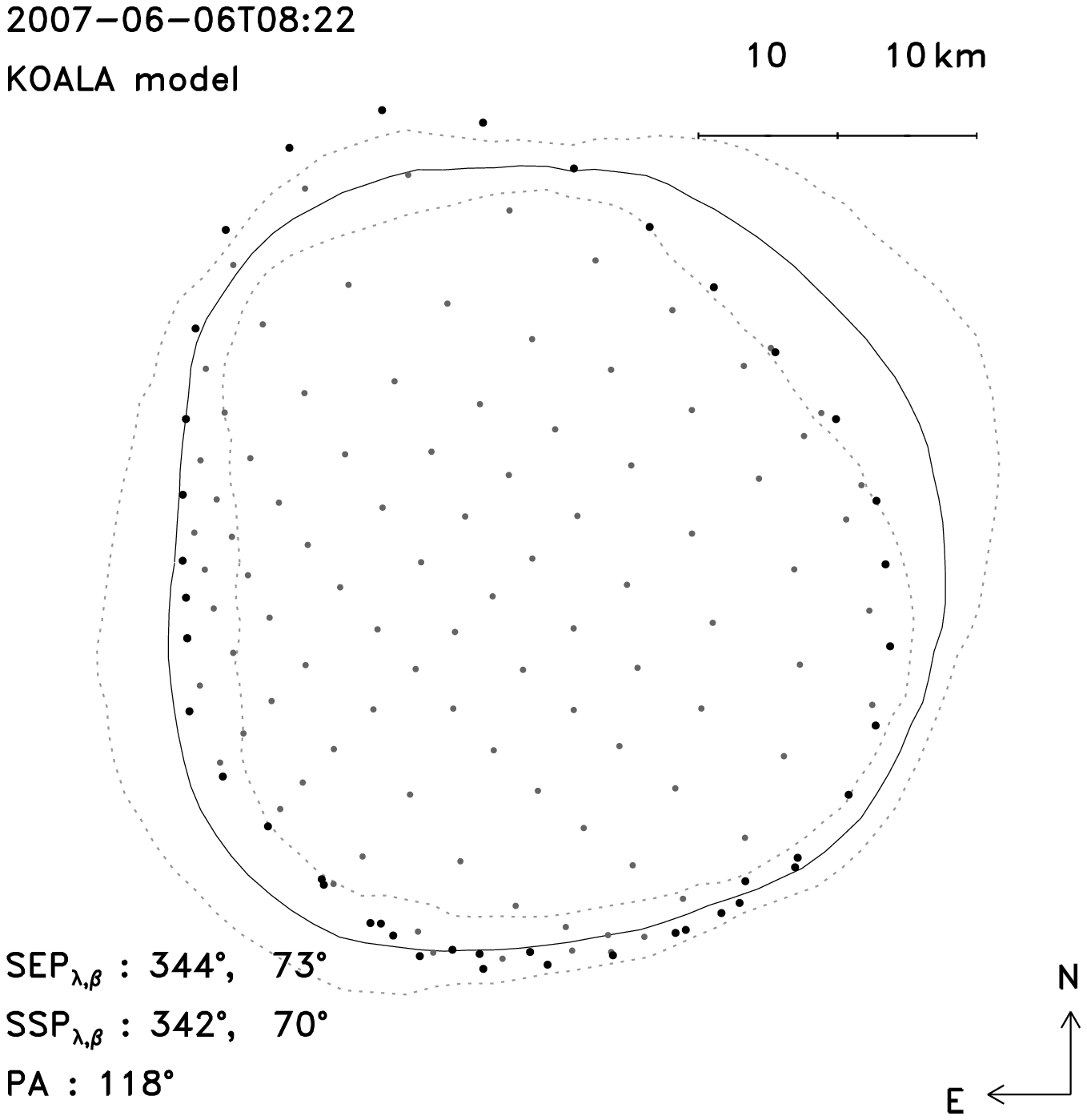}\\
    \includegraphics[width=.3\textwidth]{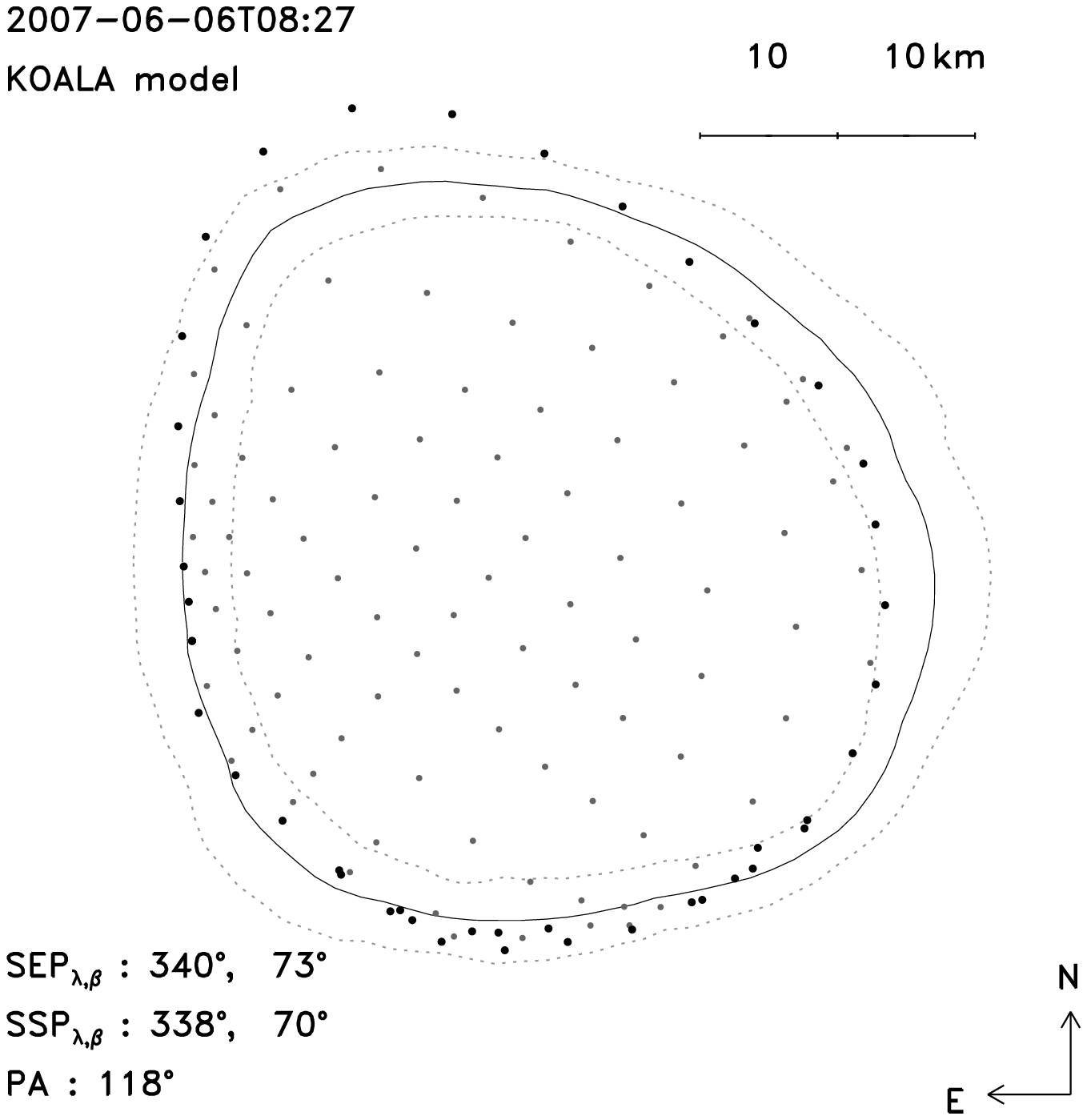}&
    \includegraphics[width=.3\textwidth]{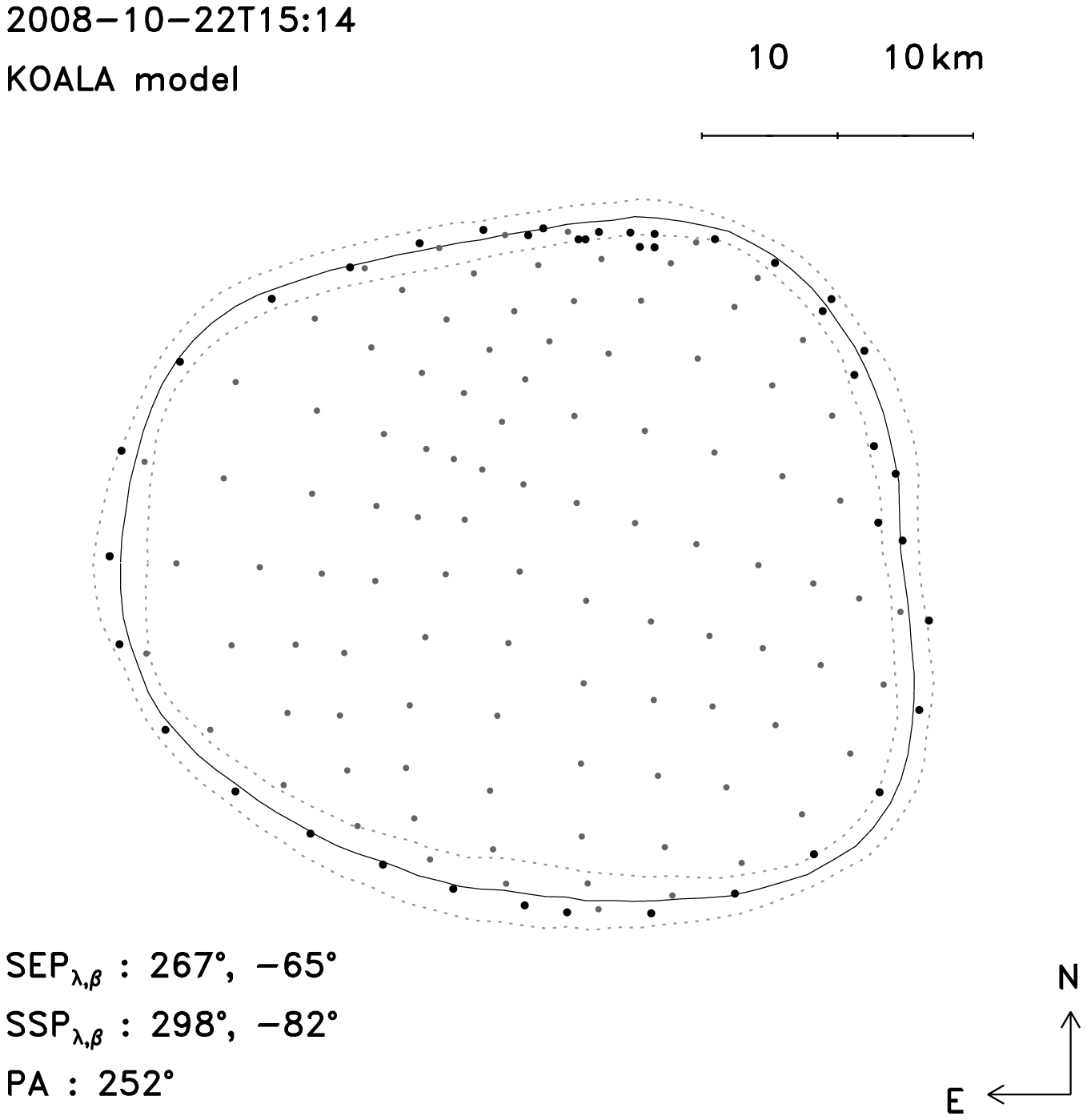}&
    \includegraphics[width=.3\textwidth]{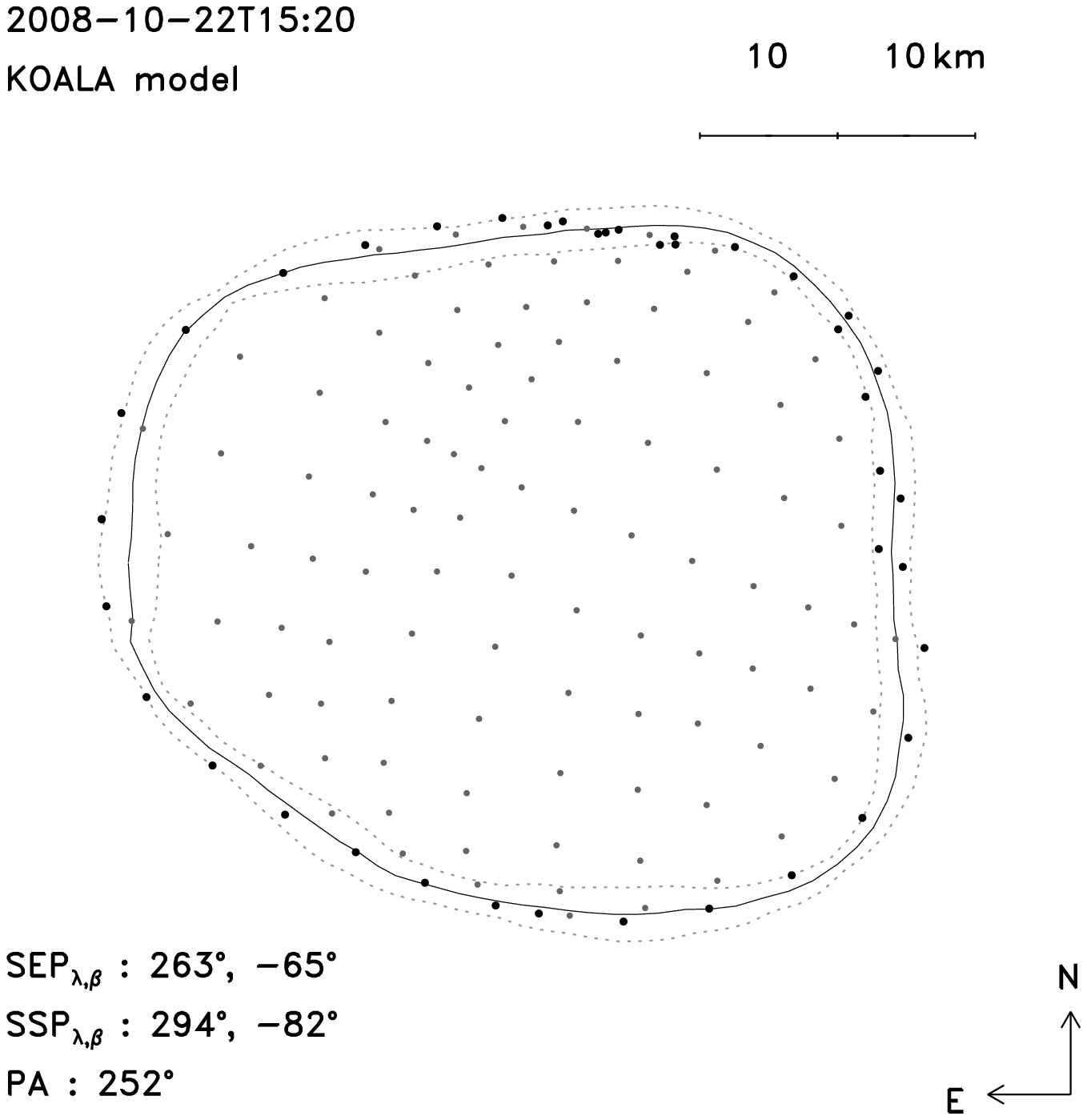}\\
    \includegraphics[width=.3\textwidth]{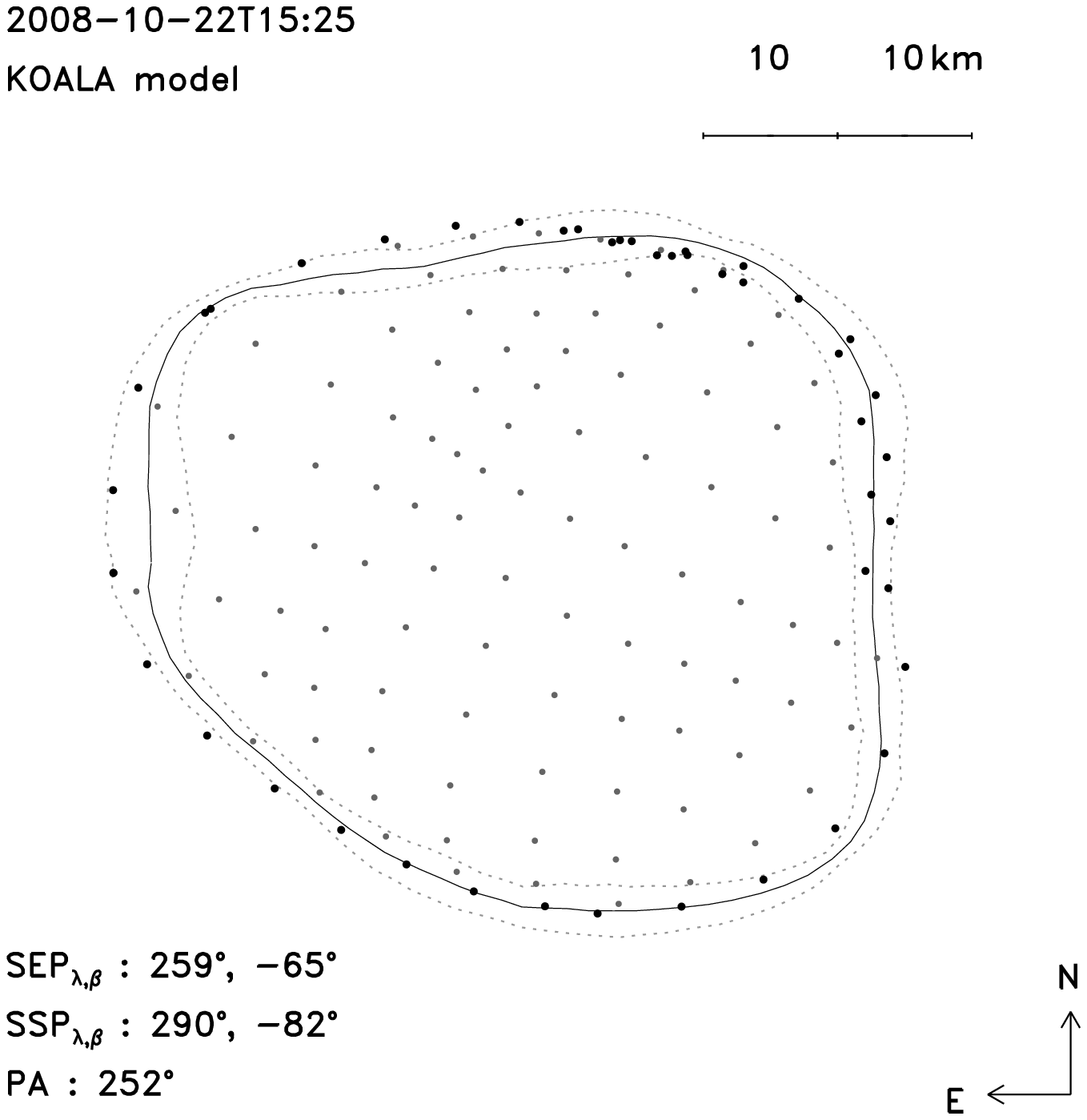}&
    \includegraphics[width=.3\textwidth]{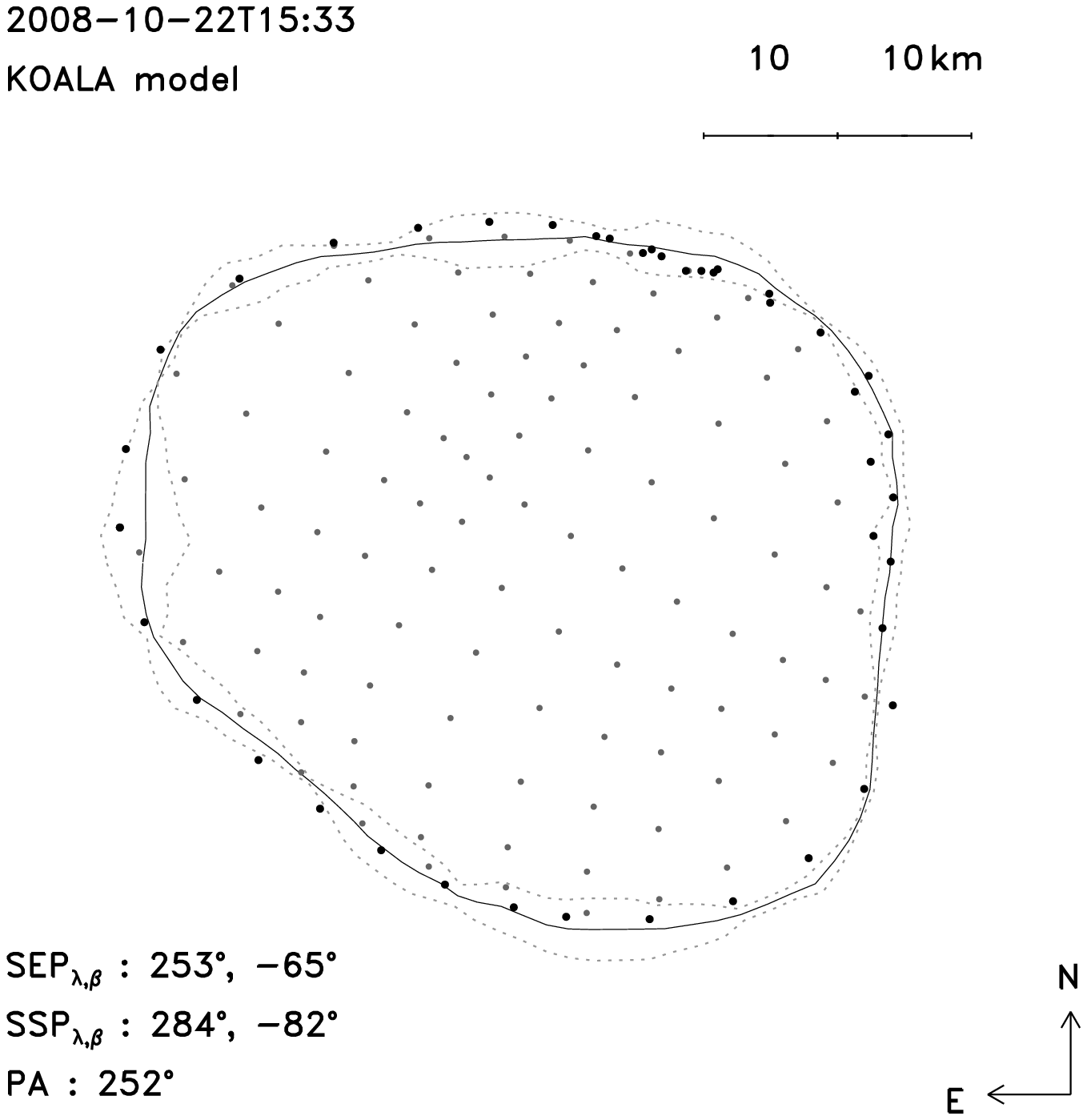}&
    \includegraphics[width=.3\textwidth]{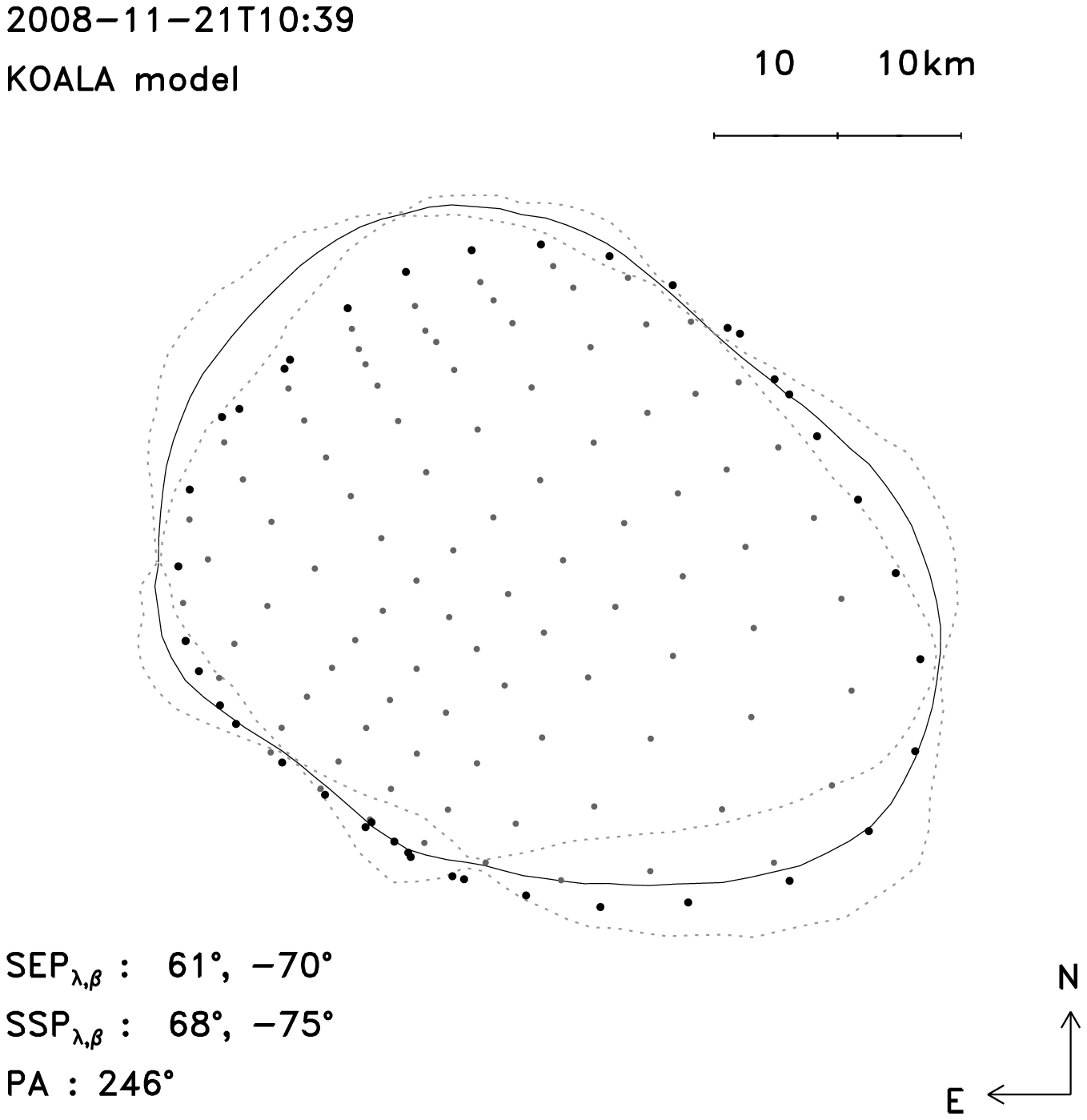}
    \end{tabular}
  }%
  \caption[Contours of (21) Lutetia from disk-resolved images]{%
    Comparison of the KOALA shape model of (21) Lutetia to the
    contours extracted on the adaptive-optics images.
    Each vertex of the shape model is represented by a grey dot,
    with the exception of limb/terminator vertices, which are drawn as
    black dots.
    The median AO-contour for each epoch is plotted as a solid grey line, and
    the 3\,$\sigma$ deviation area is delimited by the dotted grey
    lines.
    We report the
    observing time (in UT), Sub-Earth Point (SEP), Sub-Solar Point (SSP)
    coordinates and Pole Angle
    (PA: defined as the angle in the plane of the sky
    between celestial north and the
    projected asteroid spin-vector, measured counter-clockwise,
    from north through east)
    on each frame.
    \label{fig: comp: ao}
  }
\end{center}
\end{figure*}

\begin{figure*}[!t]
\begin{center}
\setcounter{subfigure}{0}
  \subfigure[First set of lightcurves (1962--1981)]{%
    \includegraphics[width=.8\textwidth]{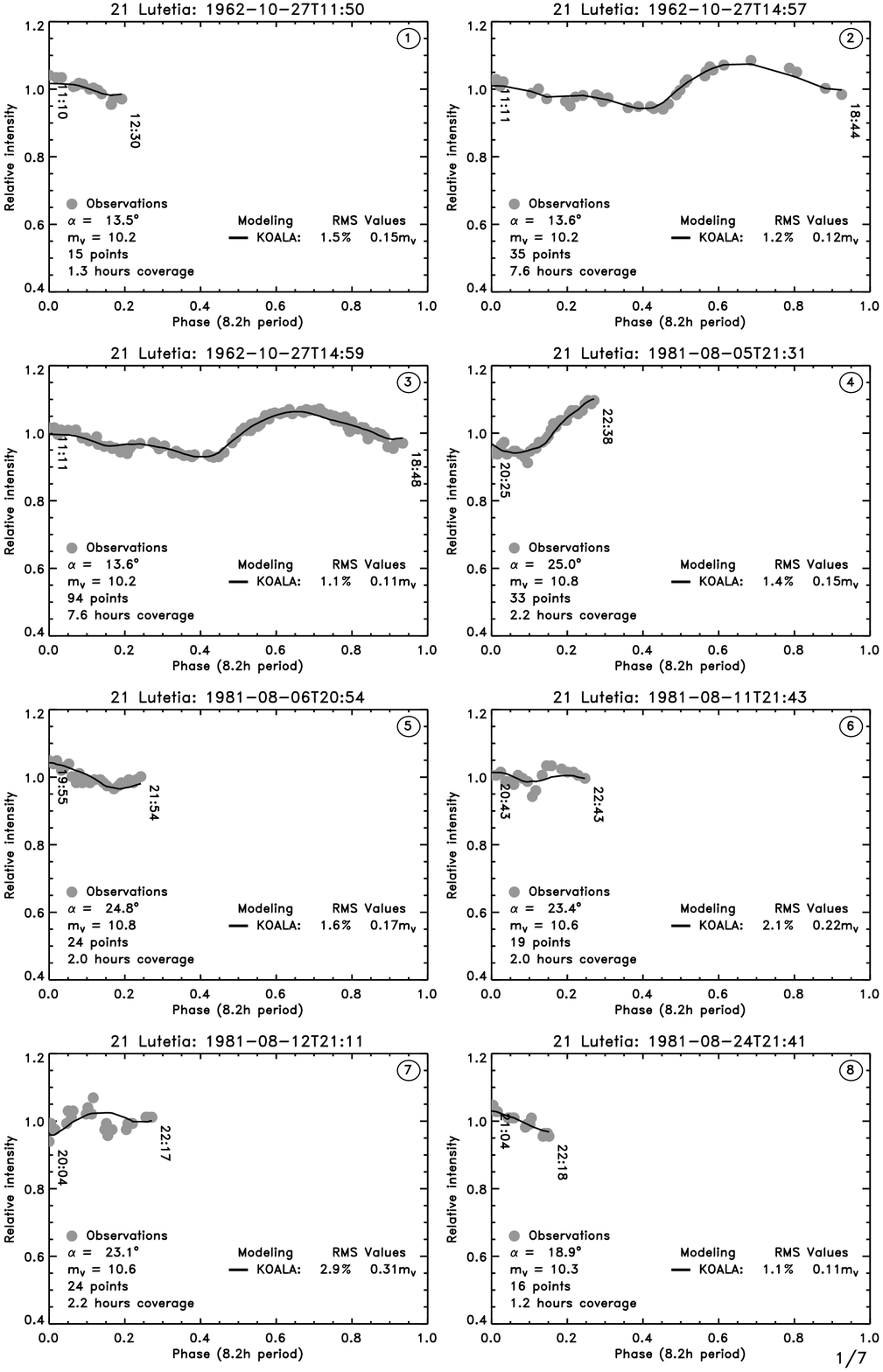}
  }%
  \caption[lightcurves of 21 Lutetia]{%
    Synthetic lightcurves obtained with
    the KOALA model
    plotted against
    the 50
    lightcurves used in the current study,
    plotted in arbitrary relative intensity.
    The observing conditions
    (phase angle $\alpha$,
    average apparent visual magnitude $m_V$,
    number of points
    and duration of the observation) of each lightcurve are reported
    on each panel, along with the synthetic lightcurve fit RMS
    (in percent and visual magnitude).
    Lightcurve observations were acquired by
    {\tiny $(1-3  )$} \citet{1963-AAS-11-Chang}
    {\tiny $(4-9  )$} \citet{1983-SvAL-9-Lupishko},
    {\tiny $(10-11)$} \citet{1984-AA-130-Zappala},
    {\tiny $(12   )$} \citet{1983-SvAL-9-Lupishko},
    {\tiny $(13   )$} \citet{1984-AA-130-Zappala},
    {\tiny $(14-15)$} \citet{1987-KFNT-3-Lupishko-a},
    {\tiny $(16   )$} \citet{1992-AA-95-Dotto},
    {\tiny $(17-21)$} \citet{1987-KFNT-3-Lupishko-b},
    {\tiny $(22   )$} \citet{1995-AAS-113-Lagerkvist},
    {\tiny $(23   )$} \citet{1987-KFNT-3-Lupishko-b},
    {\tiny $(24   )$} \citet{1995-AAS-113-Lagerkvist},
    {\tiny $(25-28)$} \citet{1998-PSS-46-Denchev},
    {\tiny $(29-32)$} \citet{2000-PSS-48-Denchev}
    {\tiny $(33-36)$} L. Bernasconi,
    {\tiny $(37   )$} R. Roy,
    {\tiny $(38   )$} \citet{2008-AA-479-Carvano},
    {\tiny $(39-40)$} \citet{2007-AA-470-Nedelcu},
    {\tiny $(41)$} OSIRIS on Rosetta \citep{2009-DPS-40-Faury},
    {\tiny $(42)$} \citet{2010-AA-Belskaya},
    {\tiny $(43-46)$} F. Colas, F. Vachier, A. Kryszczynska and M. Polinska,
    {\tiny $(47   )$} R. Poncy, 
    {\tiny $(48-49)$} R. Naves, 
    {\tiny $(50   )$} P. Wiggins
}
  \label{fig: comp: lc}

\end{center}
\end{figure*}

    \indent The shape of asteroid (21) Lutetia is well described by
    a wedge of Camembert cheese (justifying the Parisian name of
    Lutetia),
    as visible in Fig.~\ref{fig: model}.
    The shape model derived here suggests the presence of several
    large concavities 
    on the surface of Lutetia, presumably resulting from large
    cratering events.\\
    \indent The major feature (\#1, see Fig.~\ref{fig: model})
    is a large depression
    situated close to the north pole around (10\degr,+60\degr),
    suggesting the presence
    of one or several craters,
    and giving a flat-topped shape to Lutetia.
    \citet{2006-AA-447-Mueller} and
    \citet{2008-AA-479-Carvano} found
    the surface of the northern hemisphere
    to be rougher than in the southern hemisphere,
    possibly due to the
    presence of large crater(s).
    Two other large features are possible:
    the second largest feature (\#2)
    lies at (300\degr, -25\degr), 
    and the third (\#3) at (20\degr, -20\degr).\\
    \indent This shape model provides a very good
    fit to
    disk-resolved images (Fig.~\ref{fig: comp: ao})
    and optical lightcurves
    (Fig.~\ref{fig: comp: lc}).
    The root mean square (RMS) deviations for the two modes of data
    are, 
    3.3 km (0.3 pixel) for imaging
    and
    0.15 magnitude (1.7 \% relative deviation)
    for lightcurves.
    The overall shape compellingly matches the \textsl{convex} shape
    derived by \citet{2003-Icarus-164-Torppa}, and the pole solution
    derived here lies
    18 degrees from the synthetic solution from 
    \citet{2007-Icarus-192-Kryszczynska}\footnote{\href{http://vesta.astro.amu.edu.pl/Science/Asteroids/}
      {http://vesta.astro.amu.edu.pl/Science/Asteroids/}}, based mainly
    on indirect determinations.\\
    \indent An ellipsoid approximation to the 3D shape
    model has dimensions 
    $124 \times 101 \times 80$ km
    (we estimate the 1\,sigma uncertainties to be about
    $\pm$~$ 5 \times 5 \times 15$ km).
    We note here that dimension along 
    the shortest ($c$) axis of Lutetia is much
    more poorly constrained here than the $a$ and $b$ axes. 
    Indeed, all the disk-resolved images were obtained with high
    Sub-Earth Point latitudes
    ($|$SEP$_\beta| \geq 65$\degr, ``\textsl{pole-on}'' views)
    and we, therefore, have limited knowledge of the size of Lutetia
    along its rotation axis.
    Hence, shape models of Lutetia with $b/c$ axes ratio ranging from
    1.1 to 1.3 are not invalidated by our observations
    (all the values and figures presented here are for the model with
    $b/c=1.2$). 
    Higher values of $b/c$ decrease the quality of the fit,
    and although lower values are possible (\citet{2010-AA-Belskaya}
    even suggested $b/c$ should be smaller than 1.1),
    the algorithm begins to break down: 
    (a) spurious localized features
    appear (generated by the lack of shape constraints along
    meridians),
    and
    (b) the spin axis begins to show large
    departures from the short axis and would
    be dynamically unstable.
    A complete discussion on the size and density
    of (21) Lutetia is presented in \citet{2010-AA--Drummond}.
    To better constrain the $c$-dimension, we
    combine the best attributes of our KOALA
    model and our triaxial ellipsoid model to
    create a hybrid shape model, 
    with ellipsoid-approximated dimensions
    124 $\times$ 101 $\times$ 93 km, and thus having an
    average diameter of 105 $\pm$ 5 km. \\
    \indent We list in Table~\ref{tab: spin}
    the spin solution we find. The high precision (3 ms) on
    sidereal period results from the long time-line (47 years) of
    lightcurve observations.
    This solution yields an obliquity of
    95\degr, Lutetia being
    being tilted with respect to its orbital plane, similar to Uranus.
    Consequently, the northern/southern hemispheres of Lutetia experience
    long seasons, alternating between constant illumination
    (summer) and constant darkness (winter)
    while the asteroid orbits around the Sun.
    This has strong implications for the Rosetta flyby, as described in
    the following section.
%
%

%


\setcounter{subfigure}{0}

\begin{figure*}[!t]
\begin{center}
\setcounter{subfigure}{1}
    \subfigure[Second set of Lutetia contours]{%
    \begin{tabular}{ccc}
    \includegraphics[width=.3\textwidth]{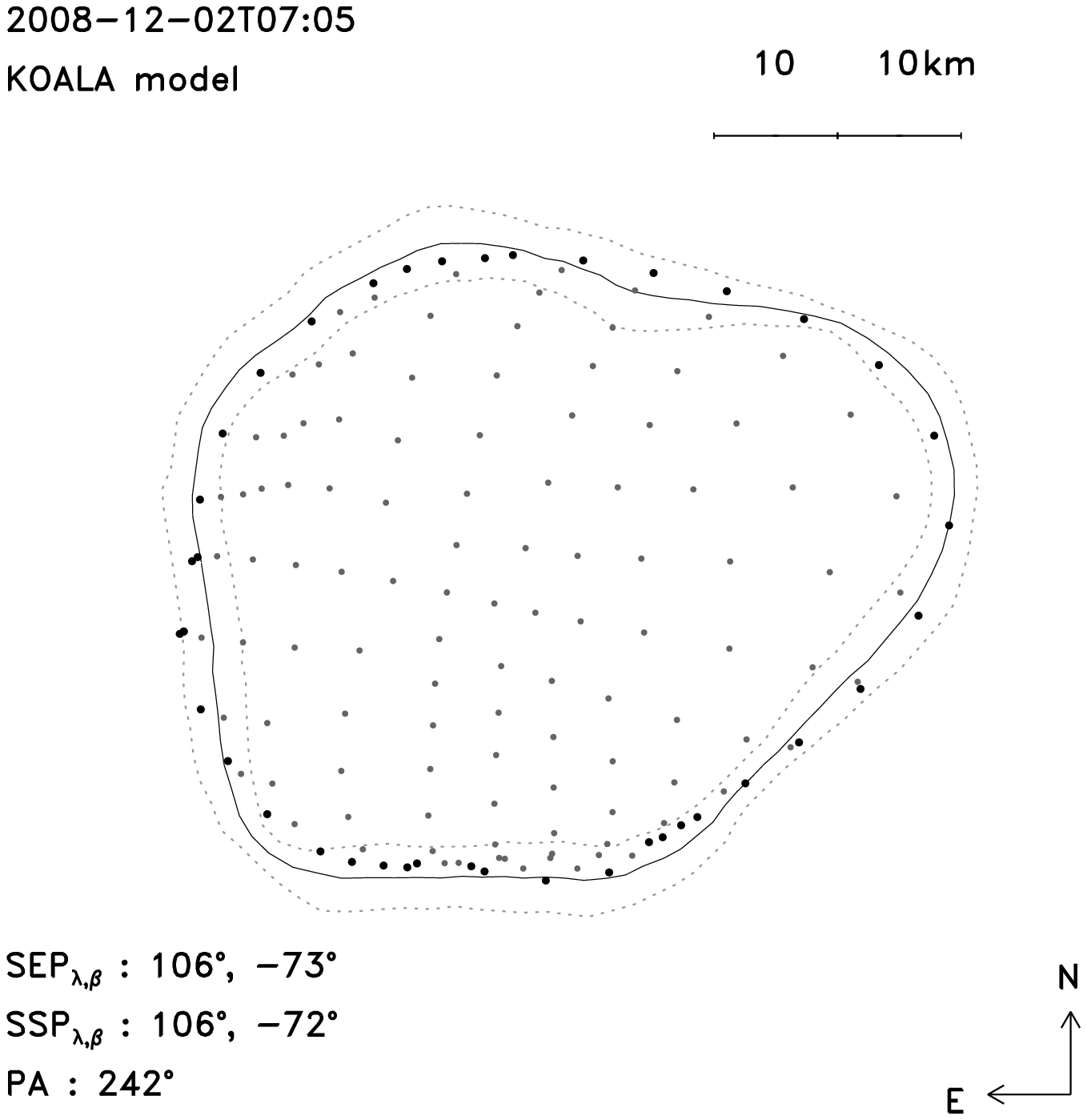}&
    \includegraphics[width=.3\textwidth]{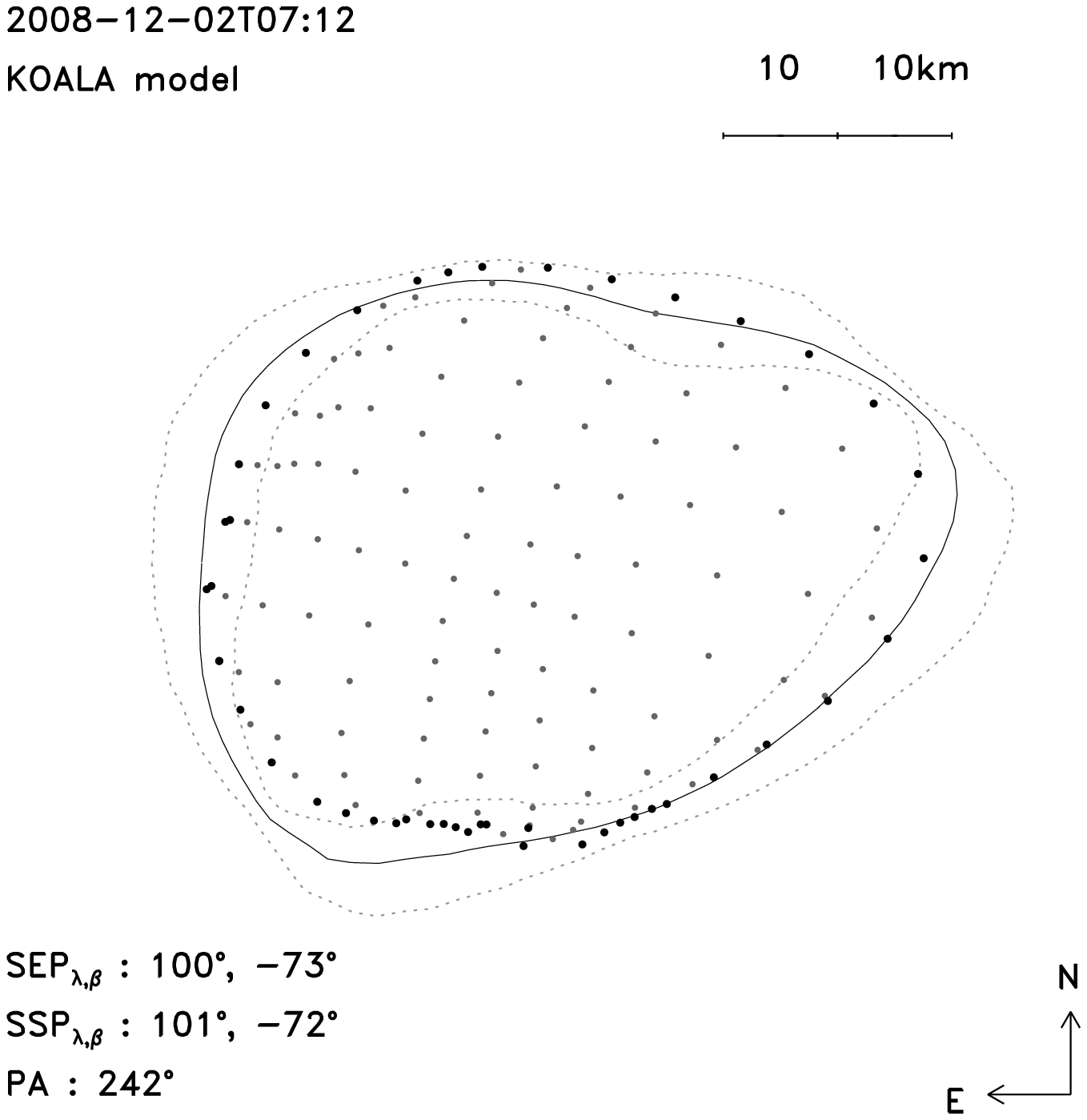}&
    \includegraphics[width=.3\textwidth]{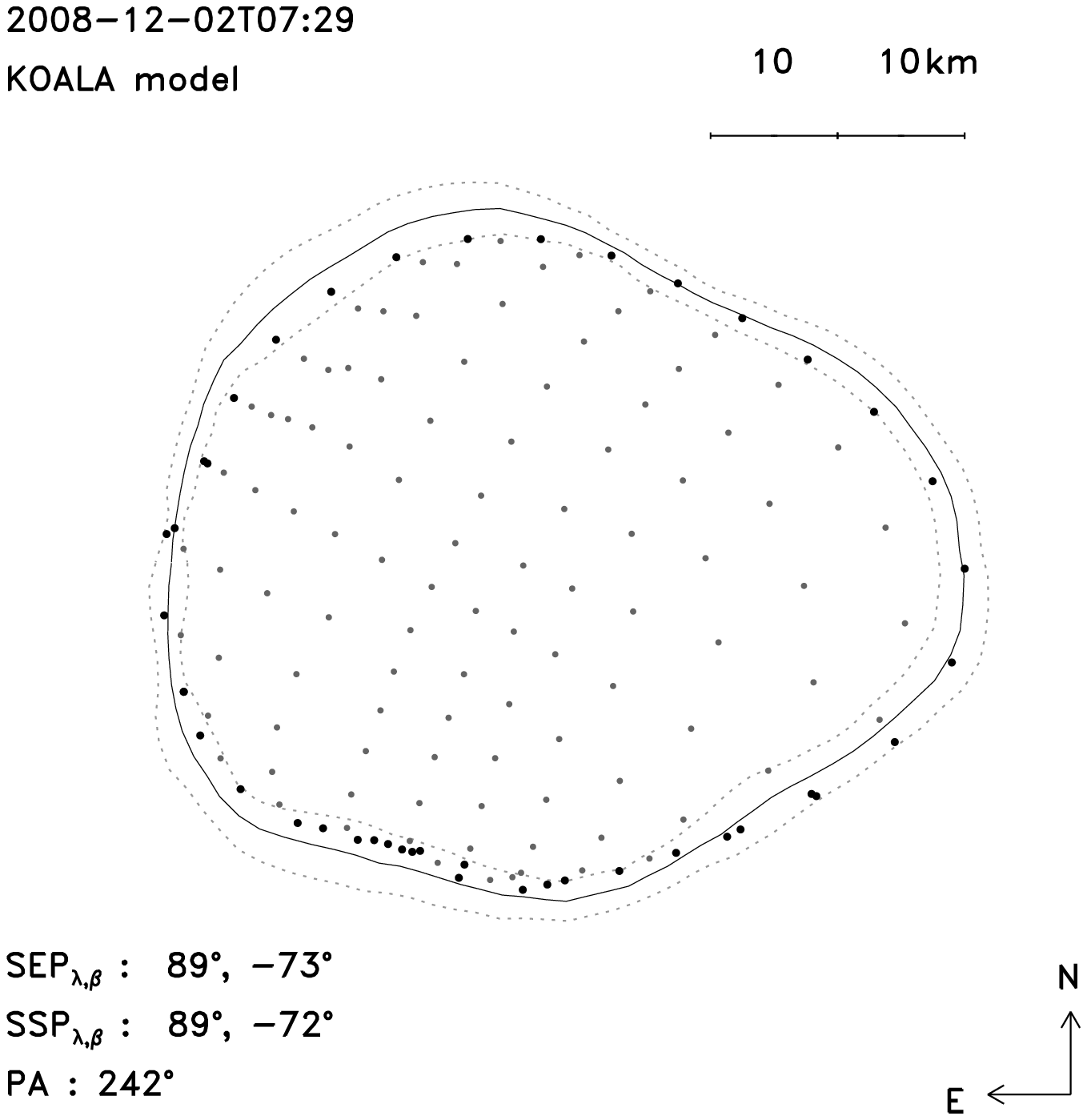}\\
    \includegraphics[width=.3\textwidth]{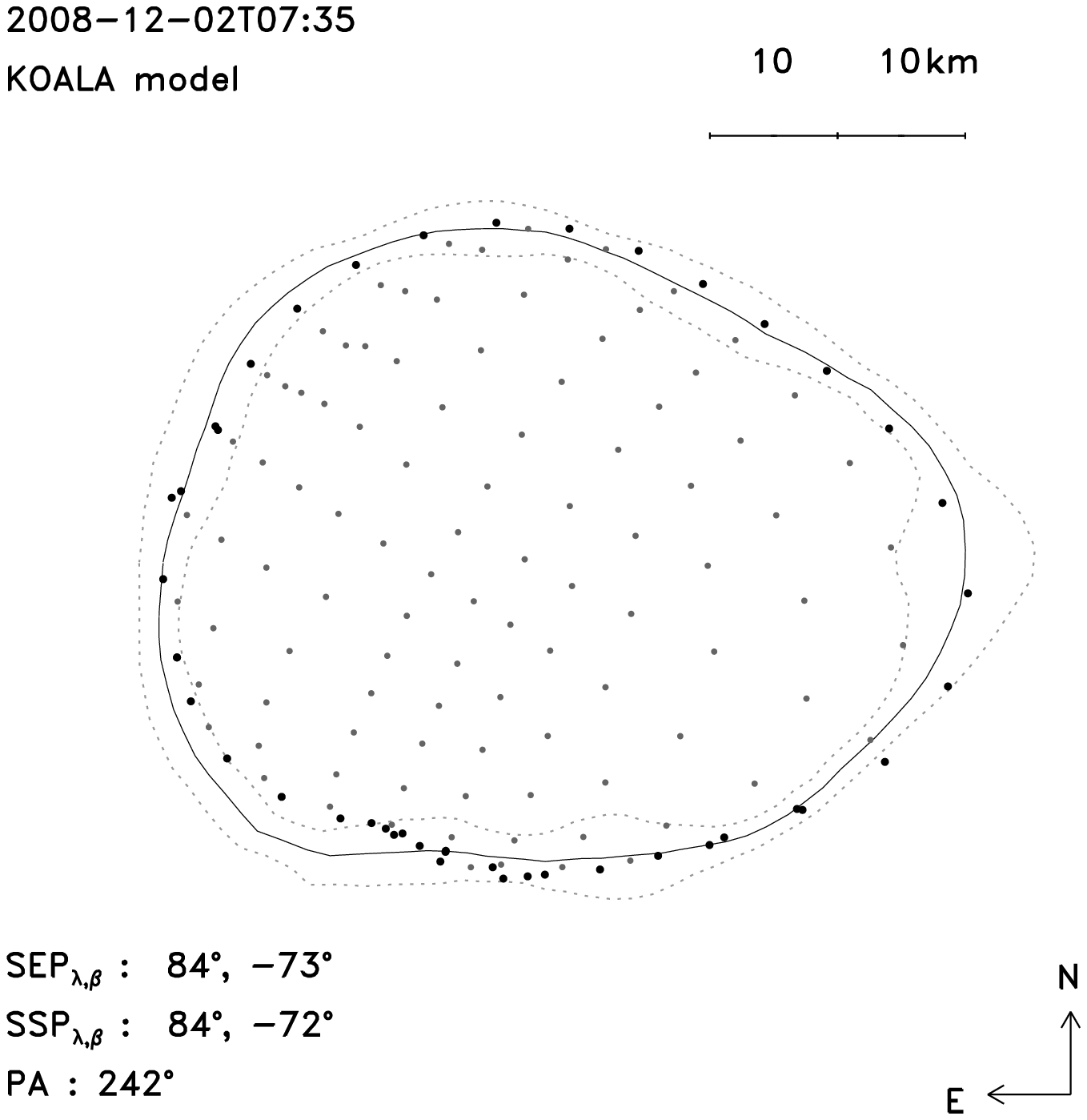}&
    \includegraphics[width=.3\textwidth]{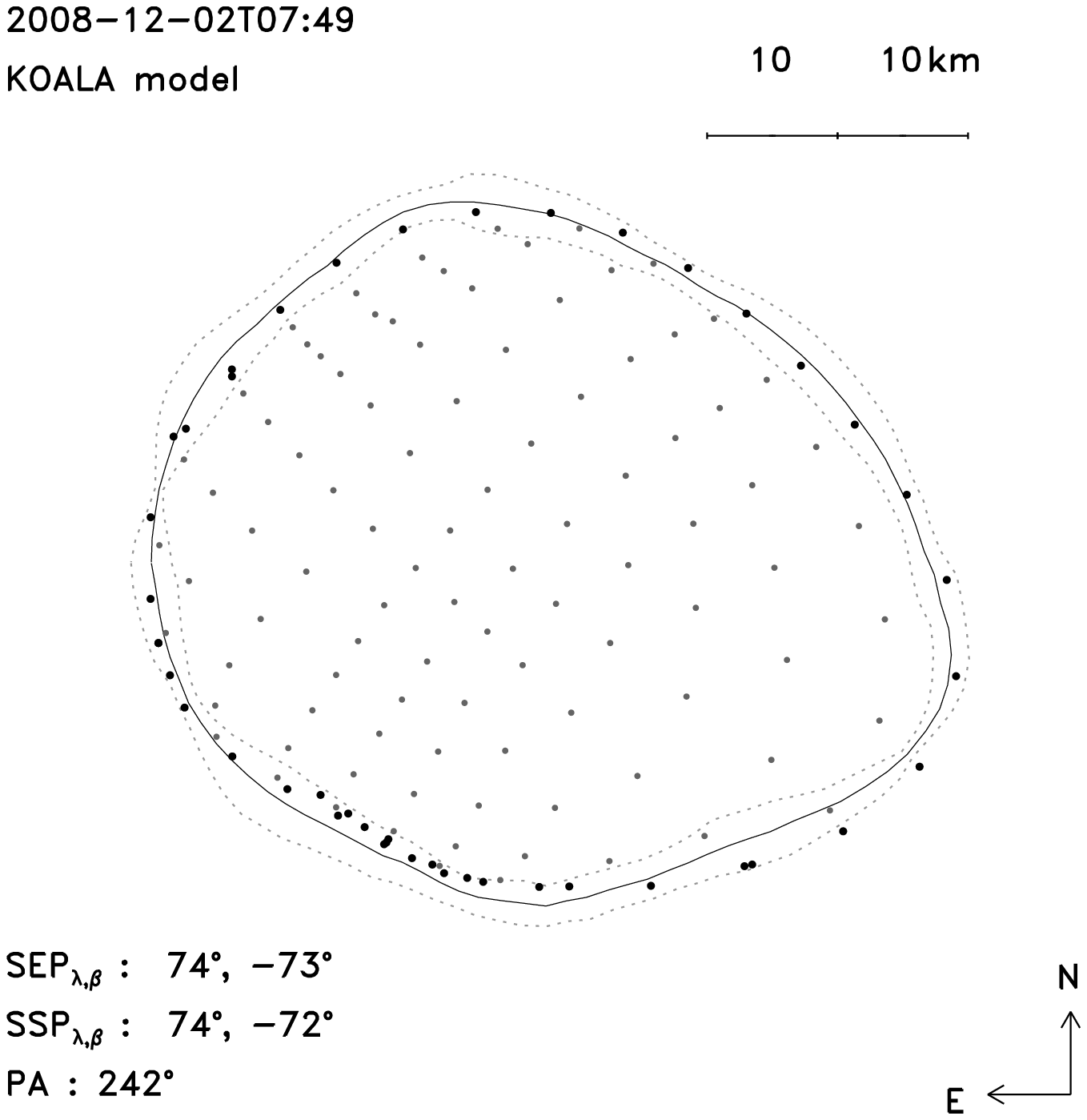}&
    \includegraphics[width=.3\textwidth]{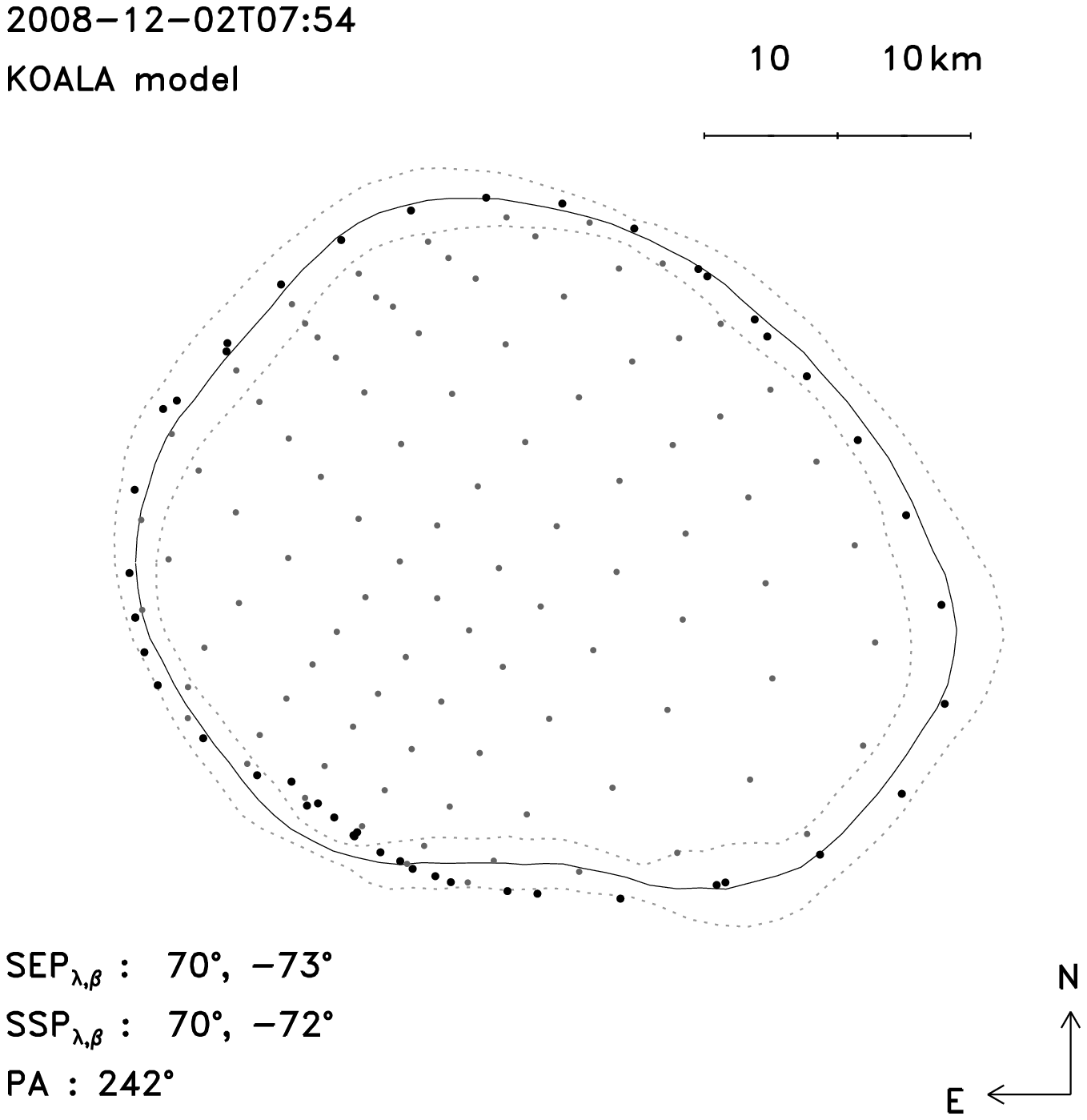}\\
    \includegraphics[width=.3\textwidth]{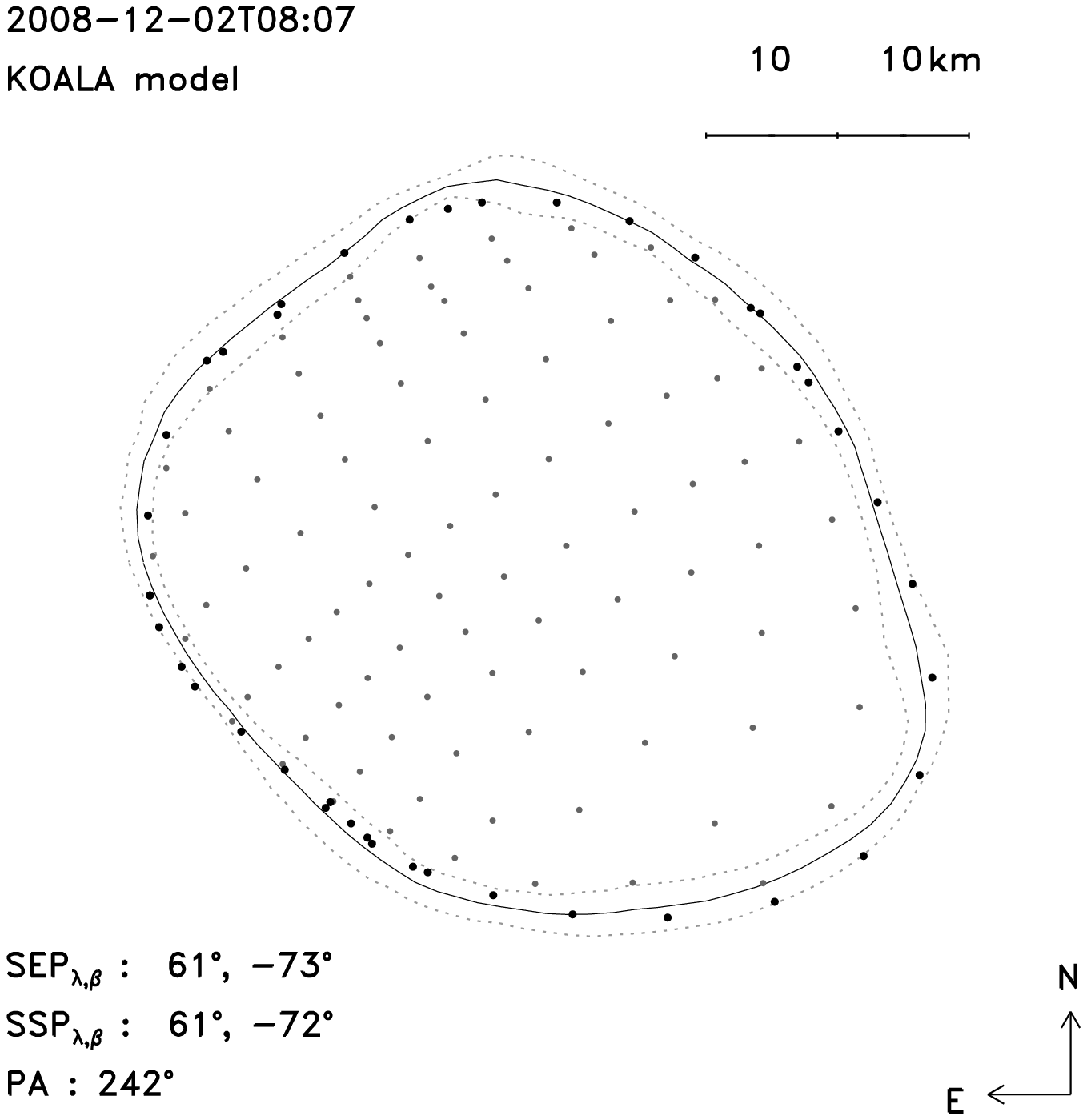}&
    \includegraphics[width=.3\textwidth]{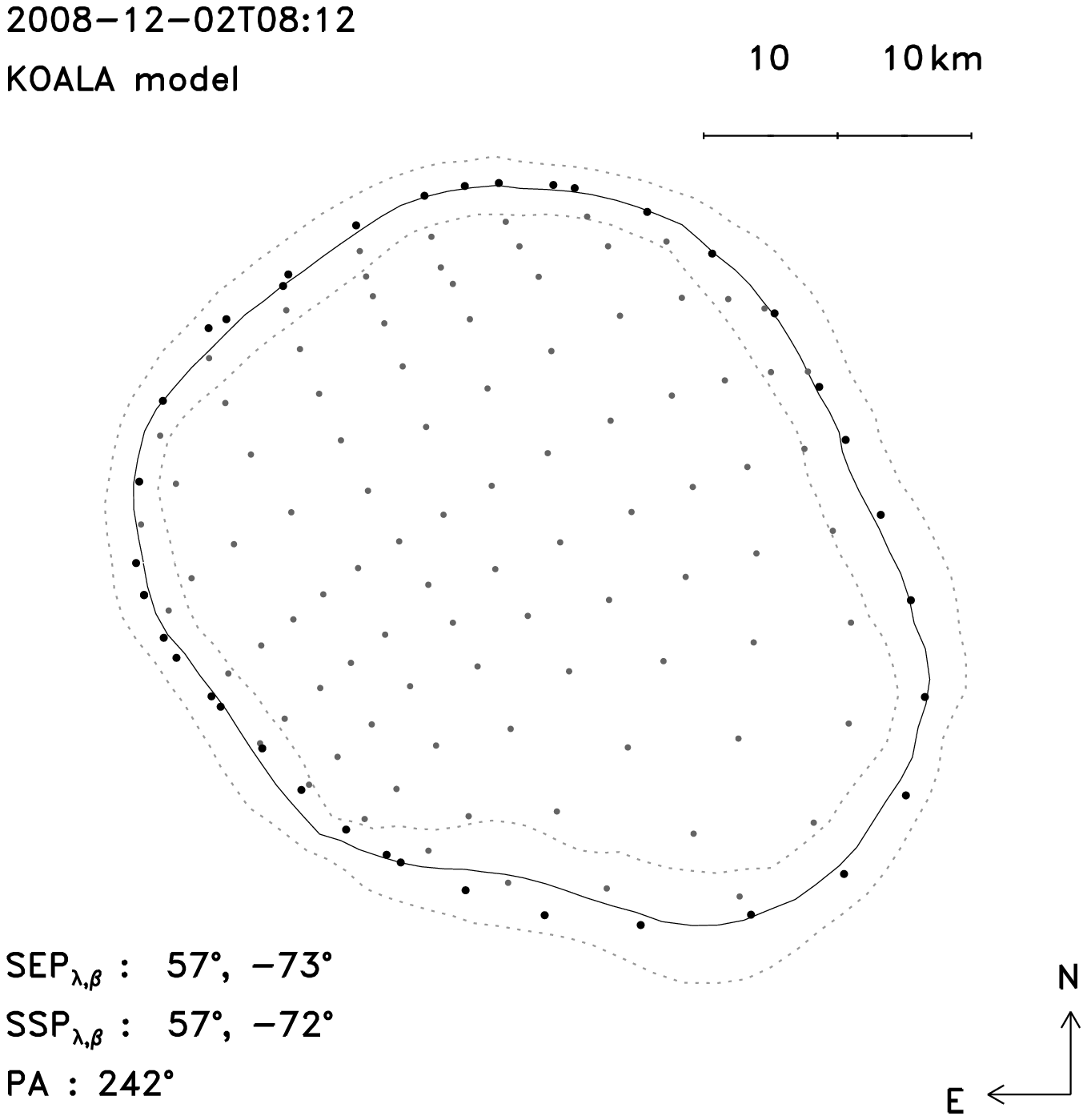}&
    \includegraphics[width=.3\textwidth]{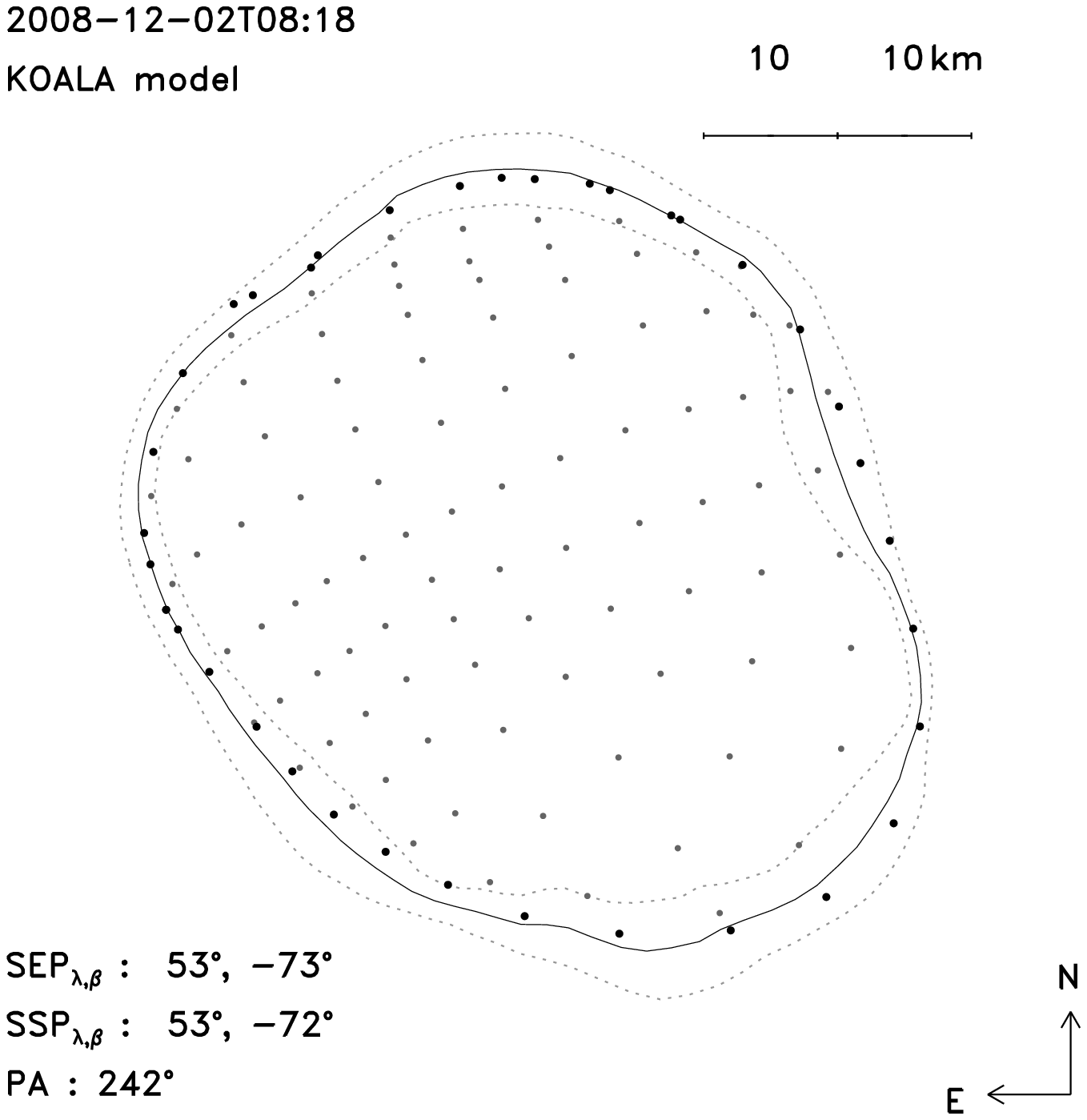}\\
    \includegraphics[width=.3\textwidth]{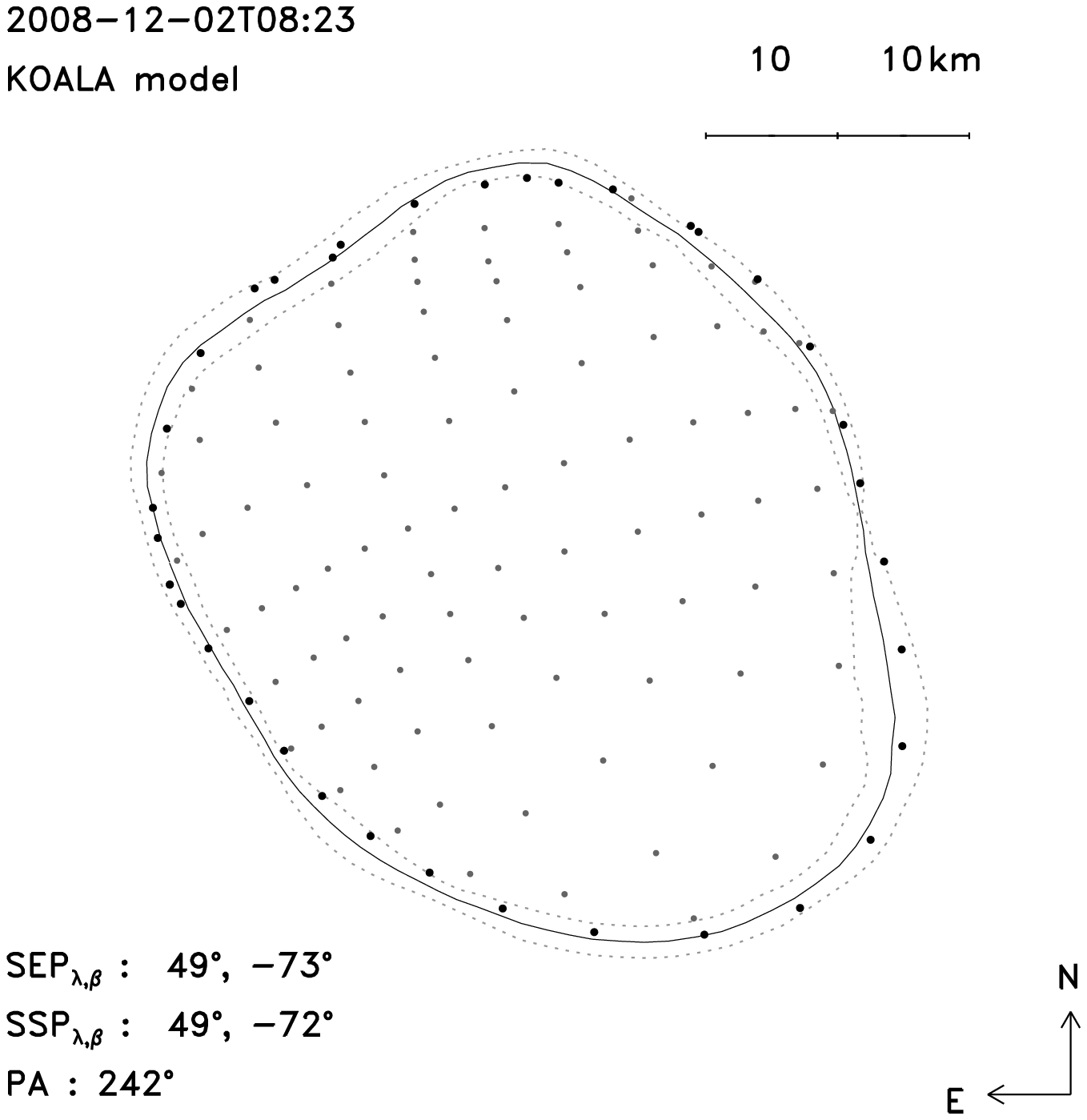}&
    \includegraphics[width=.3\textwidth]{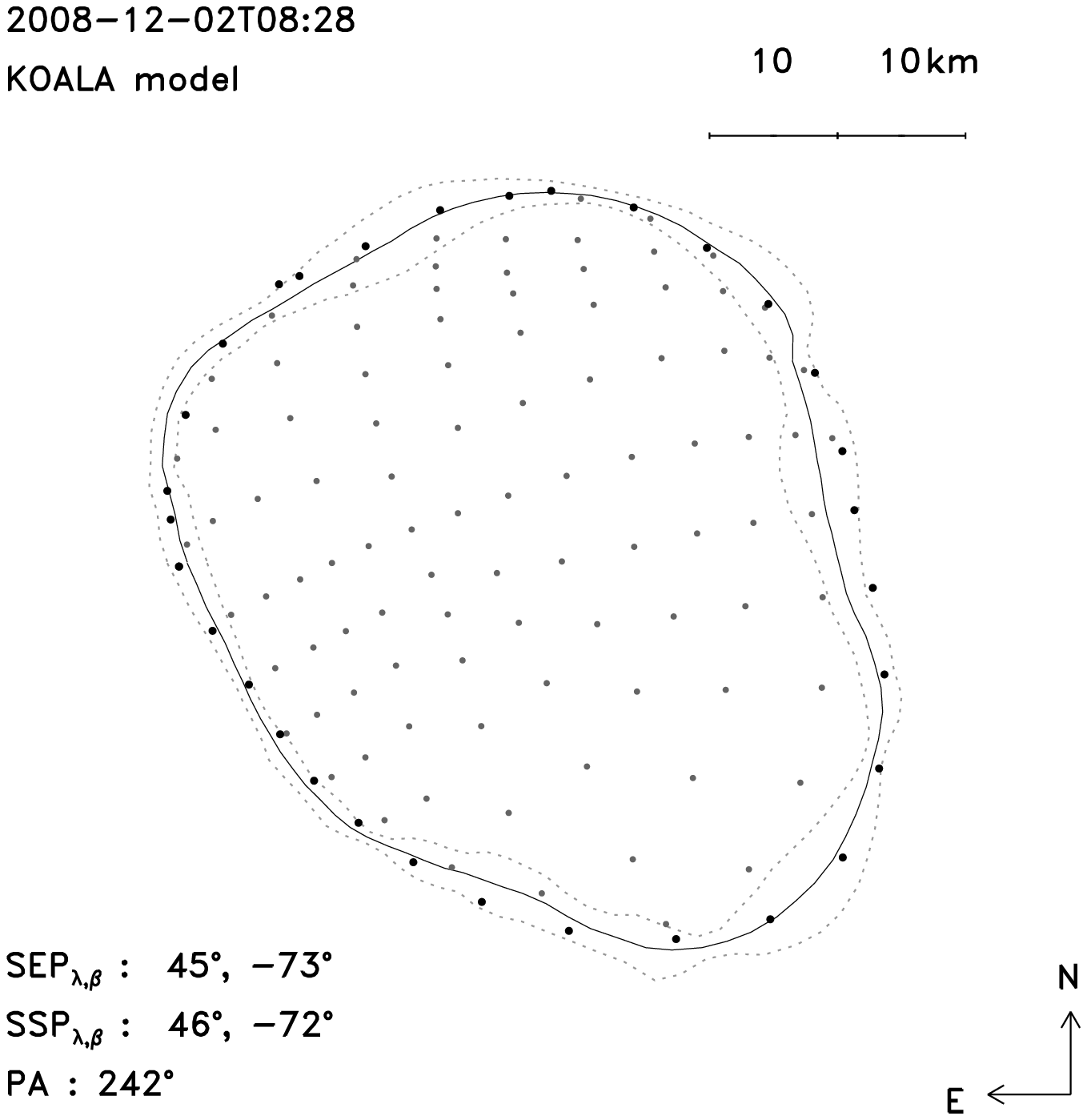}&
    \includegraphics[width=.3\textwidth]{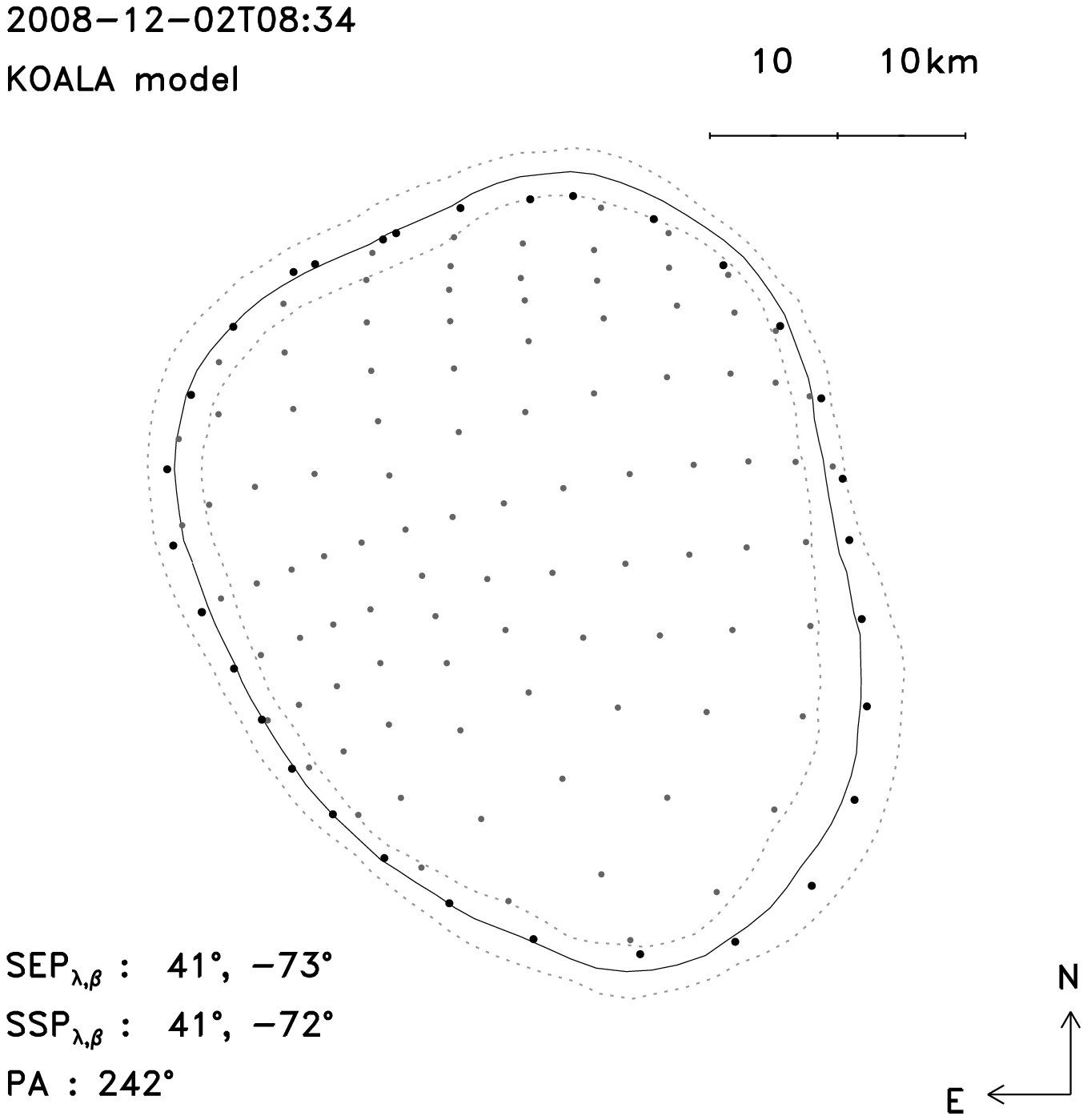}
    \end{tabular}
    }
\end{center}
  \label{fig: comp: ao2}
\end{figure*}

\begin{figure*}[!t]
\begin{center}
\setcounter{subfigure}{2}
    \subfigure[Third set of Lutetia contours]{%
    \begin{tabular}{ccc}
    \includegraphics[width=.3\textwidth]{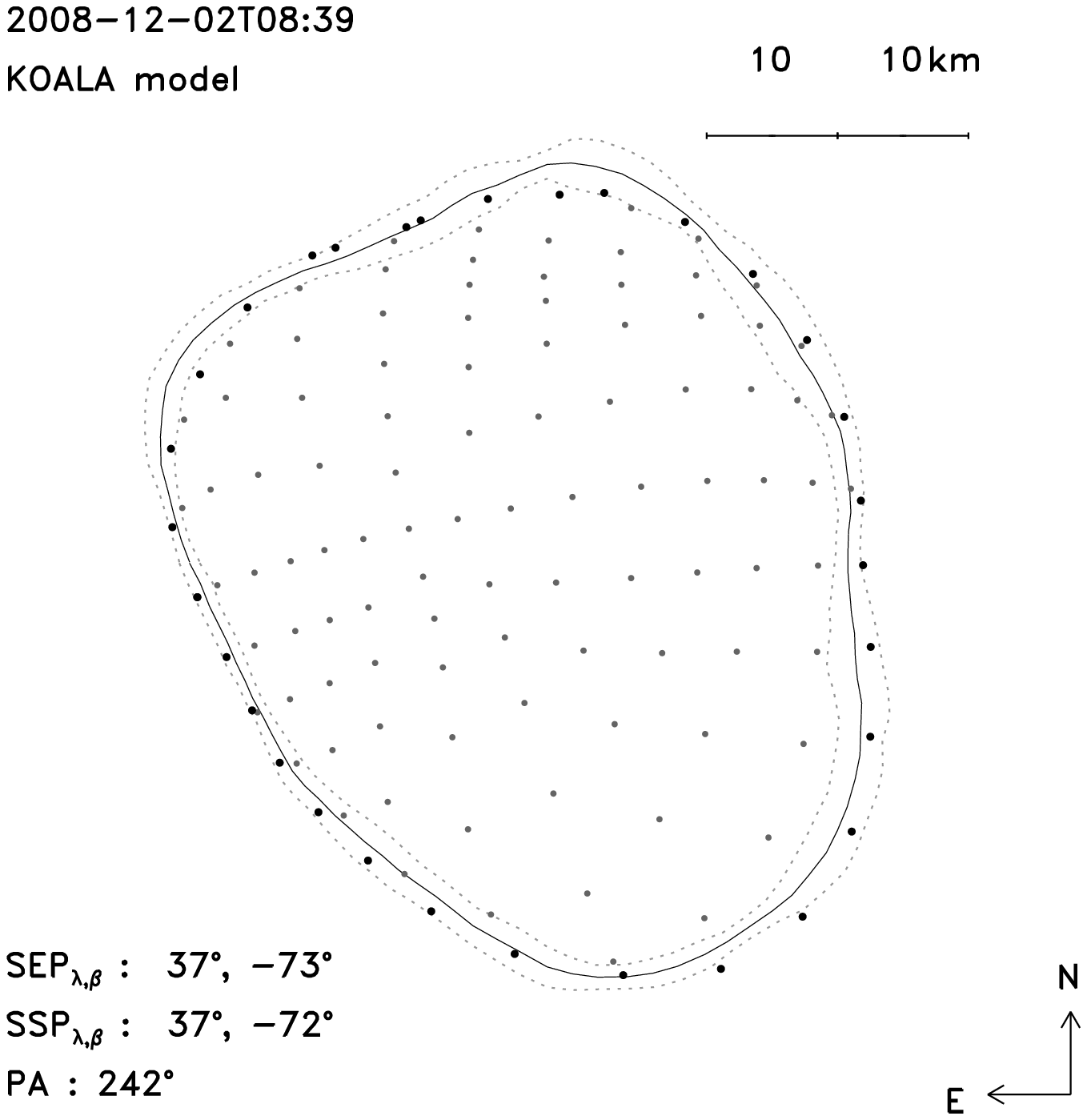}&
    \includegraphics[width=.3\textwidth]{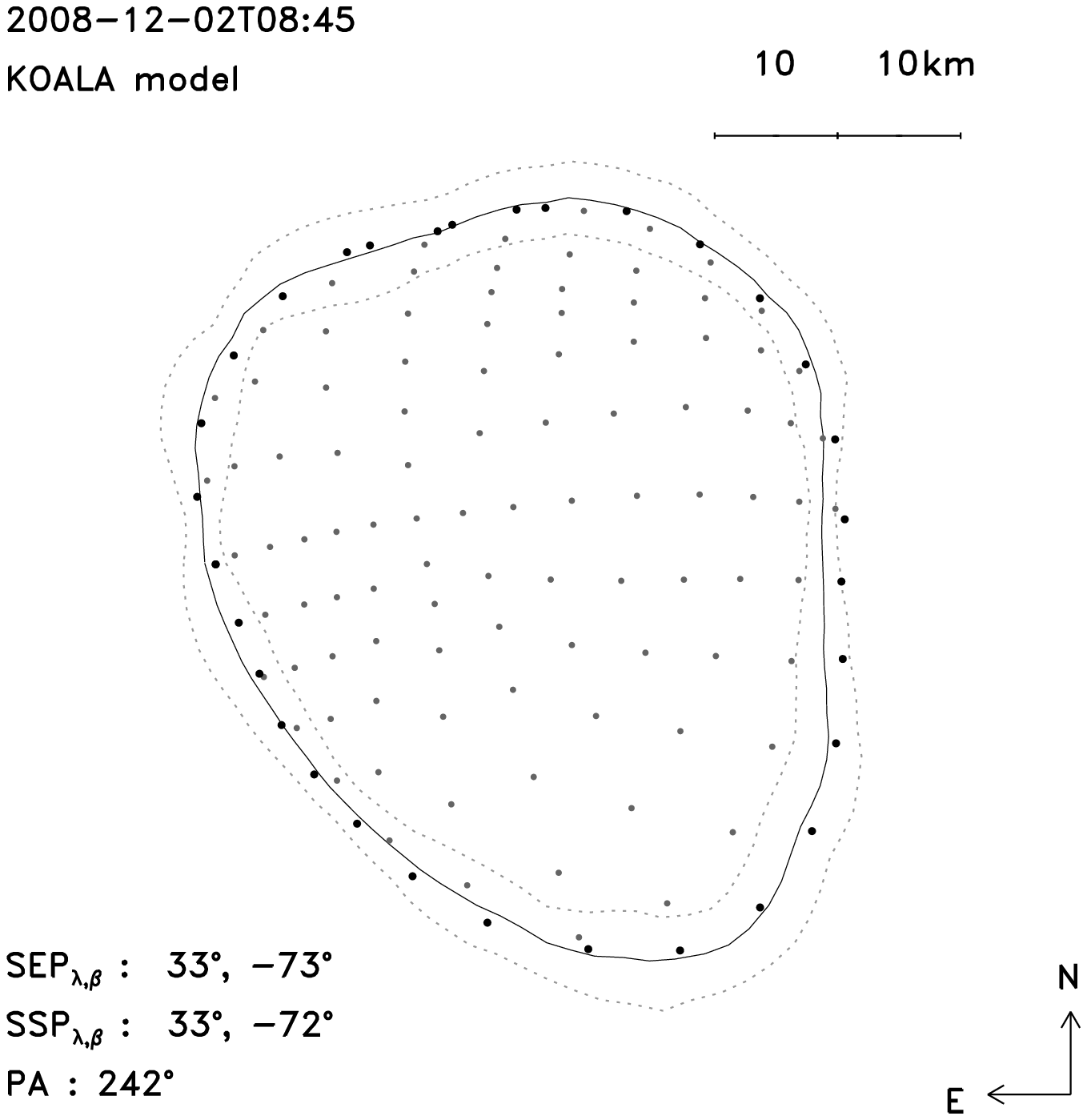}&
    \includegraphics[width=.3\textwidth]{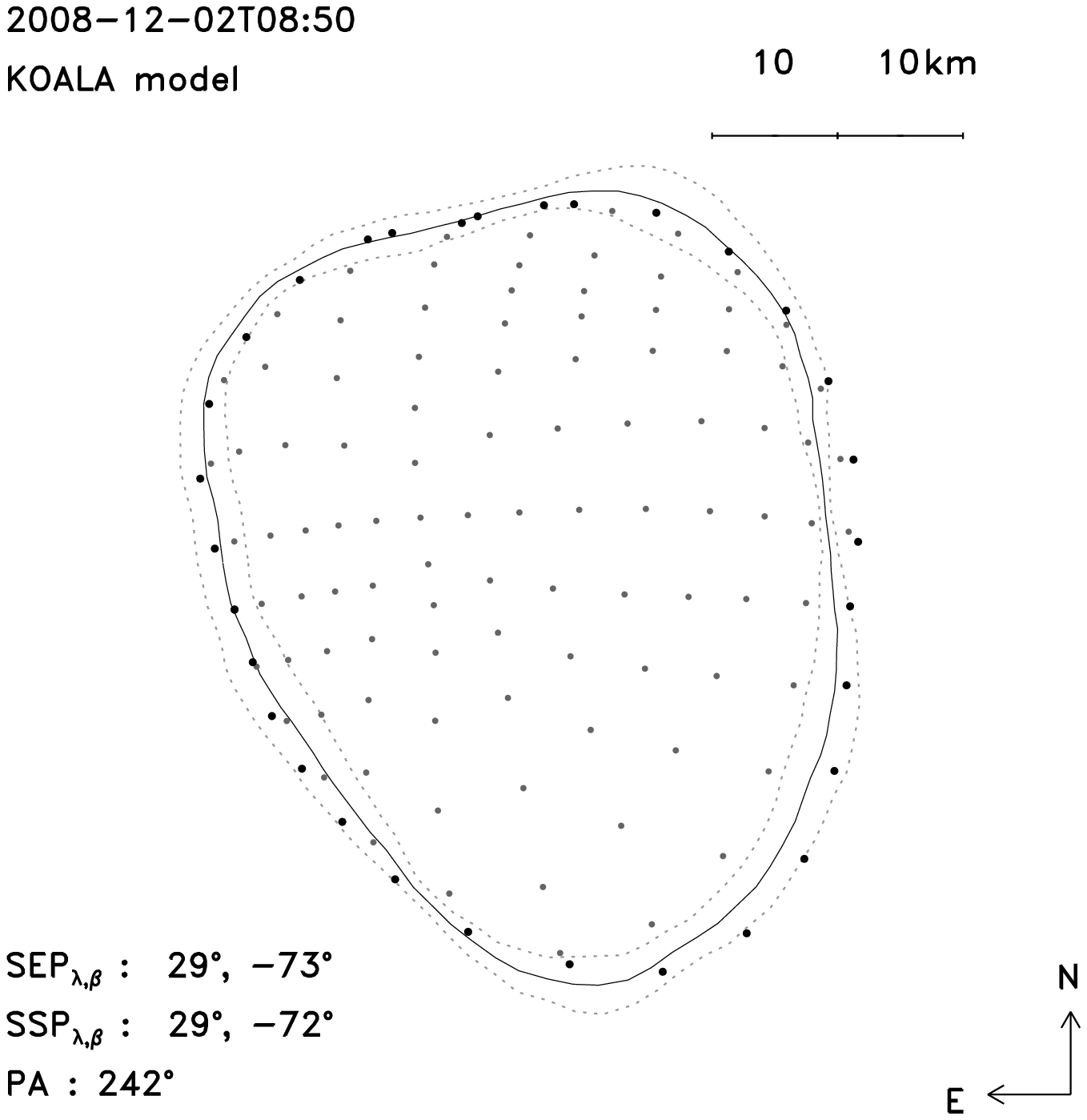}\\
    \includegraphics[width=.3\textwidth]{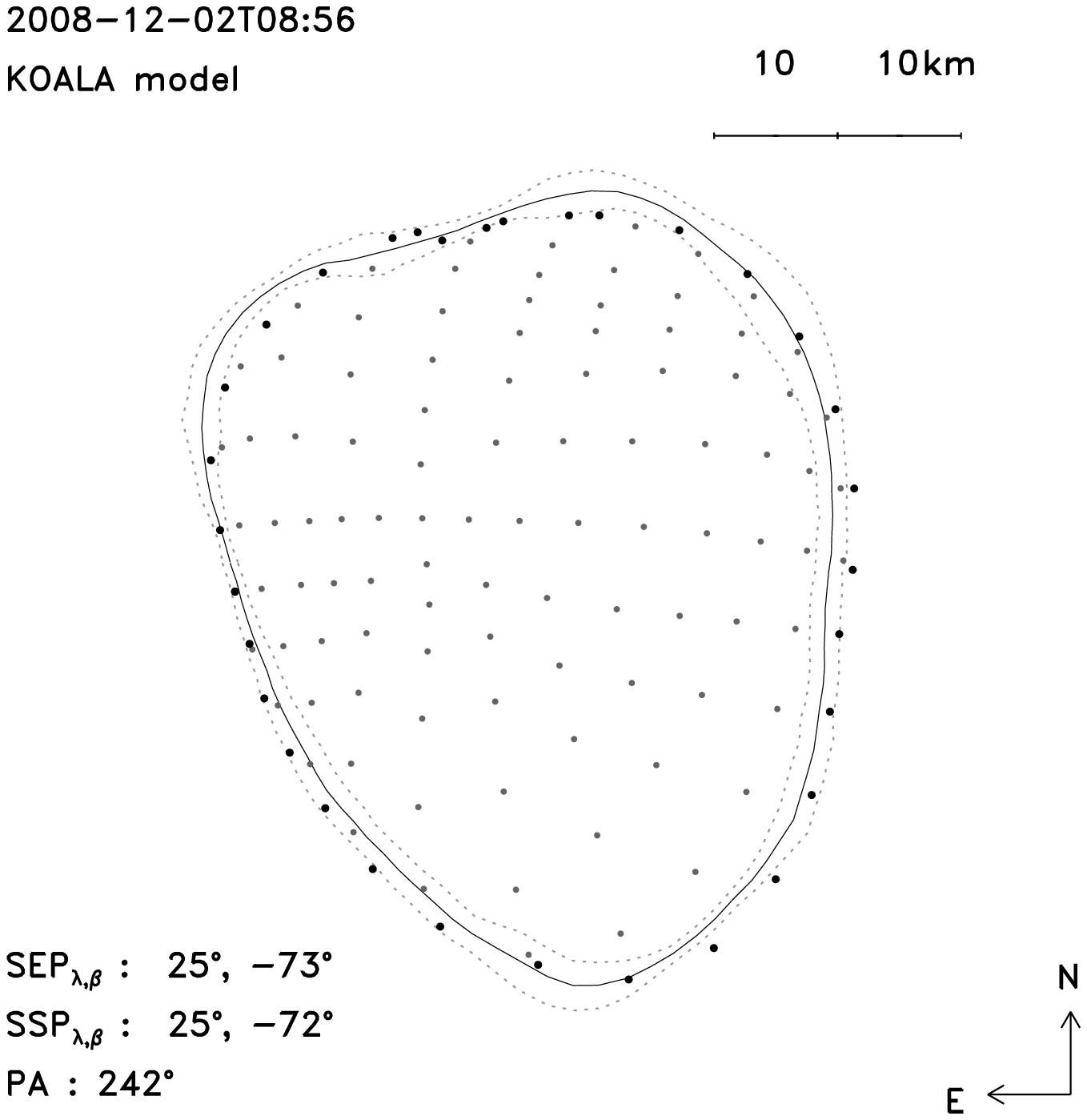}&
    \includegraphics[width=.3\textwidth]{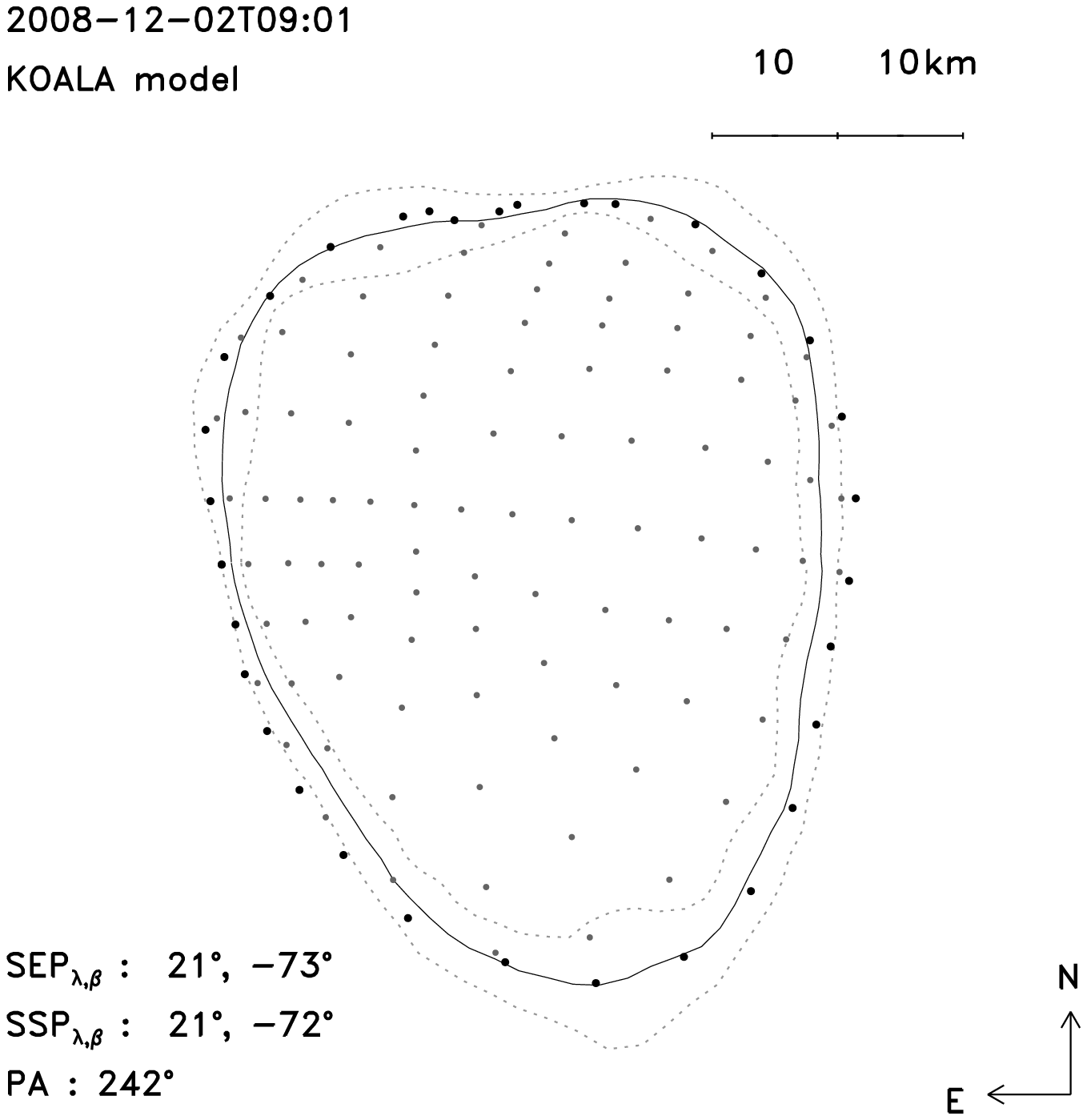}&
    \includegraphics[width=.3\textwidth]{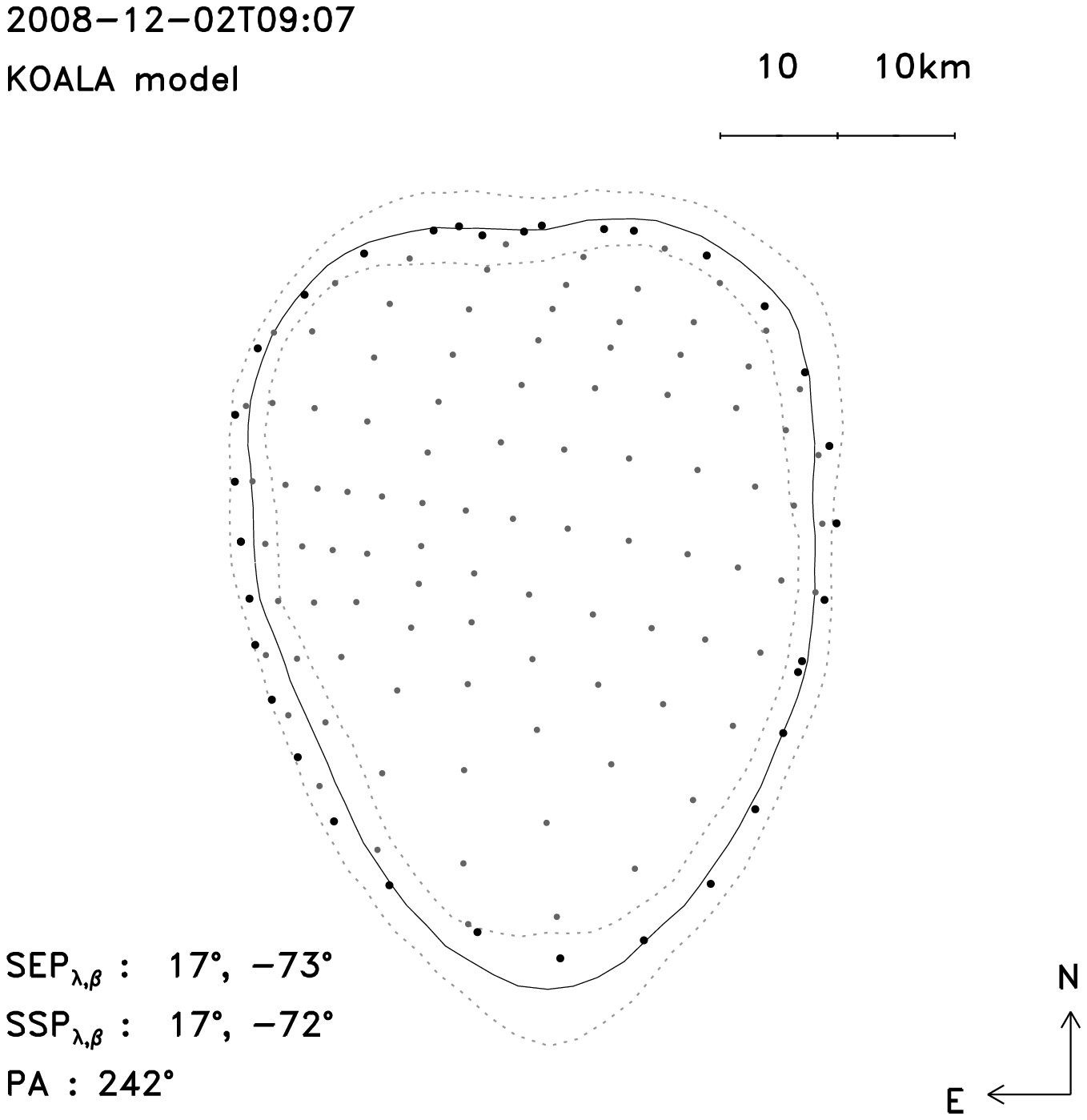}\\
    \includegraphics[width=.3\textwidth]{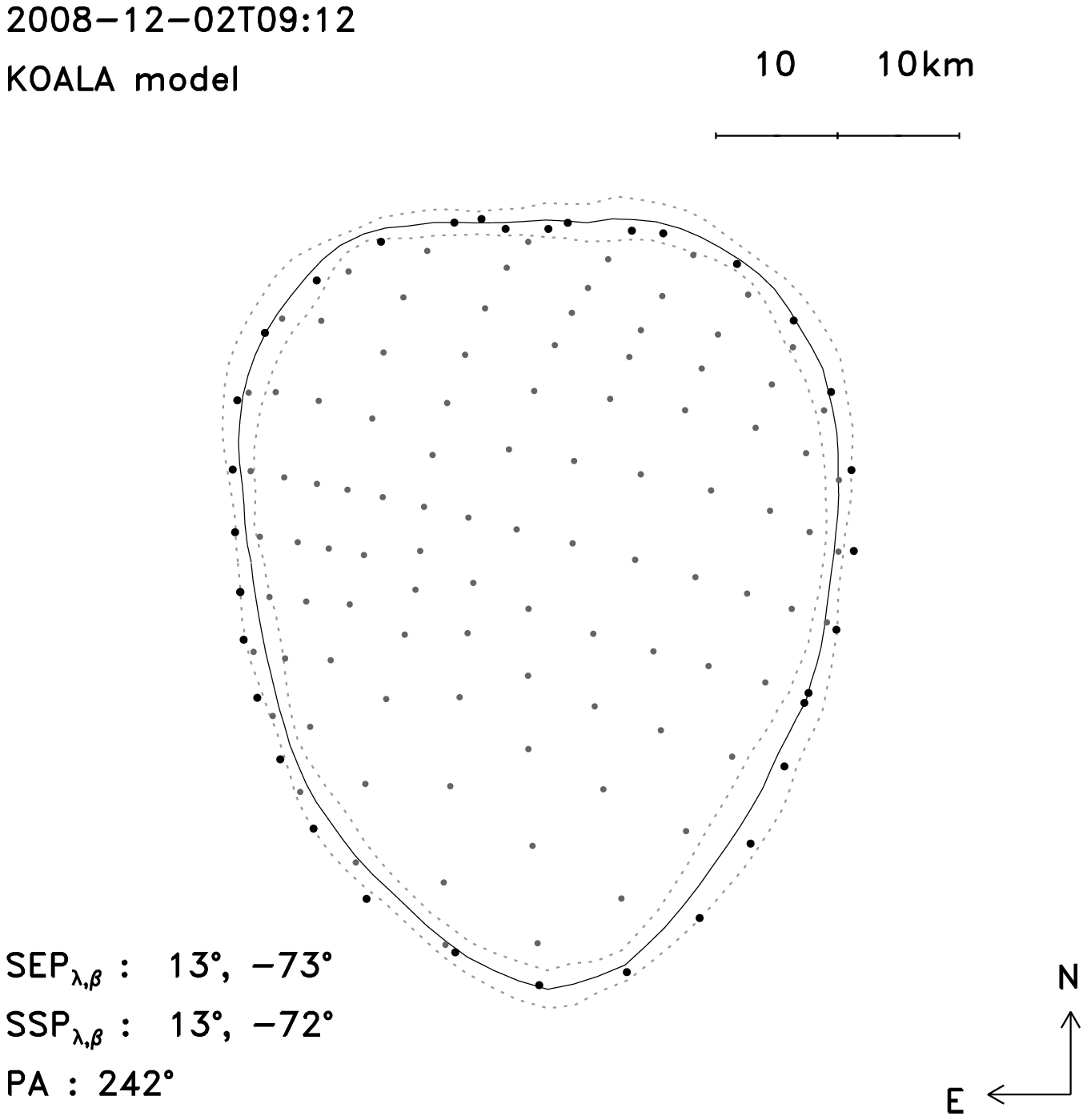}&
    \includegraphics[width=.3\textwidth]{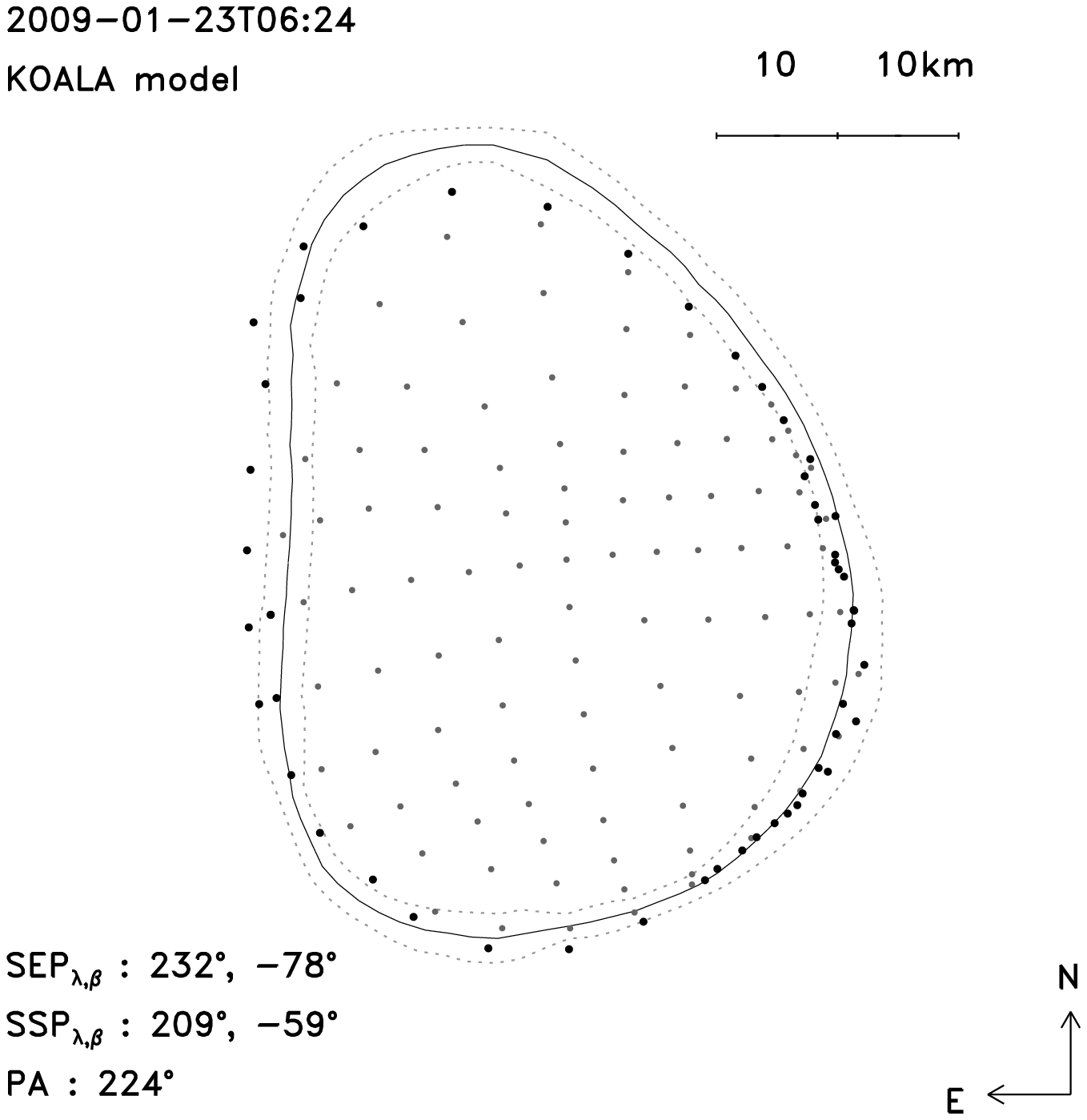}&
    \includegraphics[width=.3\textwidth]{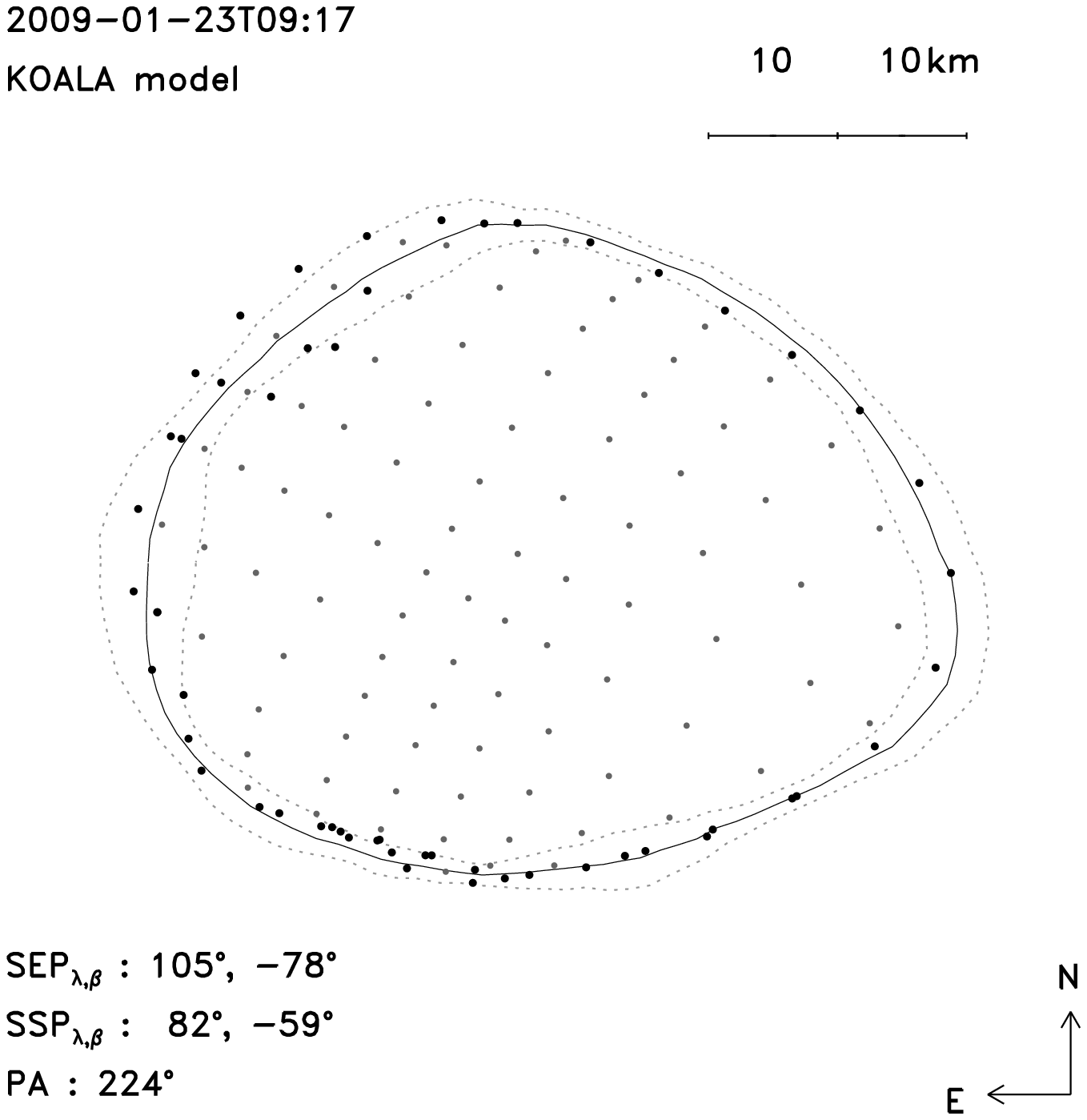}\\
    \includegraphics[width=.3\textwidth]{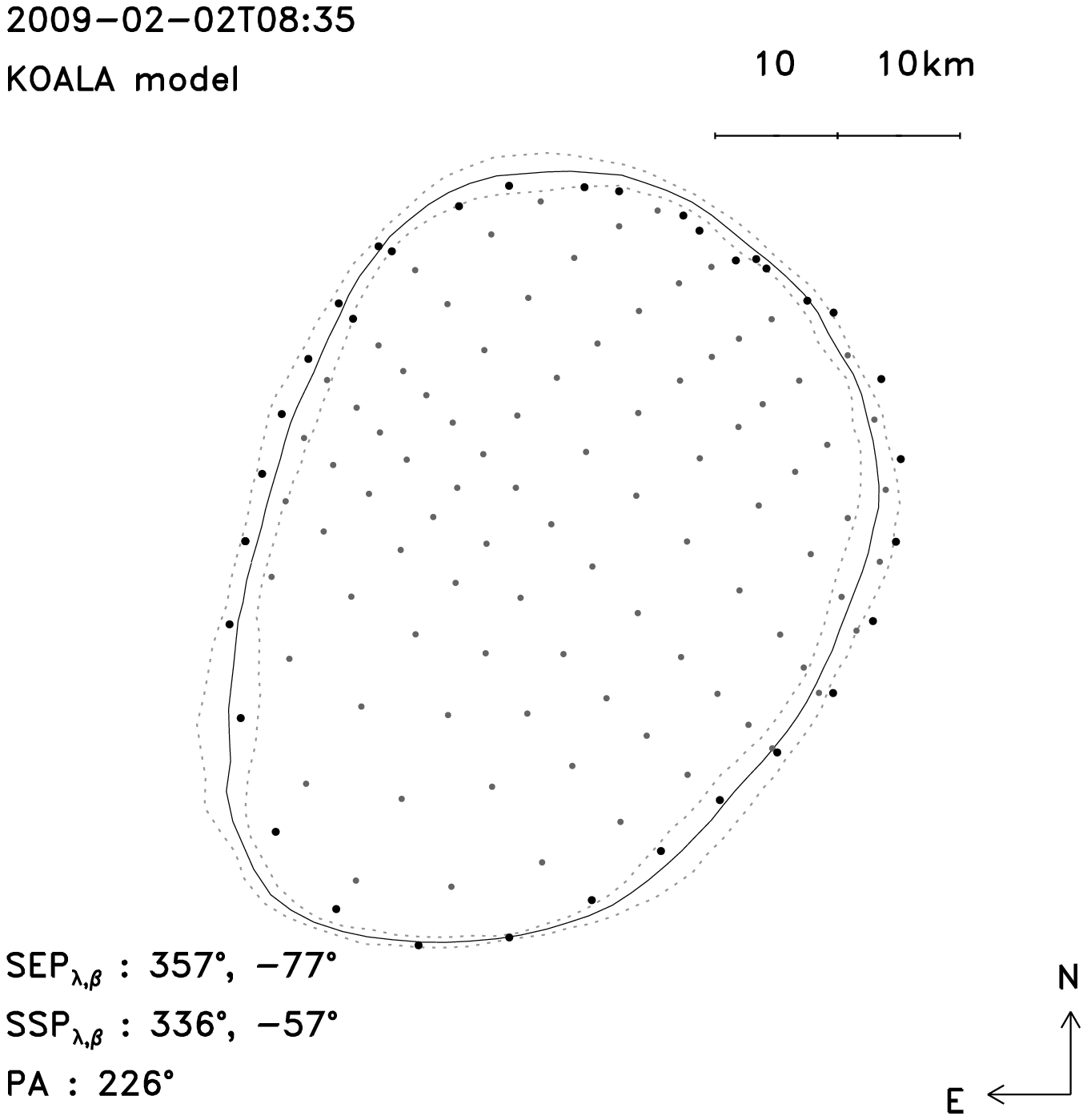}&
    \includegraphics[width=.3\textwidth]{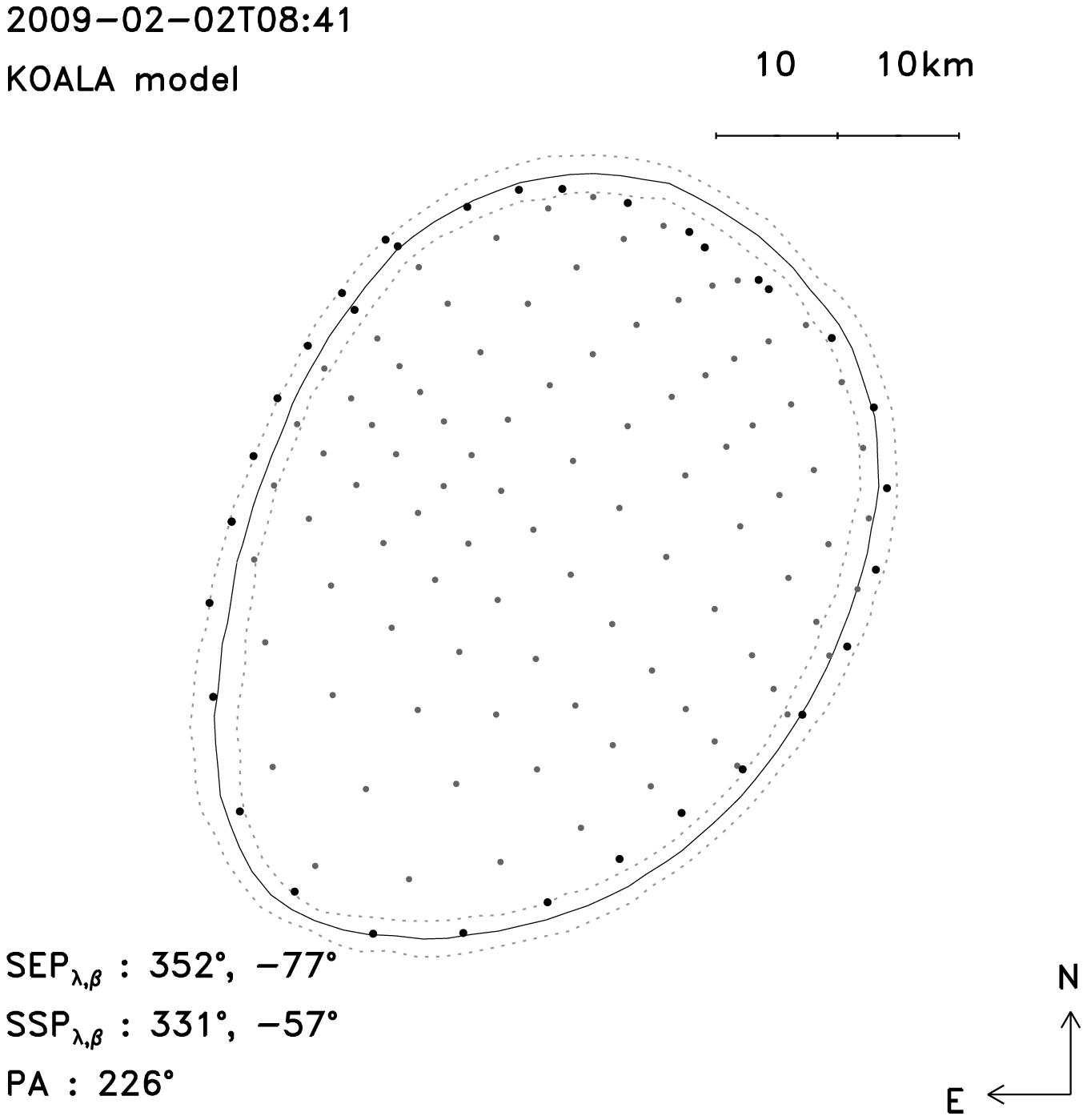}&
    \includegraphics[width=.3\textwidth]{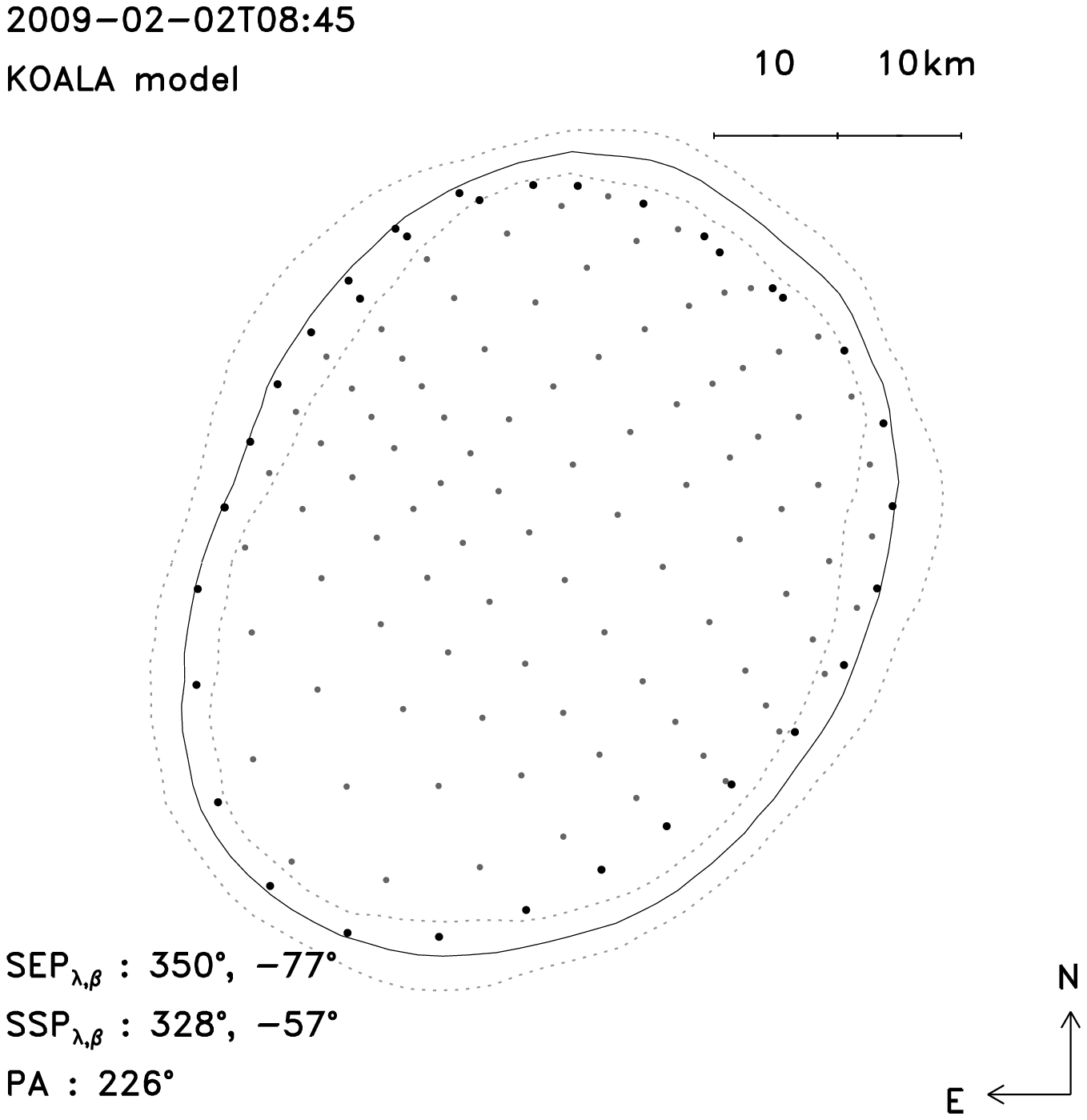}
    \end{tabular}
    }
\end{center}
  \label{fig: comp: ao3}

\end{figure*}

\begin{figure*}[!t]
\begin{center}
  \setcounter{subfigure}{1}
    \centering
    \subfigure[Second set of lightcurves (1981--1985)]{%
      \includegraphics[width=.8\textwidth]{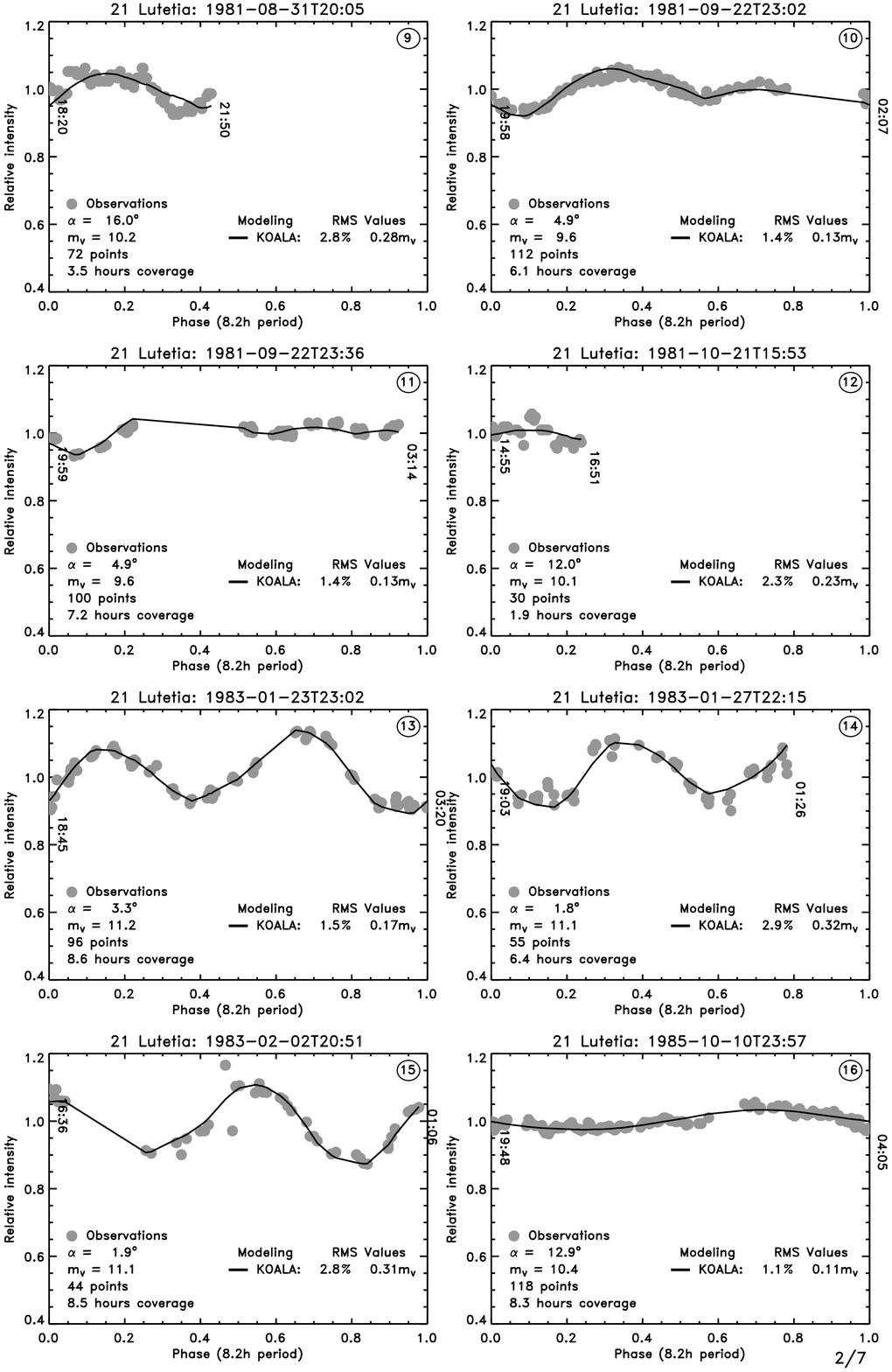}
    }%
\end{center}
\end{figure*}

\begin{figure*}[!t]
\begin{center}
  \setcounter{subfigure}{2}
    \centering
    \subfigure[Third set of lightcurves (1985--1991)]{%
      \includegraphics[width=.8\textwidth]{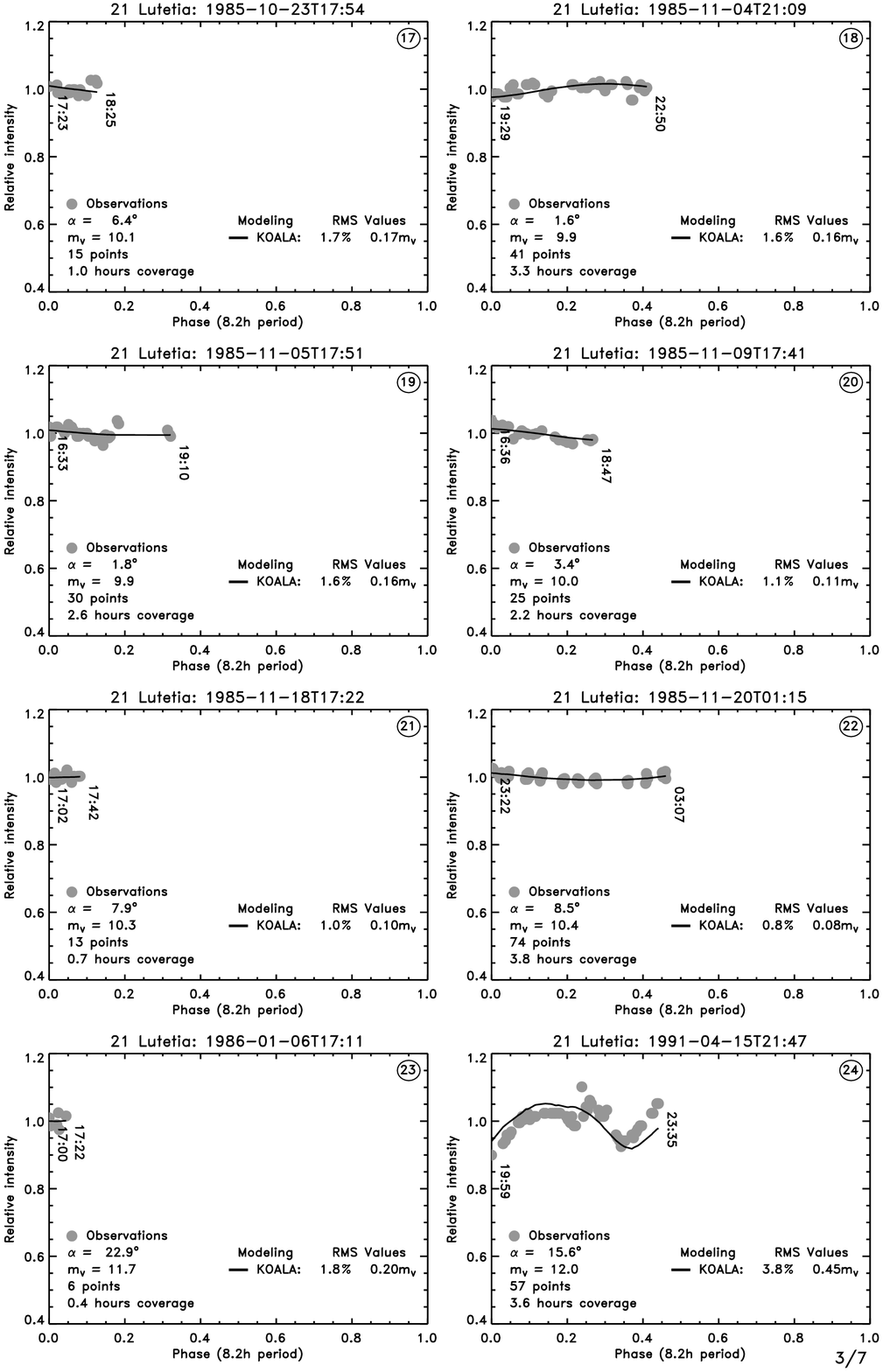}
    }%
\end{center}
\end{figure*}

\begin{figure*}[!t]
\begin{center}
  \setcounter{subfigure}{3}
    \centering
    \subfigure[Fourth set of lightcurves (1995--1998)]{%
      \includegraphics[width=.8\textwidth]{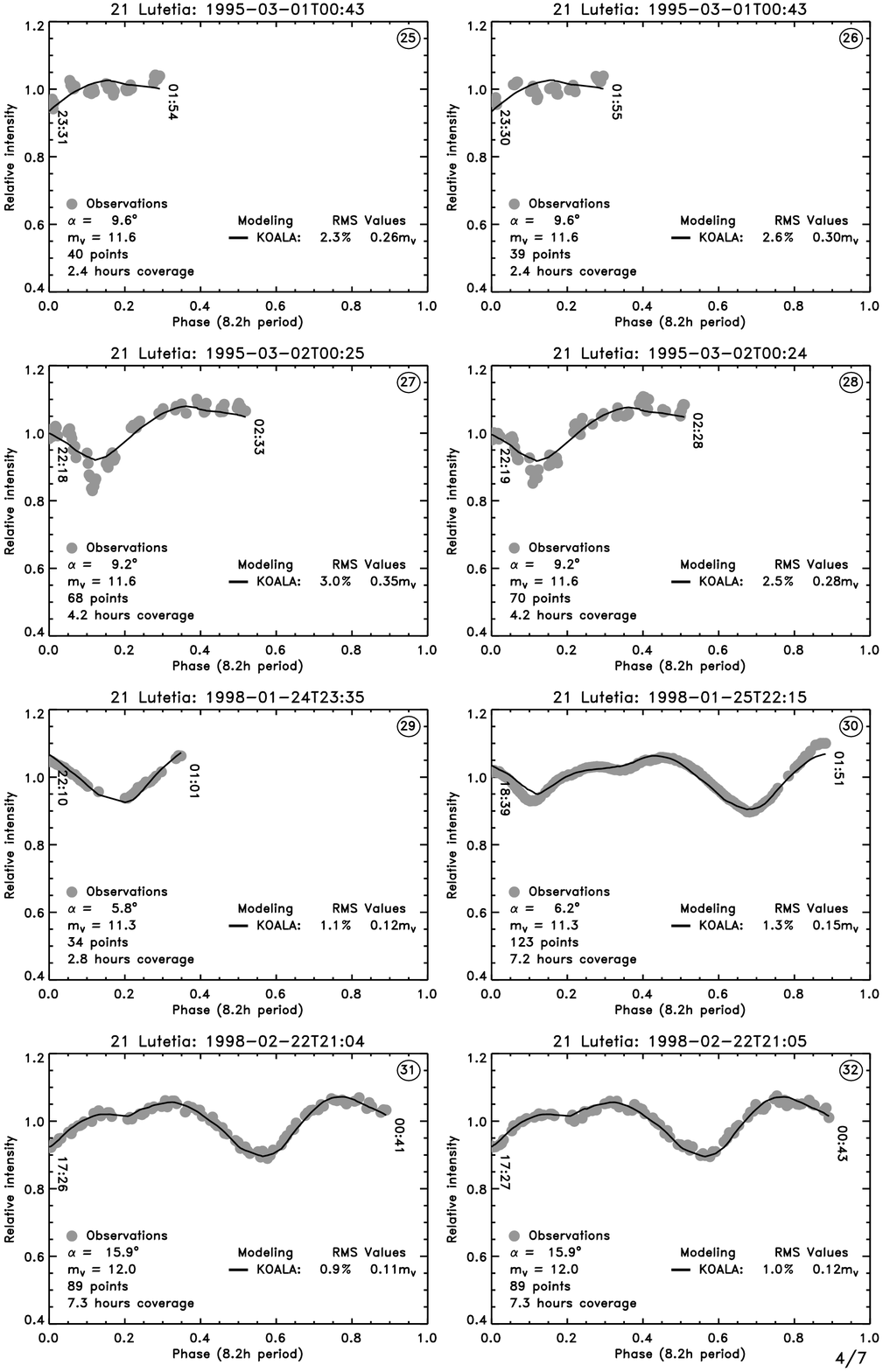}
    }%
\end{center}
\end{figure*}

\begin{figure*}[!t]
\begin{center}
  \setcounter{subfigure}{4}
    \centering
    \subfigure[Fifth set of lightcurves (2003--2006)]{%
      \includegraphics[width=.8\textwidth]{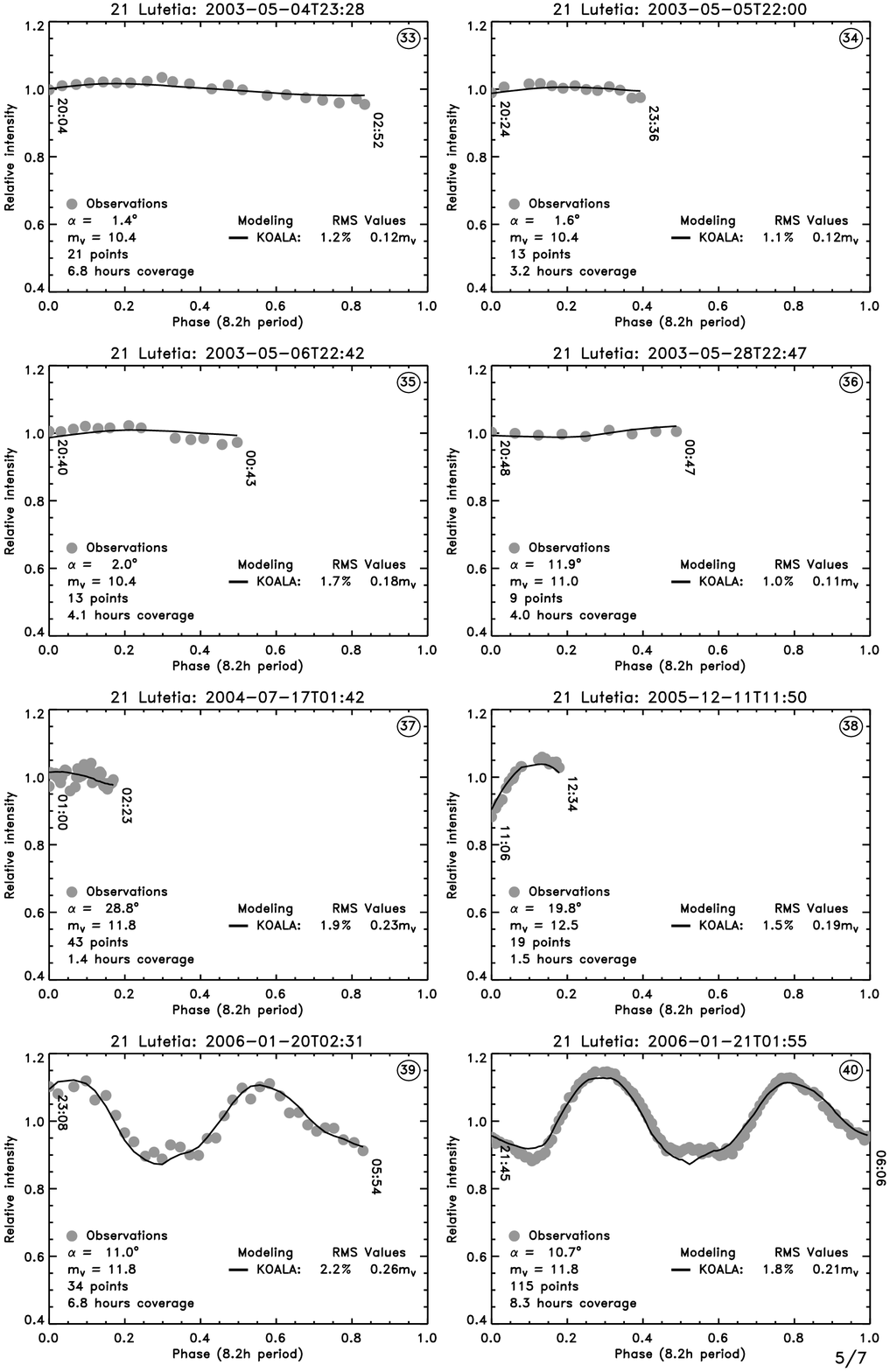}
    }%
\end{center}
\end{figure*}

\begin{figure*}[!t]
\begin{center}
  \setcounter{subfigure}{5}
    \centering
    \subfigure[Sixth set of lightcurves (2007-2010)]{%
      \includegraphics[width=.8\textwidth]{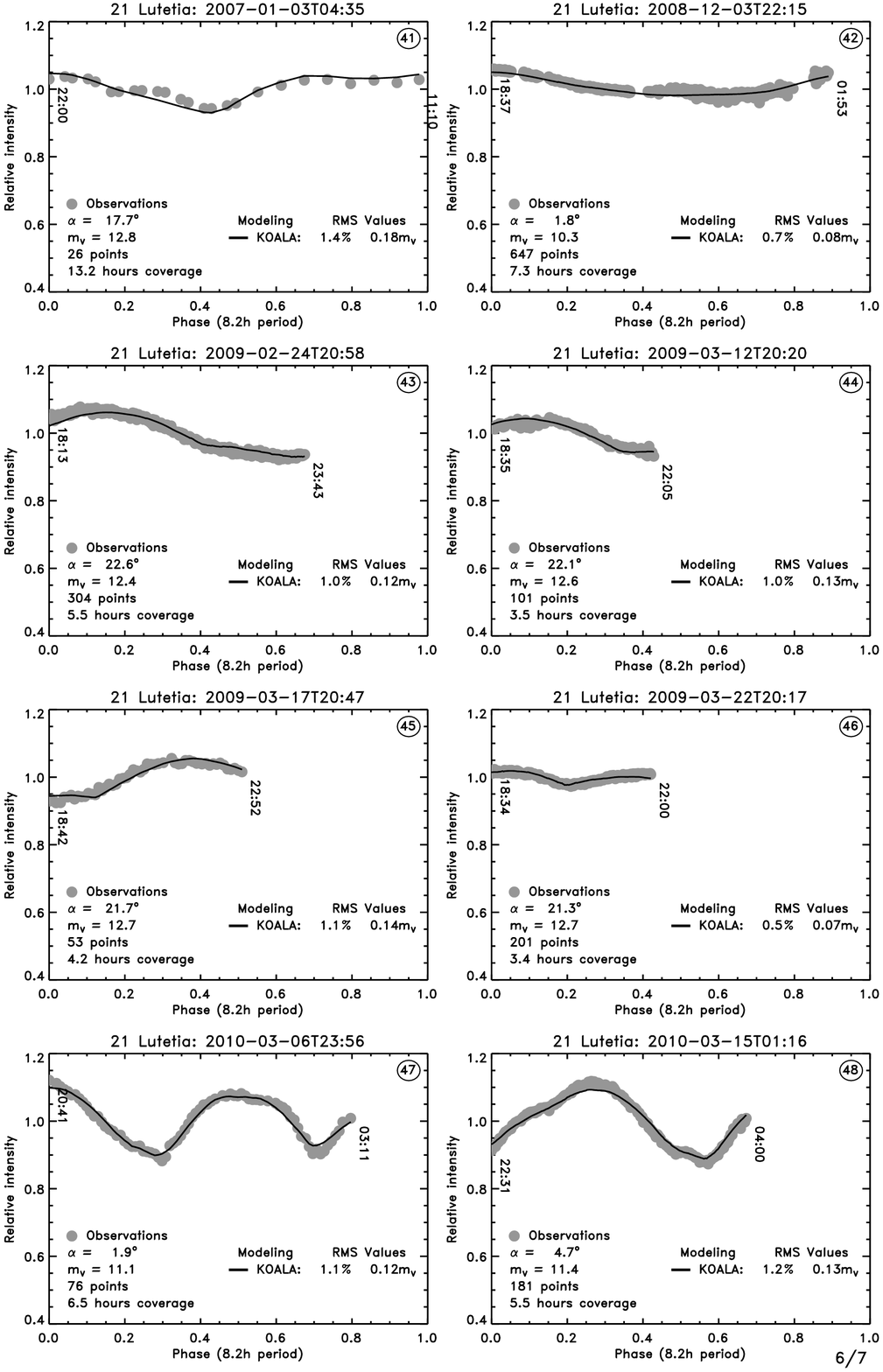}
    }%
\end{center}
\end{figure*}

\begin{figure*}[!t]
\begin{center}
  \setcounter{subfigure}{5}
    \centering
    \subfigure[Seventh set of lightcurves (2010)]{%
      \includegraphics[width=.8\textwidth]{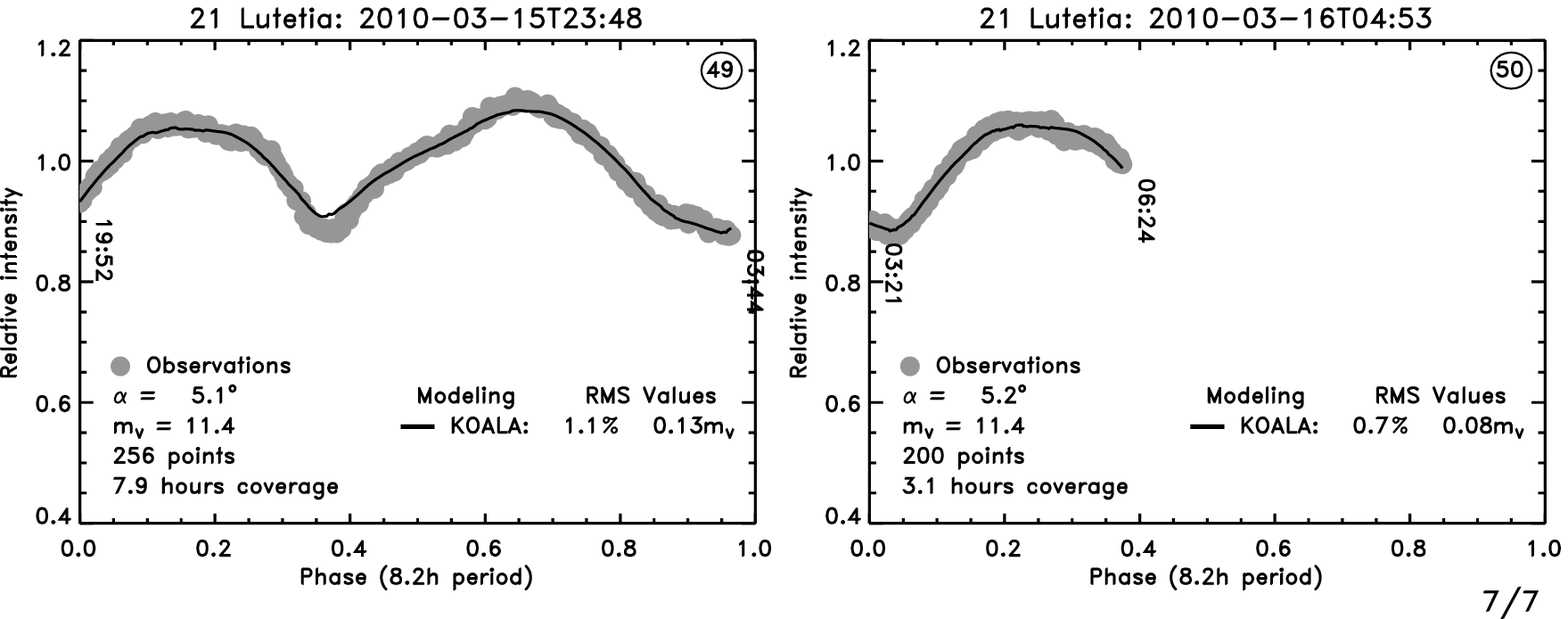}
    }%
\end{center}
\end{figure*}

\begin{table*}
\caption{Characteristics of the pole solution obtained with KOALA.
  The pole solution is in agreement
  with the purely triaxial-ellipsoid
  results (7\degr~of arc difference)
  presented by \citet{2010-AA--Drummond}.
  We list both $W_0$ and $t_0$ to orient the shape model in space at
  any time from the IAU definition 
  \citep[see][]{2007-CeMDA-98-Seidelmann} or the definition usually used in
  inversion techniques \citep[following][]{2001-Icarus-153-Kaasalainen-b}
}
\label{tab: spin}
\centering
\begin{tabular}{ccccc}
\hline\hline
 $P_s$ & ECJ2000   & EQJ2000 &  $W_0$   &   $t_0$\\
  (h)  & ($\lambda_0$,$\beta_0$ in \degr) & ($\alpha_0$, $\delta_0$ in \degr) &  (\degr) &  (JD)\\
\hline
8.168\,270 $\pm$ 0.000\,001 &
(52,-6) $\pm$ 5 &
(52,12) $\pm$ 5 &
94 $\pm$ 2 &
2444822.35116 \\
\hline
\end{tabular}
\end{table*}

\section{Rosetta flyby of (21) Lutetia\label{sec: flyby}}
\begin{figure*}
  \centering
  \includegraphics[width=.95\textwidth]{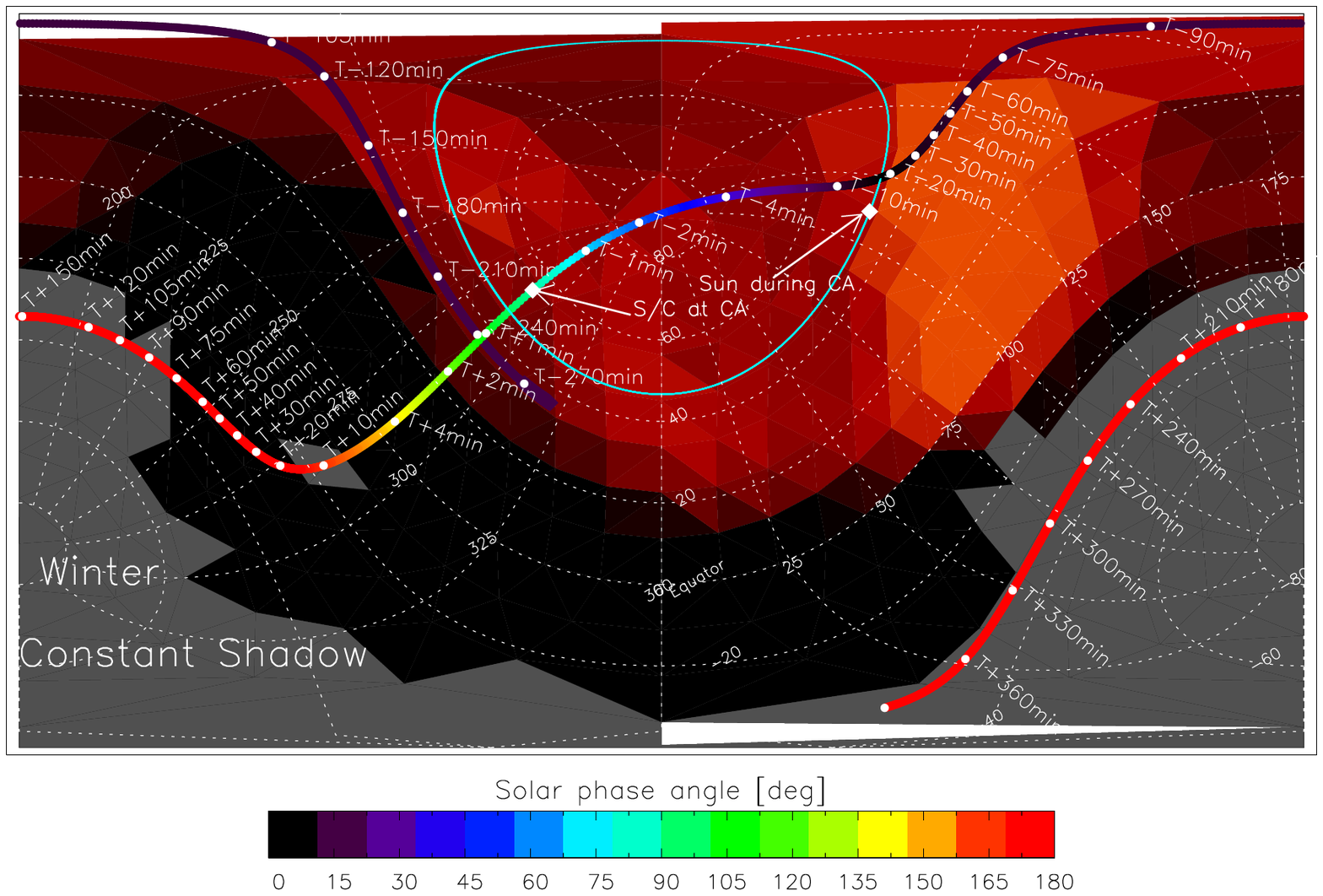}
  \caption[flyby geometry]{%
  Oblique Mercator projection of the
  Sub-Rosetta Point (SRP) and Sub-Solar Point (SSP)
  paths during the Lutetia encounter on 2010 July 10 by the Rosetta
  spacecraft.
  The grey area near the South pole represents surface
  points where the Sun is never above the local horizon at the
  encounter epoch (constant shadow area).
  The reddish shades on the surface give the
  local illumination conditions at closest
  approach (CA), with the equatorial black
  band corresponding to night time at CA.
  Brighter shades of red depict a smaller local
  solar incidence angle (Sun high in sky), while darker
  shades represent a larger solar incidence angle.
  For flyby imaging, crater measurements will be much better
  in regions of low sun (high incidence angle), while
  albedo/color will be better discernible at high sun.
  The thin blue line is the SSP path, with the
  Sun traversing this path east-to-west on
  Lutetia's surface.
  The location of the
  SRP with time (thick, multi-colored line)
  is color-coded in phase angle (see Table~\ref{tab: flyby}
  for a detailed listing of the path coordinates
  as a function of time). Positions of the
  SSP and SRP at CA are labeled for convenience.
  The actual estimate of the CA time is
  15:44 UT, but it may vary by a few tens of
  seconds, depending on trajectory-correction
  maneuvers that are applied to the spacecraft
  before the encounter. Thus, we provide
  times relative to CA, indicated in minutes.
  }
  \label{fig: flyby}
\end{figure*}

  \indent Finally, we investigate the regions of Lutetia that will be
  observed by Rosetta during the upcoming flyby on 2010 July
  10.
  We used the shape model and spin solution
  described in
  section~\ref{sec: shape}
  and the spacecraft trajectory (obtained using
  the most recent
  spice
  kernels)
  to derive the relative
  position (SPK\footnote{ORHR \& ORGR \#00091}) and
  orientation (PCK\footnote{personal kernel with spin solution from
    section~\ref{sec: shape}})
  of Rosetta and Lutetia.
  This provides
  the relative distance between Rosetta and (21) Lutetia,
  the coordinates of the Sub-Rosetta Point (SRP) and
  Sub-Solar Point (SSP),
  the illuminated fraction of Lutetia surface,
  and
  the Solar phase angle as function of time.\\
  \indent At the time of the flyby,
  the northern hemisphere will be in constant sunlight
  (SSP$_\beta$ will be +52\degr),
  while regions below -35\degr~latitude will be in a constant shadow
  (see Table~\ref{tab: flyby} and
  Fig.~\ref{fig: flyby}).
  Therefore,
  extreme southern latitudes of Lutetia
  will not be observable from Rosetta
  in optical wavelengths,
  preventing precise
  shape reconstruction of the southern regions.
  Therefore, size determination along the rotation axis will probably
  have to rely on thermal observations conducted with MIRO
  \citep{2007-SSRv-128-Gulkis} 
  (the observation plan for the flyby
  includes a slew along the shadowed
  regions of the asteroid).\\ 
  \indent Rosetta will approach Lutetia with a SRP$_\beta$ close
  to +48\degr, and a nearly constant phase angle of $\sim$10\degr,
  observing Lutetia as it rotates around its spin axis.
  The solar phase angle will
  then decrease slowly while SRP$_\beta$ will increase.
  The lowest solar phase angle (0.7\degr) will
  occur at 1040 seconds (17min) before
  closest approach (CA).
  A few minutes before CA, the spacecraft will fly over the North pole at a
  maximum latitude of about +84\degr,
  allowing the putative large-scale depression
  reported here to be observed.
  CA will then occur at 79\degr~phase angle over
  +48\degr~latitude, close to the terminator.
  At that time, the  relative speed between Rosetta and Lutetia
  will be about 15 km/s and the distance will reach its
  minimum at 3063 km. This implies an apparent size of Lutetia of
  about 2 degrees at CA, which corresponds approximately to
  the field of view of the 
  Narrow Angle Camera (NAC) of the OSIRIS instrument
  \citep{2007-SSRv-128-Keller}.\\
  \indent The SRP will then move rapidly into the Southern
  hemisphere. A few tens of seconds after CA, the day-to-night thermal
  transition will be observed between latitudes +30\degr~and
  +40\degr,
  over 280\degr~longitude, at rapidly increasing phase angles.
  One hour after CA, the SRP will finally enter into the ``\textsl{seasonal}''
  shadow area between -20\degr~and -40\degr~latitude, at very high
  phase angles ($\geq 150$\degr).
  Differences in the thermal emissions coming from
  both regions (night and winter) should be detectable with MIRO
  \citep{2007-SSRv-128-Gulkis}.
  The distance will then increase
  rapidly while the phase angle will reach an almost constant value of
  about 170\degr.\\

\begin{table*}
\caption[Flyby Circumstances]{%
  Sub-Rosetta Point (SRP) coordinates (longitude $\lambda$, latitude
  $\beta$), 
  Sub-Solar Point (SSP) coordinates and 
  phase angle ($\alpha$) as a function of
  the nominal flyby schedule (UT time $t$) and 
  relatively to the instant of Closest Approach (CA), listed in bold. 
}
\label{tab: flyby}
\centering
\begin{tabular}{crrrrrrr}
\hline
  \multicolumn{1}{c}{Time $t$} &
  \multicolumn{1}{c}{$t$-CA} &
  \multicolumn{1}{c}{Distance} &
  \multicolumn{1}{c}{SRP$_\lambda$} &
  \multicolumn{1}{c}{SRP$_\beta$} &
  \multicolumn{1}{c}{SSP$_\lambda$} &
  \multicolumn{1}{c}{SSP$_\beta$} &
  \multicolumn{1}{c}{$\alpha$} \\
  \multicolumn{1}{c}{(UT)} &
  \multicolumn{1}{c}{(min)} &
  \multicolumn{1}{c}{(km)} &
  \multicolumn{1}{c}{(\degr)} &
  \multicolumn{1}{c}{(\degr)} &
  \multicolumn{1}{c}{(\degr)} &
  \multicolumn{1}{c}{(\degr)} &
  \multicolumn{1}{c}{(\degr)} \\
\hline
\hline
  11:14:35 & -270 & 242\,906 & 309.5 &  36.2 & 310.1 & 46.6 &  10.4 \\
  11:44:35 & -240 & 215\,918 & 287.5 &  36.3 & 288.1 & 46.6 &  10.3 \\
  12:14:35 & -210 & 188\,930 & 265.4 &  36.5 & 266.0 & 46.6 &  10.2 \\
  12:44:35 & -180 & 161\,941 & 243.4 &  36.6 & 244.0 & 46.6 &  10.0 \\
  13:14:35 & -150 & 134\,953 & 221.4 &  36.8 & 221.9 & 46.6 &   9.8 \\
  13:44:35 & -120 & 107\,968 & 199.3 &  37.2 & 199.9 & 46.6 &   9.5 \\
  13:59:35 & -105 &  94\,477 & 188.3 &  37.4 & 188.9 & 46.6 &   9.2 \\
  14:14:35 &  -90 &  80\,990 & 177.3 &  37.7 & 177.9 & 46.6 &   8.9 \\
  14:29:35 &  -75 &  67\,506 & 166.3 &  38.1 & 166.9 & 46.6 &   8.5 \\
  14:44:35 &  -60 &  54\,028 & 155.3 &  38.8 & 155.8 & 46.6 &   7.9 \\
  14:54:35 &  -50 &  45\,049 & 148.0 &  39.4 & 148.5 & 46.6 &   7.2 \\
  15:04:35 &  -40 &  36\,079 & 140.7 &  40.4 & 141.1 & 46.6 &   6.2 \\
  15:14:35 &  -30 &  27\,126 & 133.4 &  42.0 & 133.8 & 46.6 &   4.6 \\
  15:24:35 &  -20 &  18\,216 & 126.1 &  45.2 & 126.5 & 46.6 &   1.5 \\
  15:34:35 &  -10 &   9\,468 & 119.1 &  54.3 & 119.1 & 46.6 &   7.7 \\
  15:40:35 &   -4 &   4\,689 & 116.9 &  75.8 & 114.7 & 46.6 &  29.2 \\
  15:42:35 &   -2 &   3\,513 & 282.3 &  85.0 & 113.2 & 46.6 &  48.3 \\
  15:43:35 &   -1 &   3\,150 & 288.8 &  71.0 & 112.5 & 46.6 &  62.3 \\
  {\bf 15:44:35} &{\bf CA} &{\bf 3\,016} & {\bf 289.4} & {\bf 54.7} & {\bf 111.8} & {\bf 46.6} & {\bf 78.7} \\
  15:45:35 &    1 &   3\,140 & 289.2 &  38.3 & 111.0 & 46.6 &  95.1 \\
  15:46:35 &    2 &   3\,496 & 288.7 &  24.2 & 110.3 & 46.6 & 109.2 \\
  15:48:35 &    4 &   4\,666 & 287.5 &   5.0 & 108.8 & 46.6 & 128.4 \\
  15:54:35 &   10 &   9\,443 & 283.5 & -16.7 & 104.4 & 46.6 & 150.1 \\
  16:04:35 &   20 &  18\,192 & 276.3 & -25.9 &  97.1 & 46.6 & 159.2 \\
  16:14:35 &   30 &  27\,104 & 269.0 & -29.1 &  89.7 & 46.6 & 162.4 \\
  16:24:35 &   40 &  36\,058 & 261.7 & -30.7 &  82.4 & 46.6 & 164.0 \\
  16:34:35 &   50 &  45\,029 & 254.4 & -31.6 &  75.0 & 46.6 & 165.0 \\
  16:44:35 &   60 &  54\,009 & 247.0 & -32.3 &  67.7 & 46.6 & 165.6 \\
  16:59:35 &   75 &  67\,486 & 236.0 & -32.9 &  56.7 & 46.6 & 166.3 \\
  17:14:35 &   90 &  80\,969 & 225.0 & -33.4 &  45.6 & 46.6 & 166.7 \\
  17:29:35 &  105 &  94\,455 & 214.0 & -33.7 &  34.6 & 46.6 & 167.0 \\
  17:44:35 &  120 & 107\,944 & 203.0 & -33.9 &  23.6 & 46.7 & 167.2 \\
  18:14:35 &  150 & 134\,925 & 181.0 & -34.2 &   1.6 & 46.7 & 167.6 \\
  18:44:35 &  180 & 161\,912 & 158.9 & -34.4 & 339.5 & 46.7 & 167.8 \\
  19:14:35 &  210 & 188\,903 & 136.9 & -34.6 & 317.5 & 46.7 & 167.9 \\
  19:44:35 &  240 & 215\,894 & 114.9 & -34.7 & 295.5 & 46.7 & 168.0 \\
  20:14:35 &  270 & 242\,885 &  92.8 & -34.8 & 273.4 & 46.7 & 168.1 \\
  20:44:35 &  300 & 269\,874 &  70.8 & -34.9 & 251.4 & 46.7 & 168.2 \\
  21:14:35 &  330 & 296\,862 &  48.7 & -34.9 & 229.3 & 46.7 & 168.2 \\
  21:44:35 &  360 & 323\,849 &  26.7 & -35.0 & 207.3 & 46.7 & 168.3 \\
\hline
\end{tabular}
\end{table*}

\section{Conclusions\label{sec: conclu}}
  \indent We have reported disk-resolved imaging
  observations of (21) Lutetia obtained with the W. M. Keck
  and Very Large Telescope
  observatories in 2007, 2008, and 2009.
  We have derived the shape and spin of (21) Lutetia using the
  Knitted Occultation, Adaptive-optics, and Lightcurve Analysis
  (KOALA) method, which is based on combining these AO images with
  optical lightcurves gathered from over four decades.\\
  \indent The shape of (21) Lutetia is well described by a Camembert
  wedge, and our shape model suggests the presence of several
  concavities near its 
  north pole and around its equator.
  The spin axis of Lutetia
  is tilted with respect to its orbital plane, much like Uranus,
  implying strong seasonal effects on
  its surface.
  At the time of the Rosetta flyby, (21) Lutetia's northern hemisphere
  will be illuminated while the southern hemisphere will be in
  long-term darkness, hindering the size determination from Rosetta.\\
  \indent The next opportunity to observe Lutetia's shortest dimension,
  impacting its volume determination,
  will occur in 2011 July, one year after Rosetta flyby, when the sub-Earth point 
  will be close to its equator (SEP$_\beta$ of +31\degr).
  During this time, observations using
  large telescopes equipped with adaptive-optics will allow refinement of Lutetia's
  short dimension and thus improve the
  volume determination.
  This ground-based support will be essential to take advantage of the
  high-precision mass determination provided by the spacecraft
  deflection observed during the flyby.

\section*{Acknowledgments}
  This research has made use of IMCCE's Miriade VO tool and 
  NASA's Astrophysics Data System.
  The authors wish to recognize and acknowledge the very significant
  cultural role and reverence that the summit of Mauna Kea has always
  had within the indigenous Hawaiian community. We are most fortunate
  to have the opportunity to conduct observations from this mountain.
  This work was supported, in part,
  by the NASA Planetary Astronomy and NSF
  Planetary Astronomy Programs (Merline PI).
  We are grateful for telescope time made
  available to us by S. Kulkarni and M. Busch
  (Cal Tech) for a portion of this dataset.  We
  also thank our collaborators on Team Keck,
  the Keck science staff, for making possible
  some of these observations, and for observing
  time granted at Gemini Observatory under NOAO time
  allocation, as part of our
  overall Lutetia campaign.

  \bibliographystyle{aa}
  \bibliography{biblio}

\end{document}